\documentclass[]{article}
\usepackage{graphicx}
\usepackage{color}
\usepackage{amsfonts}
\def\ligne#1{\hbox to\hsize{#1}}
\def\leurre{\noindent\leftskip0pt\small\baselineskip 10pt}
\newtheorem{thm}{\textrm{\sc Theorem}}

\newtheorem{fig}{\textrm{Figure}}
\newtheorem{tab}{\textrm{Table}}
\def\boxempty{\hbox{\vbox{\hsize=7pt\offinterlineskip
\ligne{
\vrule height 7pt depth 0pt width 0.6pt
\vbox to 7pt{\hsize=5.8pt
\hrule height 0pt depth 0.6pt width 5.8pt
\vfill
\hrule height 0.6pt depth 0pt width 5.8pt
}\hskip-0.5pt
\vrule height 7pt depth 0pt width 0.6pt
}}
}}
\def\trep{\hrule height 1pt depth 1pt width \hsize}
\def\trfn{\hrule height 0.5pt depth 0.5pt width \hsize}
\newcounter{laform}
\author{Maurice {\sc Margenstern}}
\title{A weakly universal universal cellular automaton on the tessellation $\{9,3\}$.}
\begin{document}
\maketitle

\begin{abstract}
In this paper, we construct a weakly universal cellular automaton on the tessellation
$\{9,3\}$ which is not rotation invariant but which is truly planar. 
\end{abstract}

\section{Introduction}

   In~\cite{mmarXiv1605}, I indicate a new system of coordinates for the tilings
$\{p,3\}$ and $\{p$$-$$2,4\}$. In this paper, we apply this system in order to
construct a weakly universal cellular on the tessellation $\{9,3\}$ which is not rotation
invariant and which is truly planar. This new system of coordinates was used by
the computer program which I devised in order to check the rules given in 
Section~\ref{rules}.

    This paper can be compared to~\cite{mmarXiv1510} where a weakly universal
cellular automaton is constructed which is truly planar and which is also rotation
invariant. We can see that the relaxation of rotation invariance allows us to reduce the
number of neighbours for a cell.

   In this paper, the same model as in~\cite{mmJCA2016} and~\cite{mmarXiv1512} is
used. In particular, we use the model of the latter paper: it simplifies a bit
the task of implementing the model used in this paper.

In Section~\ref{scenar}, I reproduce the presentation of the model which was given
in~\cite{mmarXiv1512}. In Section~\ref{rules}, we give the
rules of the automaton, insisting in the way we defined these rules in a context where
rotation invariance is no more required, which gives a definite advantage as indicated by
the result itself:

\begin{thm}\label{letheo}
There is a weakly universal cellular automaton on the tessellation $\{9,3\}$
which is 
truly planar and which has two states.
\end{thm}

Presently, we turn to the proof of this result.

\section{The scenario of the simulation}
\label{scenar}

    For the convenience of the reader, we reproduce the section written 
in~\cite{mmarXiv1512}.

    We sketchily remember that we simulate a register machine by a railway circuit.
Such circuit assembles infinitely many portions of straight lines, quarters of circles 
and switches. There are three kinds of switches, see~\cite{stewart,mmbook3}, the fixed
switch, the flip-flop and the memory switch. In any case, a switch consists of three 
portions of straight tracks, say $a$, $b$ and~$c$, which meet at a point which we call
the \textbf{centre} of the switch. In an \textbf{active} passage, the locomotive arrives
through~$a$ and leaves the switch either through~$b$ or through~$c$. In a \textbf{passive}
crossing, the locomotive arrives either through~$b$ or through~$c$ and leaves the switch
through~$a$. Table~\ref{switchtab} displays the working of the switches. Note that the 
flip-flop must always be crossed actively.

    As in previous papers, the flip-flop and the memory switch are decomposed into
simpler ingredients which we call sensors and control devices. This reinforce the
role of the tracks as their role for conveying key information is more and more important.

\newdimen\dlarge\dlarge=15pt
\newdimen\ddlarge\ddlarge=40pt
\newdimen\dddlarge\dddlarge=40pt
\def\lalligne #1 #2 #3 #4 {%
\ligne{%
\hbox to \dlarge{\hfill#1\hfill}
\hbox to \dlarge{\hfill#2\hfill}
\hbox to \ddlarge{\hfill#3\hfill}
\hbox to \dddlarge{\hfill#4\hfill}
\hfill}
}

\ligne{\hfill
\vtop{\leftskip 0pt\parindent 0pt\hsize=270pt
\begin{tab}\label{switchtab}
\leurre
Working of the switches. Note that {\tt S} stands for the side, either {\tt L}
or {\tt R}, \textit{i.e.} {\rm left-hand side} or {\rm right hand-side}
respectively. Similarly, {\tt W} which is either {\tt A} or {\tt P} indicates the
way the switch is crossed, \textit{i.e.} {\rm actively} or {\rm passively}
respectively, \textit{cross} and \textit{new sel} indicate how the crossing is 
performed and which is the new selected way respectively. 
\end{tab}
\trep
\vskip 5pt
\ligne{\hfill
\vtop{\leftskip 0pt\parindent 0pt\hsize=110pt
\lalligne S W {cross} {new sel} 
}
\hskip 30pt
\vtop{\leftskip 0pt\parindent 0pt\hsize=110pt
\lalligne S W {cross} {new sel} 
}
\hfill}
\vskip 5pt
\trfn
\vskip 5pt
\ligne{\hfill
\vtop{\leftskip 0pt\parindent 0pt\hsize=110pt
\ligne{\hfill\tt fixed switch \hfill}
\vskip 3pt
\lalligne L  A {$a\rightarrow b$} {$a\rightarrow b$}
\lalligne {\hfill}  P {$b\rightarrow a$} {$a\rightarrow b$} 
\lalligne {\hfill}  {\hfill} {$c\rightarrow a$} {$a\rightarrow b$} 
\lalligne R  A {$a\rightarrow c$} {$a\rightarrow c$} 
\lalligne {\hfill}  P {$b\rightarrow a$} {$a\rightarrow c$} 
\lalligne {\hfill}  {\hfill} {$c\rightarrow a$} {$a\rightarrow c$} 
\vskip 5pt
\ligne{\hfill \tt flip-flop \hfill}
\vskip 3pt
\lalligne L  A {$a\rightarrow b$} {$a\rightarrow c$} 
\lalligne R  A {$a\rightarrow c$} {$a\rightarrow b$}
}
\hskip 30pt
\vtop{\leftskip 0pt\parindent 0pt\hsize=110pt
\ligne {\hfill\tt memory switch \hfill} 
\vskip 3pt
\lalligne L  A {$a\rightarrow b$} {$a\rightarrow b$}
\lalligne {\hfill}  P {$b\rightarrow a$} {$a\rightarrow b$}
\lalligne {\hfill}  {\hfill} {$c\rightarrow a$} {$a\rightarrow c$}
\lalligne R  A {$a\rightarrow c$} {$a\rightarrow c$}
\lalligne {\hfill}  P {$b\rightarrow a$} {$a\rightarrow b$}
\lalligne {\hfill}  {\hfill} {$c\rightarrow a$} {$a\rightarrow c$}
}
\hfill}
\vskip 5pt
\trfn
}
\hfill}
\vskip 10pt
Here too, tracks are blank cells marked by appropriate black cells we call
\textbf{milestones}. We carefully study 
this point in Sub-section~\ref{tracks}.  Later, in Sub-sections~\ref{roundabout} 
and~\ref{forcontrol}, we look at the changes introduced with respect to~\cite{mmarXiv1510}.

\subsection{The tracks}\label{tracks}

In this implementation, the tracks are represented in a 
way which is a bit different from that of~\cite{mmarXiv1510}. The present implementation
is given by Figure~\ref{elemtrack}. Here, we explicitly indicate the numbering of
the sides in a cell which will be used through the paper. We fix a side which will
be, by definition, side~1 of the considered cell. Then, all the other sides are numbered
starting from this one and growing one by one while counter-clockwise turning around the
cell. Note that, in our setting, the same side, which is shared by two cells, can receive
two different numbers in the cells which share it. An example of this situation
is given in Figure~\ref{elemtrack}: in the central cell, side~1 is side~8 in the
neighbour of the central cell sharing this side. In Sub-section~\ref{trackrules}, we go 
back to the construction of the tracks starting from the elements indicated
in Figure~\ref{elemtrack}. 
\vskip 10pt
\vtop{
\ligne{\hfill
\includegraphics[scale=1]{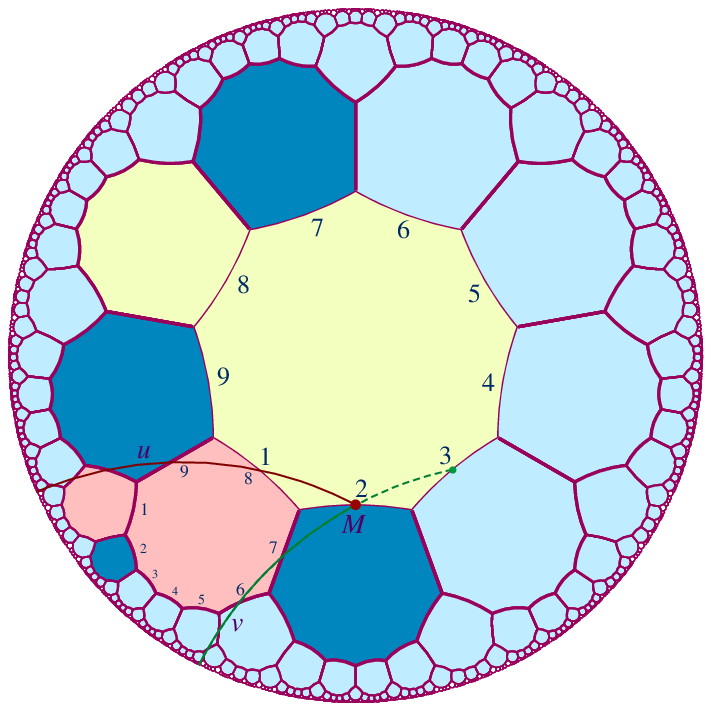}
\hfill}
\vspace{-10pt}
\ligne{\hfill
\vtop{\leftskip 0pt\parindent 0pt\hsize=160pt
\begin{fig}\label{elemtrack}
\leurre
Element of the tracks.
\end{fig}
}
\hfill}
}
\vskip 5pt
As can be seen in Figures of Section~\ref{rules}, the locomotive is implemented as
a single black cell: it has the same colour as the milestones of the tracks. Only the 
position of the locomotive with respect to the milestones allow us to distinguish it 
from the milestones. As clear from the next sub-section, we know that besides this
\textbf{simple locomotive}, the locomotive also occurs as a \textbf{double one} in some
portions of the circuit. 

The circuit also makes use of signals which are implemented in the form of a simple 
locomotive. So that at some point, it may happen that we have three
simple locomotives travelling on the circuit: the locomotive and two auxiliary signals 
involved in the working of some switch. For aesthetic reasons, the black colour which 
is opposed to the blank is dark blue in the figures.

\subsection{The round-about}
\label{roundabout}

    The round-about replaces the crossing, a railway structure, by a structure inspired by
road traffic. At a round-about where two roads are crossing, if you want to keep the 
direction arriving at the round-about, you need to leave the round-about at the second 
road. Figure~\ref{roundaboutfig} illustrates this features, a figure from~\cite{mmarXiv1510}
which is repeated here for convenience.

Now, this strategy requires that the cellular automaton knows how to count up to two. 
As in~\cite{mmarXiv1510}, we use the three auxiliary structures which are represented in 
Figure~\ref{roundaboutfig} by a rhombus, a small circle and a bit larger one,
the doubler, the fixed switch and the selector respectively. We study the fixed switch
in Sub-subsection~\ref{fix}, and the other structures in Sub-subsection~\ref{doublesel}. 

When the locomotive arrives close to the round-about, it first meets the doubler: it 
transforms the simple locomotive into a double one which consists of two contiguous 
black cells occupying blank cells of the tracks. Then, the locomotive may arrive at 
the fixed switch, depending on whether it arrived from~\textbf{A} or from~ \textbf{B}. 
Then, the double locomotive arrives at the first selector: the structure recognizes a 
double locomotive. It kills one of its black cells and the surviving simple 
locomotive is sent 
further on the round-about. When it meets the second selector, the structure recognizes 
a simple locomotive. Accordingly, it sends it on the track which leaves the round-about 
at that point.

\vtop{
\ligne{\hskip 110pt
\includegraphics[scale=0.6]{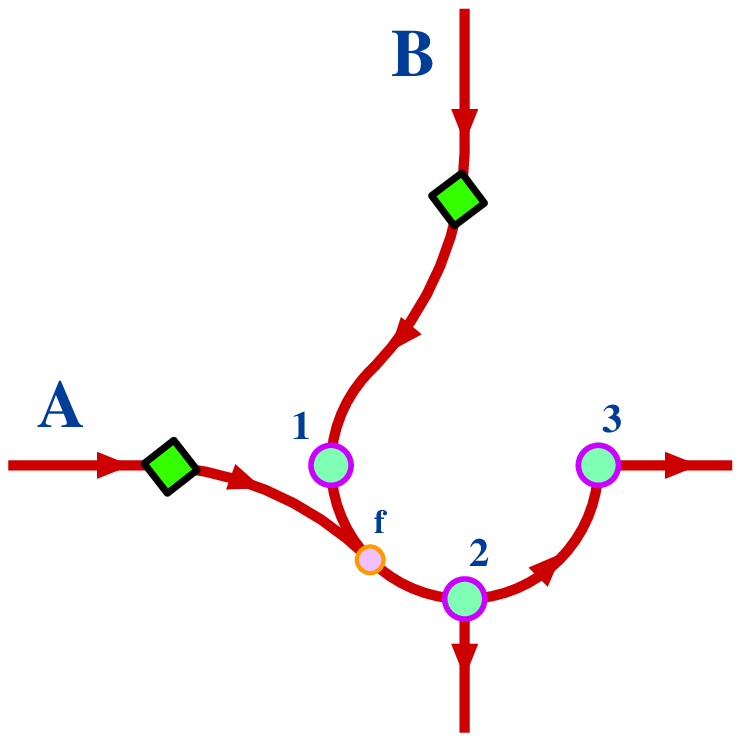}
\hfill}
\vspace{-15pt}
\ligne{\hfill
\vtop{\leftskip 0pt\parindent 0pt\hsize=200pt
\begin{fig}\label{roundaboutfig}
\leurre
Implementation scheme for the round-about.
\end{fig}
}
\hfill}
}

   In order to use the illustrations of this section and of the next one, we make use of a 
numbering of the tiles based on~\cite{mmarXiv1605}. In the figures, the central cell
is the tile whose centre is the centre of the circle in which the figure is inscribed.
The central cell is numbered by~0, denoted by~0(0). We number the sides of the tile
as already mentioned in Figure~\ref{elemtrack}. For $i\in\{1..9\}$, the cell which
shares the side~$i$ with the central cell is called \textbf{neighbour}~$i$ and it
is denoted by~$1(i)$. Number~1 in this notation is the number given to the root of the
tree attached to the sector defined from this tile. In~\cite{mmarXiv1605}, we indicate
how such a sector is defined. Here, we remind this definition for the convenience
of the reader, adapting it to the case of the tessellation~$\{9,3\}$. We invite the
reader to follow the present explanation on Figure~1. The sector attached
to the neighbour~1 of the central cell, it is the cell~1(1), is defined by two 
rays~$u$ and~$v$. The ray~$u$ starts from the mid-point~$M$ of the side~2 of~0(0) and 
passes through the mid-point of its side~1. Note that $u$ also passes through the mid-point
of the side~9 of 1(1). The ray~$v$ also starts from~$M$ and it 
passes through the mid-point of the side~7 of neighbour~1. Note that it also
passes through the mid-point of the side~6 of 1(1) and that the line supporting~$v$
also passes through the mid-point of the side~3 of~0(0). The neighbours of the 
cell~1(1) sharing its sides~$j$, $j\in\{1..5\}$ are numbered~$j$+1 and are denoted
by $j$+1(1). We say that the cell~1(1) is a $W$-cell and its sons are defined
by the rule \hbox{\tt W $\rightarrow$ BWWWW}, which means that 2(1) is a $B$-cell.
This means that the sons of~2(1) are defined by the rule \hbox{\tt B $\rightarrow$ BWWW},
where the $B$-son has two consecutive sides crossed by~$u$ in their mid-points. 
These sons of~1(1) constitute the level~1 of the tree.
The sons of~2(1), starting from its $B$-son are numbered 7, 8, 9 and~10 denoted by
7(1), 8(1), 9(1) and~10(1) respectively. By induction, the level~$n$+1 of the tree
are the sons of the cells which lie on the level~$n$. The cells are numbered from
the level~0, the root, level by level and, on each level from left to right, {\it i.e.}
starting from the ray~$u$ until the ray~$v$. What we have seen on the numbering of the
sons of~2(1) is enough to see how the process operates on the cells. From now on, we
use this numbering of the cells in the figures of the paper which represent
illustrations of the situation in the tessellation~$\{9,3\}$.

\subsubsection{The fixed switch}
\label{fix}

    As the tracks are one-way and as an active fixed switch always sends the locomotive 
in the same direction, no track is needed for the other direction: there is no active 
fixed switch. Now, passive fixed switches are still needed as just seen in the previous 
paragraph.  

\vtop{
\ligne{\hfill
\includegraphics[scale=1.3]{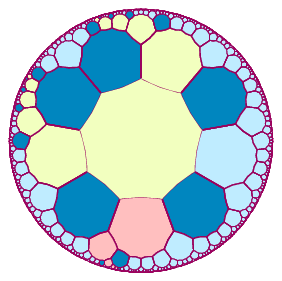}
\hfill}
\vspace{-10pt}
\ligne{\hfill
\vtop{\leftskip 0pt\parindent 0pt\hsize=270pt
\begin{fig}\label{stabfix}
The passive fixed switch in the tessellation $\{9,3\}$.
\end{fig}
}
\hfill}
\vskip 10pt
}

Figure~\ref{stabfix} illustrates the passive fixed switch when there is no 
locomotive around. We can see that it consists of elements of the tracks which are 
simply assembled in the appropriate way in order to drive the locomotive to the bottom 
direction in the picture, whatever upper side the locomotive arrived at the switch. 
The path followed by the locomotive to the switch is in yellow until the central
cell which is also yellow. The path from the left-hand side consists, in this order
of the cells 21(2), 5(2), 6(2), 2(3) and 1(3). From the right-hand side, it consists 
of the cells 21(1), 4(1), 3(1), 2(1) and~1(9). Of course, 1(3) and~1(9) are neighbours
of~0(0). The path followed by the locomotive when it leaves the cell is in pink.
It consists of the following cells in this order: 1(5), 2(5), 11(5), 12(5)and 13(5).
Note that the cell~0(0) in Figure~\ref{stabfix} has five black neighbours:
the cells 1(1), 1(2), 1(4), 1(6) and~1(8). Note that 1(1) and 1(8) are also
milestones for the cell 1(9), that 1(2) and~1(4) are milestones for 1(3) and
that 1(4) and~1(6) are milestones for 1(5).

From our description of the working of the round-about, a passive fixed switch must be 
crossed by a double locomotive as well as a simple one. Later, in 
Section~\ref{fixedswitch}, we shall check that the structure illustrated by
Figure~\ref{stabfix} allows those crossings. 

\subsubsection{The doubler and the selector}
\label{doublesel}

    The doubler is illustrated by the left-hand side picture of
 Figure~\ref{stab_doublsel}. Note that its structure is very different 
from that of the tracks or of the fixed switch. The central cell~0(0) is 
black and the path which is around consists in fact of two paths: the light green one:
1(1), 1(2), 1(3) and~1(4) and the the light blue one: 1(8), 1(7), 1(6) and~1(5).
The locomotive arrives through the yellow path: 7(1), 6(9) and 1(9). From
1(9), two locomotives appear: one in 1(1), the other in 1(8). The locomotive in 1(1)
goes along the light green path while the one in 1(8) goes along the light blue path.
Both locomotives arrive at the same time, one in 1(4) and the other in 1(5). From this 
moment, the double locomotive leaves the doubler through the pink path, namely,
in this order: 2(5) and~11(5). This doubler is very different from the doublers
constructed in the previous papers. As will be seen in Section~\ref{rules}, this is 
obtained by a particular numbering of the sides of a cell. The working of the 
configuration is the same as in~\cite{mmarXiv1510}, with this difference
that the locomotive is black.

\vtop{
\ligne{\hskip 72pt
\includegraphics[scale=1.2]{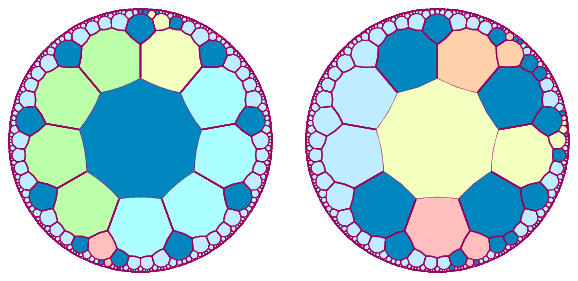}
\hfill}
\vspace{-15pt}
\ligne{\hfill
\vtop{\leftskip 0pt\parindent 0pt\hsize=230pt
\begin{fig}\label{stab_doublsel}
\leurre
To left: the doubler. To right: selector.
\end{fig}
}
\hfill}
}

The selector is illustrated by the right-hand side of Figure~\ref{stab_doublsel}. It 
is an almost symmetric picture with respect to the diameter of the disc which
passes through the common side of 1(2) and~1(3). What breaks the symmetry is the yellow
path through which the locomotive arrives, namely, in this order, the cells:
45(8), 9(8), 8(8), 7(8), 6(7) and~1(7), arriving to~0(0). At this point, the cells 1(6)
and 1(8) know whether the locomotive is simple or double. At the next top of the clock
a simple locomotive appears in both cells 1(9) and 1(5). If the locomotive
is double, the second part of it is in 0(0) and 1(8) turned to white so that
the locomotive in 1(9) is stopped. Now, the locomotive which is in 1(5) follows the pink
path, in this order: 1(5), 2(6), 7(6) and 30(5). If the locomotive is simple, 0(0) is 
white and 1(6) turned to white so that the locomotive in 1(5) is stopped. The locomotive
which is in 1(9) follows the orange path, in this order: 1(9), 2(9), 11(9), 12(9), 13(9)
and 64(9).

   Again, the numbering of the sides in a cell allows us to define those paths. In 
Subsubsection~\ref{subssel}, the rules will show that such a working will be observed.

\subsection{The fork, the controller and the controller-sensor}
\label{forcontrol}

    In this Sub-section, we look at the decomposition of two active switches: the flip-flop 
and the active part of the memory switch. We follow the same implementation as 
in~\cite{mmarXiv1510}, separating the working of the switch into two separate stages.
The fork and the controller have different implementations as in the previous papers.

First, we remind the use of those structures in order to implement a flip-flop and
an active memory switch, see the left- and the right-hand side parts respectively of
Figure~\ref{activswitches}.

When the simple locomotive arrives at the fork~$C$, 
it is duplicated into two simple locomotives, each one following its own path. 
In the active memory switch, each locomotive goes to a controller. In the flip-flop,
one of these locomotives goes on its way to the controller, and the other goes to another 
fork~$A$ which sends one locomotive to the other controller and the other, which is now a 
third locomotive, is sent to a third fork~$S$.

\vtop{
\ligne{\hfill
\includegraphics[scale=0.35]{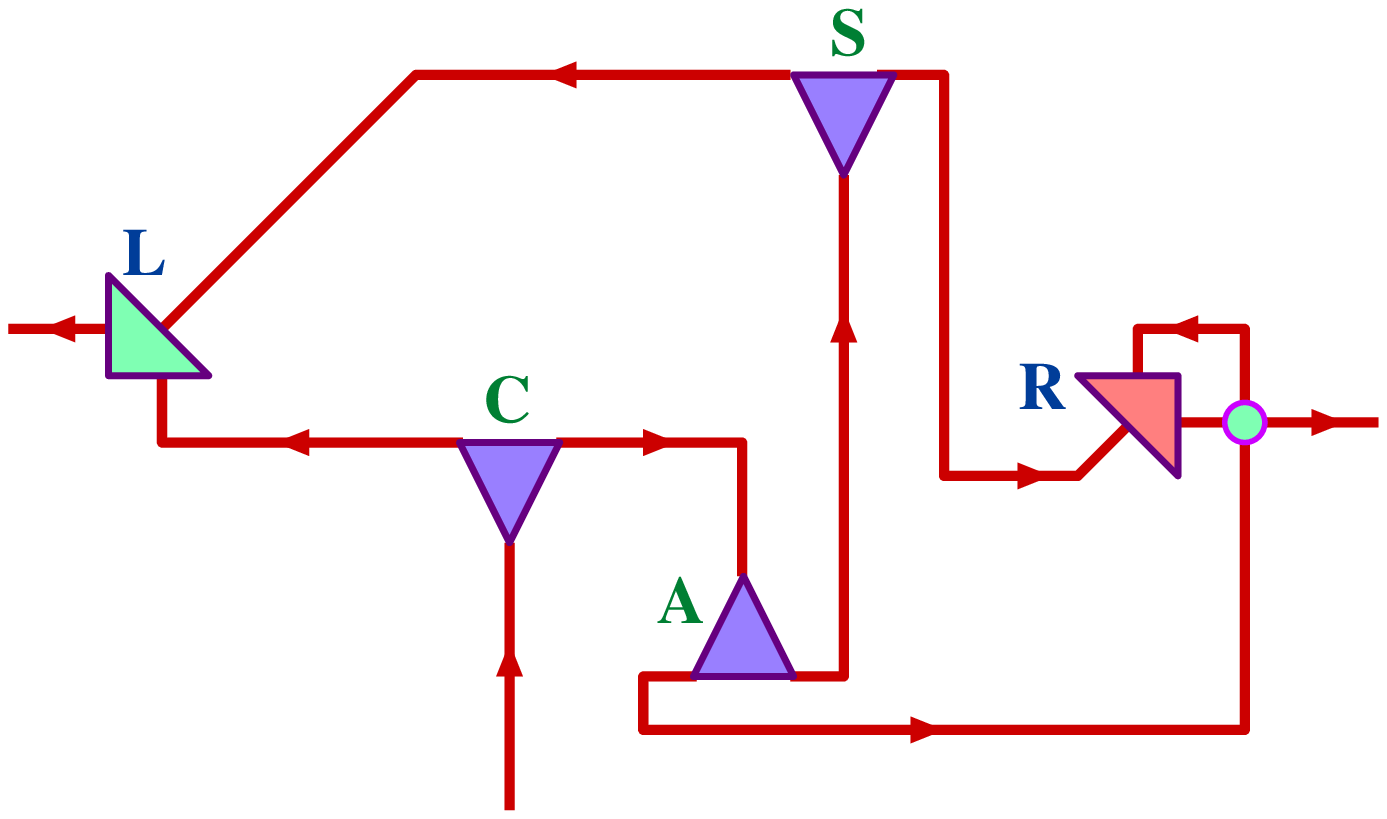}
\hskip 10pt
\includegraphics[scale=0.35]{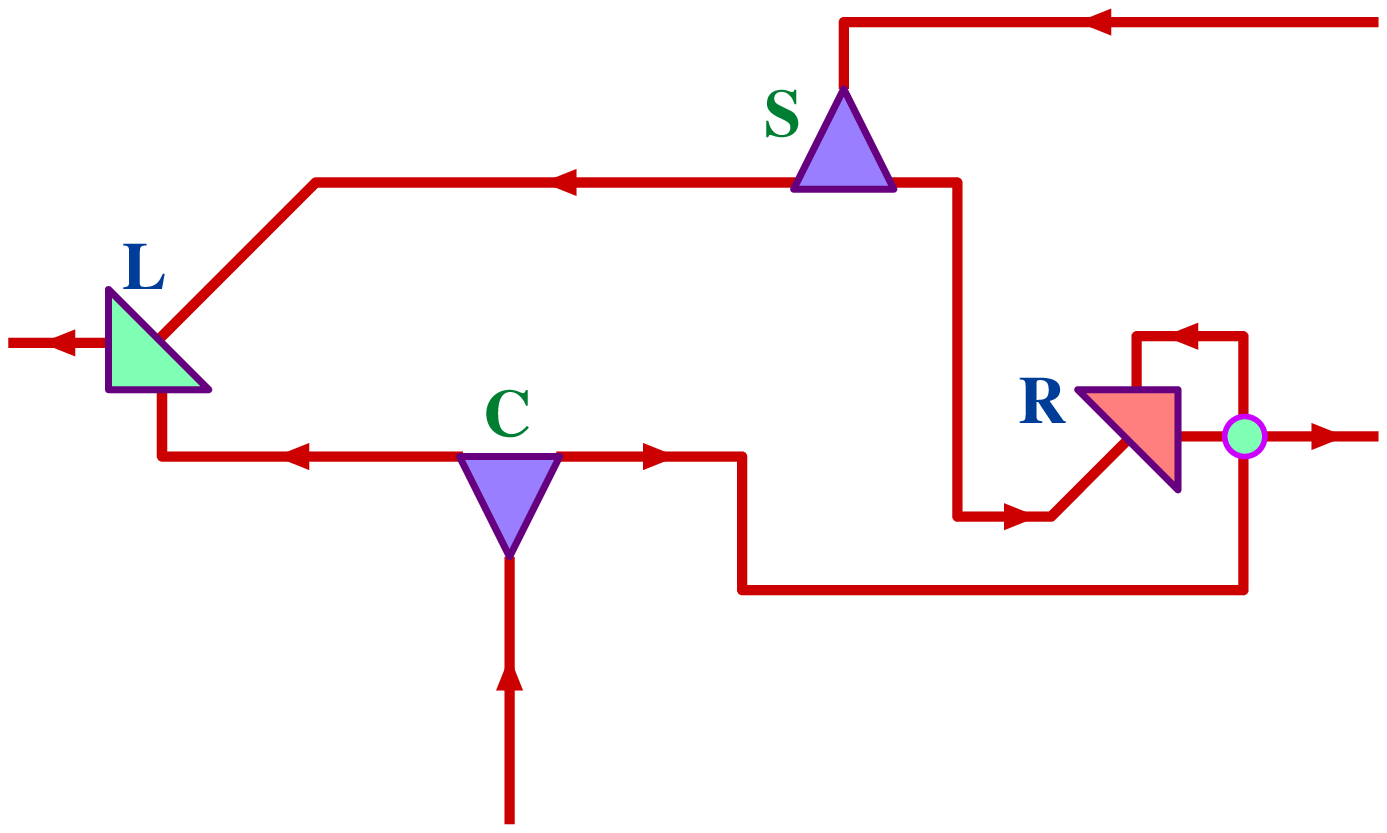}
\hfill}
\vspace{-15pt}
\ligne{\hfill
\vtop{\leftskip 0pt\parindent 0pt\hsize=300pt
\begin{fig}\label{activswitches}
\leurre
To left: the flip-flop. To right: the active memory switch.
\end{fig}
}
\hfill}
\vskip 10pt
}

The locomotives sent by~$S$ go both to a controller, one to the black controller, the other
to the white one. Now, when a locomotive sent by~$S$ arrives at a controller, it changes 
the black one into a white one and the white one into a black one. Accordingly,
what should be performed by a flip-flop is indeed performed. It is enough to manage 
things in such a way that the locomotives arriving to the controllers from~$S$ arrive 
later than those sent by~$C$ and by~$A$. In the active memory switch, the locomotive
which arrives to~$S$ is sent by the passive memory switch. In this sense, the passive 
memory switch is active while the active one is passive. 

The fork and the controller are illustrated by Figure~\ref{stab_active}: the fork lies in
the first picture of the figure, the controller is illustrated by the second and the third
pictures.

The fork appears like a doubler which would be cut in its middle part. Indeed,
the locomotive arrives to~1(9) through the same path as in the doubler: cells
7(1), 6(9) and 1(9). From the cell 1(9) one simple locomotive goes to the left through
the pink path constituted by the cells 1(1), 1(2), 6(2), 7(3), 8(3) and 9(3).
At the same time as a simple locomotive enters 1(1), a simple locomotive, coming from~1(9)
enters 1(8), going to the right through another pink path constituted by the cells
1(8), 1(7), 6(7), 7(8), 8(9) and 9(9). Accordingly, the working of the fork is
performed by such a configuration, provided that the sides~1 are fixed in an appropriate
way.

In the controller, the locomotive arrives through the yellow path defined by the cells
45(8), 9(8), 8(8), 7(8), 6(7) and 1(7) in this order. When it arrives to~0(0), the later
behaviour depends on the cell~1(3). If it is white, see the rightmost picture of
Figure~\ref{stab_active}, the locomotive is stopped at 1(7): it does not reach 0(0).
If 1(1) is black, see the second picture of Figure~\ref{stab_active}, then it goes
to~0(0) and then, it follows the pink path, the cells 1(4), 2(5) and 30(4) in this order.

\vtop{
\vskip 0pt
\ligne{\hskip 25pt
\includegraphics[scale=1]{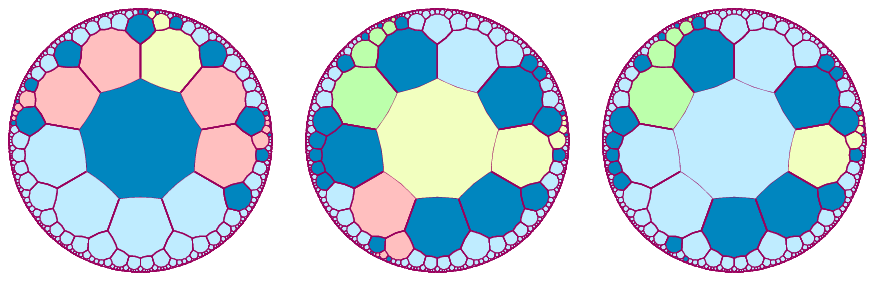}
\hfill}
\vspace{-10pt}
\ligne{\hfill
\vtop{\leftskip 0pt\parindent 0pt\hsize=320pt
\begin{fig}\label{stab_active}
\leurre
To left: the fork. To right, two configurations: the controller of the flip-flop and of the
active memory switch.
\end{fig}
}
\hfill}
\vskip 10pt
}

   Note that in both the second and the third picture of Figure~\ref{stab_active},
there is a light green path which goes to 1(2), following the cells 21(1), 5(1), 6(1),
2(2) and 1(2). If a locomotive arrives to~1(2) through this path, it is a signal sent
either by the flip-flop or the appropriate part of the passive memory switch in
order to change the selection of the controller: from black to white and to white from
black. Accordingly, when a simple locomotive arrives at 1(2), if 1(3) is black,
as in the second picture of Figure~\ref{stab_active}, then it becomes white, as
in the third picture of Figure~\ref{stab_active}. Conversely, when 1(3) is white,
if a locomotive arrives at 1(2), then 1(3) becomes black.
\vskip 10pt
   Figure~\ref{memopass} illustrates the construction of the passive memory switch with
the help of forks and controllers. That implementation is somehow different from the one
indicated in~\cite{mmarXiv1510}. It is similar to that of~\cite{mmarXiv1512}. 
However, as in those papers, the controllers of 
Figure~\ref{memopass} are not the same as those of Figure~\ref{activswitches} and they 
are different from those of~\cite{mmarXiv1510}. The main reason is that in the case
of a passage of the locomotive through the non-selected track, in~\cite{mmarXiv1510}, 
the controller let the locomotive go, it changes the selection and it sends a signal 
to the other controller in order to order it to change its selection too. So that the 
controller has to perform three tasks at once. As in~\cite{mmarXiv1512}, this implementation
slightly simplify the task of the controller. As the locomotive should anyway not be 
stopped, we can place a fork on the tracks passively arriving to the switch: this is 
the reason for placing the 
forks $S_1$ and~$S_2$ in Figure~\ref{memopass} where they are on the picture. Let us look 
at the locomotive which arrives at the fork~$S_2$ on the figure. It corresponds to the 
non-selected track. The simple locomotive is duplicated: one goes to the fixed switch 
$F_1$ and goes further. The second simple locomotive goes to the controller~$R$. As 
this controller signalizes a non-selected track, the selection is changed and the 
controller let this locomotive go on along the track which leads it to another fork~$S_3$.
That fork sends the second locomotive to~$L$ where it changes the selection from selected
to non-selected. The fork also sends a third locomotive to the active switch through
$F_2$ in order to change its selection too.

\vtop{
\vskip 5pt
\ligne{\hfill
\includegraphics[scale=0.55]{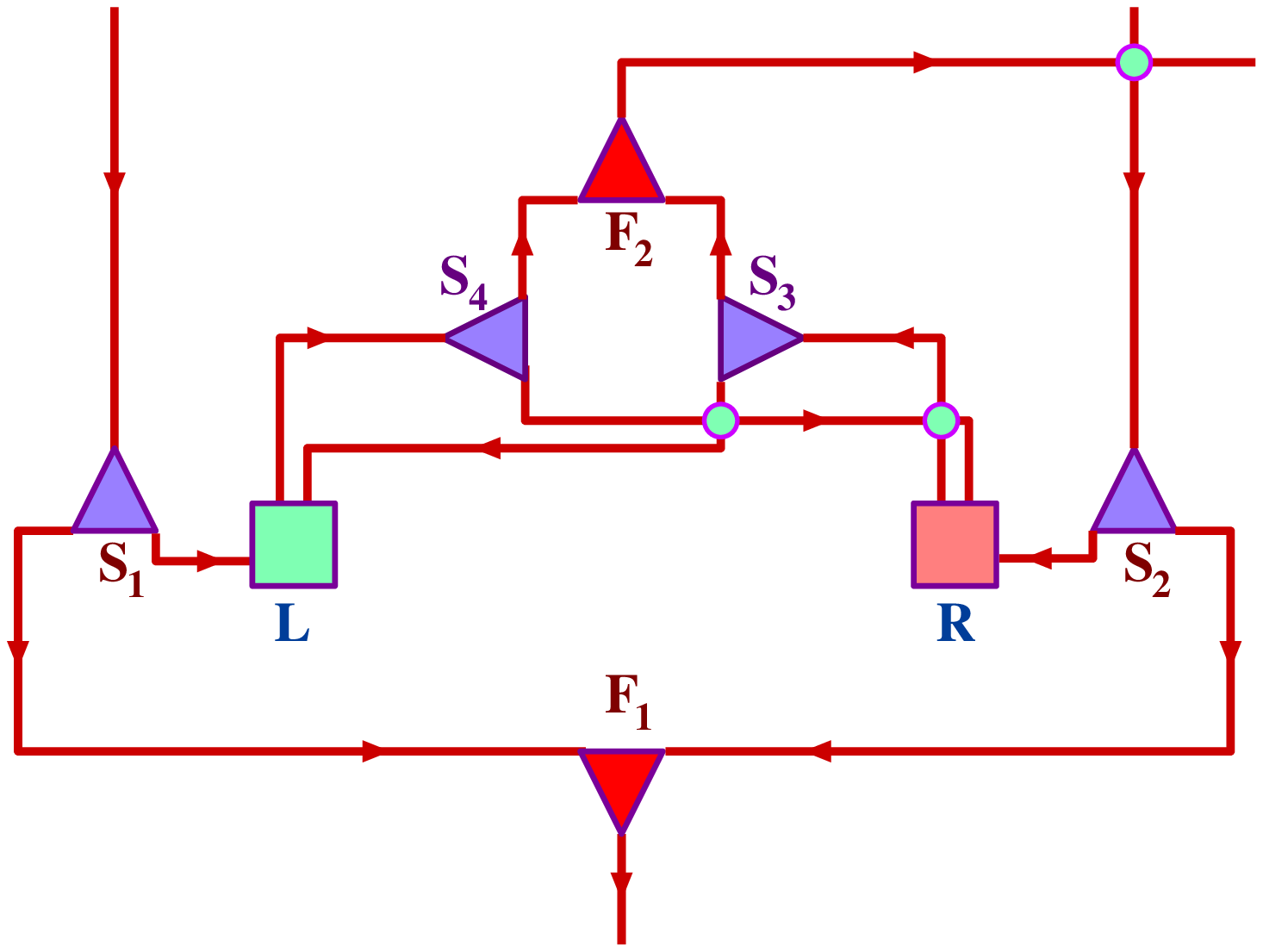}
\hfill}
\vspace{-10pt}
\ligne{\hfill
\vtop{\leftskip 0pt\parindent 0pt\hsize=260pt
\begin{fig}\label{memopass}
\leurre
Organization of the passive memory switch with forks and sensors. Note that the sensors
are not represented with the same symbol as the controllers in 
Figure~{\rm\ref{activswitches}}.
\end{fig}
}
\hfill}
\vskip 10pt
}

   Consider the case when the simple locomotive arrives to~$S_1$, the fork which 
corresponds on the figure to the selected track. Then, the locomotive is sent to~$L$ where 
it is stopped as no change should be performed.
\vskip 5pt
   Figure~\ref{stab_ctrlsgn} illustrates the controller of the passive memory switch. 
Note that the configuration is different from that of the controller of
Figure~\ref{stab_active}. This time, the locomotive arrives to the sensor through
the cell 1(6), following the yellow path constituted by the cells, in this order:
45(7), 9(7), 8(7), 7(7), 6(6) and 1(6). The sensor is the cell~1(1). If it is white, 
as in the
left-hand side picture of Figure~\ref{stab_ctrlsgn}, the locomotive goes to~0(0)
and then, it goes further on its way through the pink path which consists of the
cells 1(4), 2(5) and 7(5). The cell 0(0) is represented in the light green colour
as when the locomotive just left the cell, the colour of~1(1) changes from white
to black. The locomotive goes on its way as the other sensor must be changed
and as the active memory switch must also be changed. 

The sensor with 1(1) being black is illustrated by the right-hand side picture
of Figure~\ref{stab_ctrlsgn}. If a locomotive arrives through the yellow path, the same
as for the left-hand side picture, as 1(1) is black, the locomotive is stopped: it
does not reach the cell 0(0). In this case, which means that no change of selection
occurs, the cell 1(1) remains black. Still on the right-hand side picture, we can see
a light green path, which is similar to that of the controller, up to a rotation.
The green path is constituted by the cells 21(8), 5(8), 6(8), 2(9) and 1(9). If a 
locomotive arrives through that path at~1(9), then 1(1) turns from black to white,
corresponding to the change we have seen previously which occurred at the other sensor
of the passive memory switch.

\vskip 10pt
\vtop{
\ligne{\hfill
\includegraphics[scale=1]{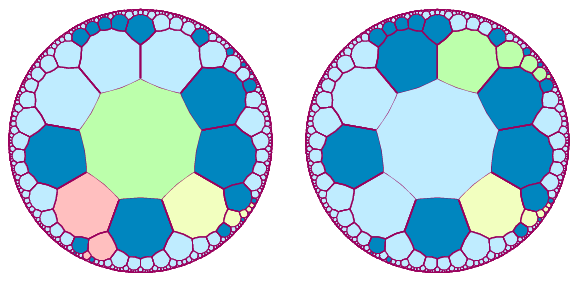}
\hfill}
\vspace{-15pt}
\ligne{\hfill
\vtop{\leftskip 0pt\parindent 0pt\hsize=200pt
\begin{fig}\label{stab_ctrlsgn}
\leurre
The controller-sensor of the passive memory switch.
\end{fig}
}
\hfill}
}

\section{Rules}
\label{rules}

    The figures of Section~\ref{scenar} help us to establish the rules. Their application
is illustrated by figures of this section which
were established with the help of a computer program which checked the 
coherence of the rules. The program also wrote the PostScript files of the pictures from 
the computation of the application of the rules to the configurations of the various type 
of parts of the circuit. The computer program also established the traces of execution 
which contribute to the checking of the application of the rules.

    We have to revisit the format of the rules and also to explain what is allowed
from the relaxation from rotation invariance. We remind the reader that a rule has the form
\hbox{\footnotesize\tt$\underline{\hbox{X}}$$_o$X$_1$..X$_9$$\underline{\hbox{X}}$$_n$},
where \hbox{\footnotesize\tt X$_o$} is the state of the cell~$c$,
\hbox{\footnotesize\tt X$_i$} is the state of the neighbour~$i$ of~$c$ 
and \hbox{\footnotesize\tt X$_n$} is the new state of~$c$ once the rule
has been applied. As the rules no more observe the rotation invariance, we may freely 
choose which is side~1 for each cell. We take this freedom from the format of the rule 
which only requires to know which is neighbour~1. In order to restrict the number of 
rules, it is decided that as a general rule, for a cell which is an element of the track, 
side~1 is the side shared by the cell and its next neighbour on the track. There can be 
exceptions when the cell is in a switch or the neighbour of the central cell in a switch.
In particular, when a cell belongs to two tracks, side~1 is arbitrarily chosen among the 
two possible cases. The milestones have their side~1 shared by an element of the track. 

    We have to keep in mind that there are two types of rules. Those
which keep the structure invariant when the locomotive is far from them, we call this
type of rules \textbf{conservative}, and those which control the motion of the locomotive.
Those latter rules, which we call \textbf{motion rules}, are the rules applied to the
cells of  the tracks as well as their milestones and, sometimes to the cells of the 
structures which may be affected by the passage of the locomotive.

\subsection{Defining tracks and their rules}
\label{trackrules}

   Table~\ref{rulesvs} provides us with rules which concern the motion of
a simple locomotive along a path. An important part of the motion
is performed around a cell: it is not a complete turn around the cell but only on five 
consecutive neighbours of the cell. The first fifteen rules are conservative rules.
Then rules~16 up to~42 deal with the motion of the locomotive in a clockwise direction
and the rules~43 up to~58 are additional rule needed by the motion in the reverse 
direction.
Figure~\ref{voieverts} illustrate the motion governed by these rules.

\def\aff #1 #2 #3 #4 {\ligne{\hfill\footnotesize\tt\hbox to 13pt{\hfill#1}
\hskip 5pt$\underline{\hbox{\tt#2}}$#3$\underline{\hbox{\tt#4}}$\hfill}\vskip-4pt
}
\def\laff #1 #2 #3 #4 {\hbox{{\footnotesize
$\underline{\hbox{\tt#1}}${\tt#2}$\underline{\hbox{\tt#3}}${}}}#4\hskip 4pt
}
\newdimen\tabruleli\tabruleli=320pt
\newdimen\tabrulecol\tabrulecol=70pt
\ligne{\hfill   
\vtop{\leftskip 0pt\parindent 0pt\hsize=\tabruleli  
\ligne{
\vtop{\leftskip 0pt\parindent 0pt\hsize=300pt  
\begin{tab}\label{rulesvs}
\leurre
Rules for a simple locomotive on the tracks.
\end{tab}
}  
\hfill}  
\vskip 2pt
\trep
\vskip 8pt
\ligne{\hfill Preliminary conservative rules\hfill}
\vskip 8pt
\ligne{\hfill  
\vtop{\leftskip 0pt\parindent 0pt\hsize=\tabrulecol
\aff {  1} {W} {WWWWWWWWW} {W}
\aff {  2} {B} {WWWWWWWWW} {B}
\aff {  3} {B} {WWWBWWBWW} {B}
\aff {  4} {B} {WWWWBWWBW} {B}
}
\hskip 5pt
\vtop{\leftskip 0pt\parindent 0pt\hsize=\tabrulecol
\aff {  5} {W} {WBWWWWBWB} {W}
\aff {  6} {W} {WBWWWBWBB} {W}
\aff {  7} {B} {WBWWWWWWW} {B}
\aff {  8} {W} {WBWWWWWWW} {W}
}
\hskip 5pt
\vtop{\leftskip 0pt\parindent 0pt\hsize=\tabrulecol
\aff {  9} {W} {WWWWWWWWB} {W}
\aff { 10} {W} {WBWWWWWWB} {W}
\aff { 11} {W} {BWWWWWWWW} {W}
\aff { 12} {W} {BWWWWWWWB} {W}
}
\hskip 5pt
\vtop{\leftskip 0pt\parindent 0pt\hsize=\tabrulecol
\aff { 13} {B} {BWWWWWWWW} {B}
\aff { 14} {W} {WBBWWWWWB} {W}
\aff { 15} {W} {WWWWWWBWW} {W}
}
\hfill
}   
\vskip 8pt
\ligne{\hfill Rules for a simple locomotive\hfill}
\vskip 8pt
\ligne{\hfill  
\vtop{\leftskip 0pt\parindent 0pt\hsize=\tabrulecol
\aff { 16} {B} {WWWBBWWBW} {B}
\aff { 17} {W} {WBWWWWBBB} {B}
\aff { 18} {B} {WBWWWBWBB} {W}
\aff { 19} {B} {WBBWWWWWW} {B}
\aff { 20} {W} {BBWWWWWWW} {W}
\aff { 21} {W} {BBWWWWWWB} {W}
\aff { 22} {B} {WWBWBWWBW} {B}
}
\hskip 10pt
\vtop{\leftskip 0pt\parindent 0pt\hsize=\tabrulecol
\aff { 23} {B} {WBWWWWBWB} {W}
\aff { 24} {W} {BBWWWBWBB} {W}
\aff { 25} {B} {WBWWBWWBW} {B}
\aff { 26} {W} {BBWWWWBWB} {W}
\aff { 27} {B} {WWWWWWBWW} {B}
\aff { 28} {B} {BWWWBWWBW} {B}
\aff { 29} {W} {WBWWWBBBB} {B}
}
\hskip 10pt
\vtop{\leftskip 0pt\parindent 0pt\hsize=\tabrulecol
\aff { 30} {B} {WWWWWWWWB} {B}
\aff { 31} {B} {WWBBWWBWW} {B}
\aff { 32} {B} {WWWWBWWBB} {B}
\aff { 33} {B} {WBWBWWBWW} {B}
\aff { 34} {B} {BWWBWWBWW} {B}
\aff { 35} {B} {WWWBWWBWB} {B}
\aff { 36} {B} {WWWBWWBBW} {B}
}
\hskip 10pt
\vtop{\leftskip 0pt\parindent 0pt\hsize=\tabrulecol
\aff { 37} {W} {WWWWBWWWW} {W}
\aff { 38} {W} {WWWBWWWWW} {W}
\aff { 39} {B} {WWWBWBBWW} {B}
\aff { 40} {B} {WWWBBWBWW} {B}
\aff { 41} {B} {WWWWBWBBW} {B}
\aff { 42} {W} {WWBWWWWWW} {W}
}
\hfill} 
\vskip 8pt
\ligne{\hfill Rules for a simple locomotive\hfill}
\vskip 8pt
\ligne{\hfill  
\vtop{\leftskip 0pt\parindent 0pt\hsize=\tabrulecol
\aff { 43} {W} {WWWWWWWBW} {W} 
\aff { 44} {W} {WBWBWWWWB} {W}
\aff { 45} {W} {WBBWBWWWB} {W}
\aff { 46} {B} {WBBWBWWWB} {W}
}\hskip 10pt
\vtop{\leftskip 0pt\parindent 0pt\hsize=\tabrulecol
\aff { 47} {W} {WBBBWWWWB} {B}
\aff { 48} {W} {BBBWWWWWB} {W}
\aff { 49} {B} {WWWWBWWWW} {B}
\aff { 50} {W} {BBBWBWWWB} {W}
}\hskip 10pt
\vtop{\leftskip 0pt\parindent 0pt\hsize=\tabrulecol
\aff { 51} {B} {WBWBWWWWB} {W}
\aff { 52} {B} {WWWBWWWWW} {B}
\aff { 53} {W} {BBWBWWWWB} {W}
\aff { 54} {W} {WBBBBWWWB} {B}
}\hskip 10pt
\vtop{\leftskip 0pt\parindent 0pt\hsize=\tabrulecol
\aff { 55} {W} {WWWBBWWWW} {W}
\aff { 56} {B} {WWWBBWWWW} {B}
\aff { 57} {B} {WWWWBBWBW} {B}
\aff { 58} {B} {BBWWWWWWW} {B}
}
\hfill}
\vskip 9pt
\trfn
}   
\hfill} 
\vskip 10pt
\newdimen\largez\largez=20pt
\def\HH#1{\hbox to\largez{\hfill #1\hfill}}
\def\Rr#1{{\color{red}{#1}}}

   As explained in Section~\ref{scenar}, the construction of the tracks is very important.
In this subsection, we give a \textit{local view} on a configuration which
will allow us to construct any track, see Figure~\ref{stab_tracks}. The configuration is
said \textit{idle} as no locomotive is around it. 
We say that Figure~\ref{stab_tracks} gives a local view only
because we can only see a very tiny part of the hyperbolic plane. We have to remember
that the central cell is not the centre of the hyperbolic plane, such a centre 
does not exist, but the cell on which we focus our attention. 

Between any pair $(A,B)$ of tiles of the tessellation, there is a path 
which joins them: it is a finite sequence of tiles $\{T_i\}_{i\in[0..k]}$ with
$A=T_0$, $B=T_k$ and for $i<k$, $T_i$ and $T_{i+1}$ share a side. There are shortest path
between~$A$ and~$B$ and \cite{mmbook2} shows us how to construct such a path in
linear time in the coordinates of~$A$ and~$B$. Next, assembling pieces of
Figure~\ref{stab_tracks} will allow us to construct a path from~$A$ to~$B$.

\vtop{
\vskip 0pt
\ligne{\hfill
\includegraphics[scale=1]{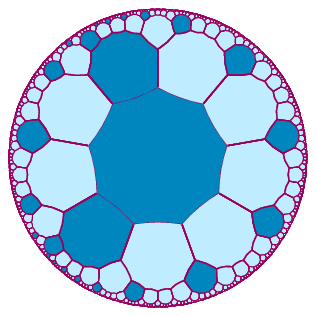}\hfill}
\vspace{-10pt}
\begin{fig}\label{stab_tracks}
\leurre
Idle configuration of the tracks used for testing the rules.
\end{fig}
}

Table~\ref{execvdm} shows us an execution of the rules of Table~\ref{rulesvs}. 
The first two rows of Figure~\ref{voieverts} illustrate this motion. In the table,
the rules are indicated by their number in the tables for the rules. In red, we 
indicate the rules which change the state of the cell, both for white to black and
for black to white. 

\vtop{
\begin{tab}\label{execvdm}
Execution of the rules $1$ up to~$42$.
\end{tab}
\vskip 2pt
\trep
\vskip 8pt
\ligne{\hfill\HH{}
\HH{{5$_1$} }\HH{{4$_1$} }\HH{{3$_1$} }\HH{{2$_1$} }\HH{{1$_9$} }\HH{{1$_8$} }
\HH{{1$_7$} }\HH{{1$_6$} }\HH{{1$_5$} }\HH{{2$_5$} }\HH{{6$_4$} }\HH{{5$_4$} }\HH{{4$_4$} }\hfill}
\ligne{\hfill\HH{1}
\HH{24}\HH{\Rr{23}}\HH{\Rr{17}}\HH{5}\HH{6}\HH{5}
\HH{5}\HH{5}\HH{6}\HH{5}\HH{5}\HH{5}\HH{6}\hfill}

\ligne{\hfill\HH{2}
\HH{6}\HH{26}\HH{\Rr{23}}\HH{\Rr{17}}\HH{6}\HH{5}
\HH{5}\HH{5}\HH{6}\HH{5}\HH{5}\HH{5}\HH{6}\hfill}

\ligne{\hfill\HH{3}
\HH{6}\HH{5}\HH{26}\HH{\Rr{23}}\HH{\Rr{29}}\HH{5}
\HH{5}\HH{5}\HH{6}\HH{5}\HH{5}\HH{5}\HH{6}\hfill}

\ligne{\hfill\HH{4}
\HH{6}\HH{5}\HH{5}\HH{26}\HH{\Rr{18}}\HH{\Rr{17}}
\HH{5}\HH{5}\HH{6}\HH{5}\HH{5}\HH{5}\HH{6}\hfill}

\ligne{\hfill\HH{5}
\HH{6}\HH{5}\HH{5}\HH{5}\HH{24}\HH{\Rr{23}}
\HH{\Rr{17}}\HH{5}\HH{6}\HH{5}\HH{5}\HH{5}\HH{6}\hfill}

\ligne{\hfill\HH{6}
\HH{6}\HH{5}\HH{5}\HH{5}\HH{6}\HH{26}
\HH{\Rr{23}}\HH{\Rr{17}}\HH{6}\HH{5}\HH{5}\HH{5}\HH{6}\hfill}

\ligne{\hfill\HH{7}
\HH{6}\HH{5}\HH{5}\HH{5}\HH{6}\HH{5}
\HH{26}\HH{\Rr{23}}\HH{\Rr{29}}\HH{5}\HH{5}\HH{5}\HH{6}\hfill}

\ligne{\hfill\HH{8}
\HH{6}\HH{5}\HH{5}\HH{5}\HH{6}\HH{5}
\HH{5}\HH{26}\HH{\Rr{18}}\HH{\Rr{17}}\HH{5}\HH{5}\HH{6}\hfill}

\ligne{\hfill\HH{9}
\HH{6}\HH{5}\HH{5}\HH{5}\HH{6}\HH{5}
\HH{5}\HH{5}\HH{24}\HH{\Rr{23}}\HH{\Rr{17}}\HH{5}\HH{6}\hfill}

\ligne{\hfill\HH{10}
\HH{6}\HH{5}\HH{5}\HH{5}\HH{6}\HH{5}
\HH{5}\HH{5}\HH{6}\HH{26}\HH{\Rr{23}}\HH{\Rr{17}}\HH{6}\hfill}

\ligne{\hfill\HH{11}
\HH{6}\HH{5}\HH{5}\HH{5}\HH{6}\HH{5}
\HH{5}\HH{5}\HH{6}\HH{5}\HH{26}\HH{\Rr{23}}\HH{\Rr{29}}\hfill}
\vskip 9pt
\trfn
\vskip 15pt
}

Using the figure and the table of execution, we can see that the motion of the
locomotive mainly uses rule~5, \laff {W} {WBWWWWBWB} {W} {,}
rule~17, \laff {W} {WBWWWWBBB} {B} {,} rule~23, \laff {B} {WBWWWWBWB} {W} {} and
rule~26, \laff {W} {BBWWWWBWB} {W} {.} The milestones of the cell of the track
are the neighbours~2, 7 and~9: rule~5 is the conservative rule for this cell. 

\vtop{
\vskip 0pt
\ligne{\hfill
\includegraphics[scale=0.55]{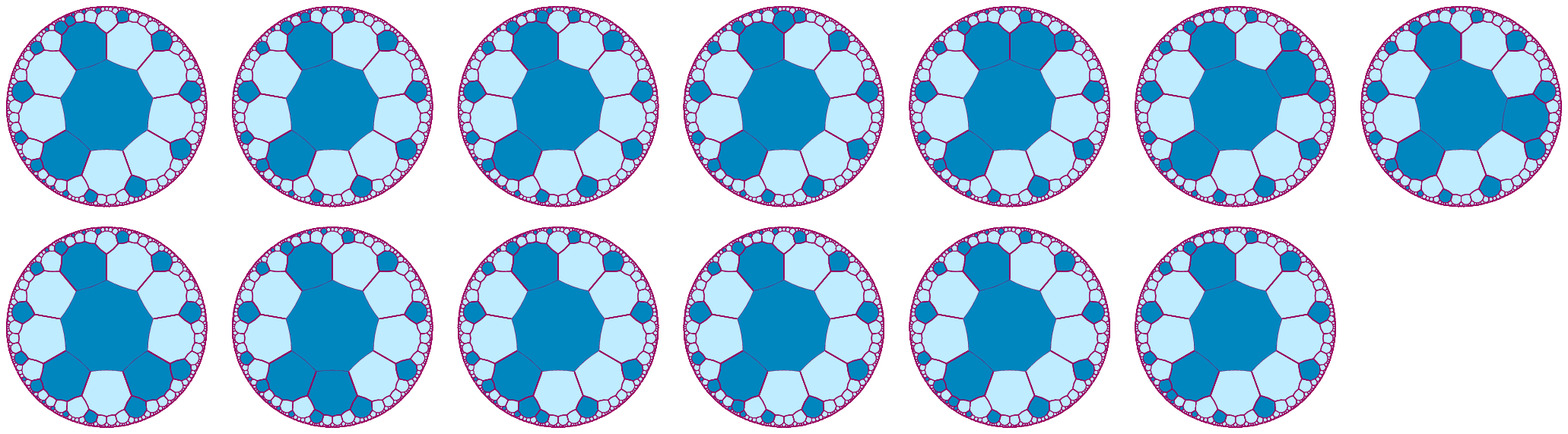} 
\hfill}
\vspace{-10pt}
\ligne{\hfill
\includegraphics[scale=0.55]{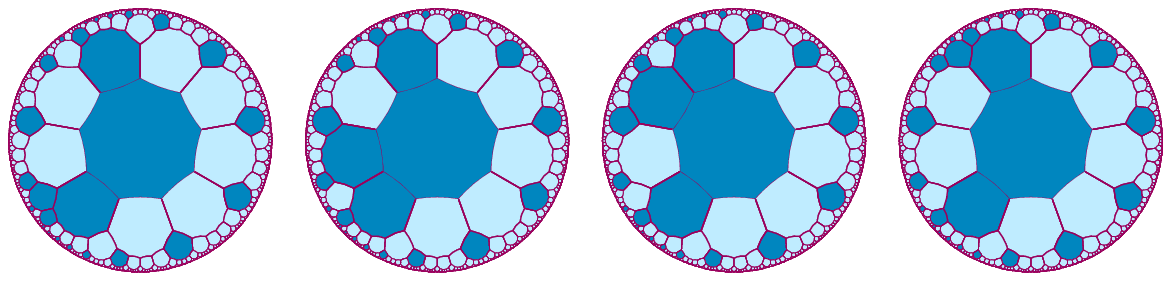} 
\hfill}
\vspace{-10pt}
\ligne{\hfill
\includegraphics[scale=0.55]{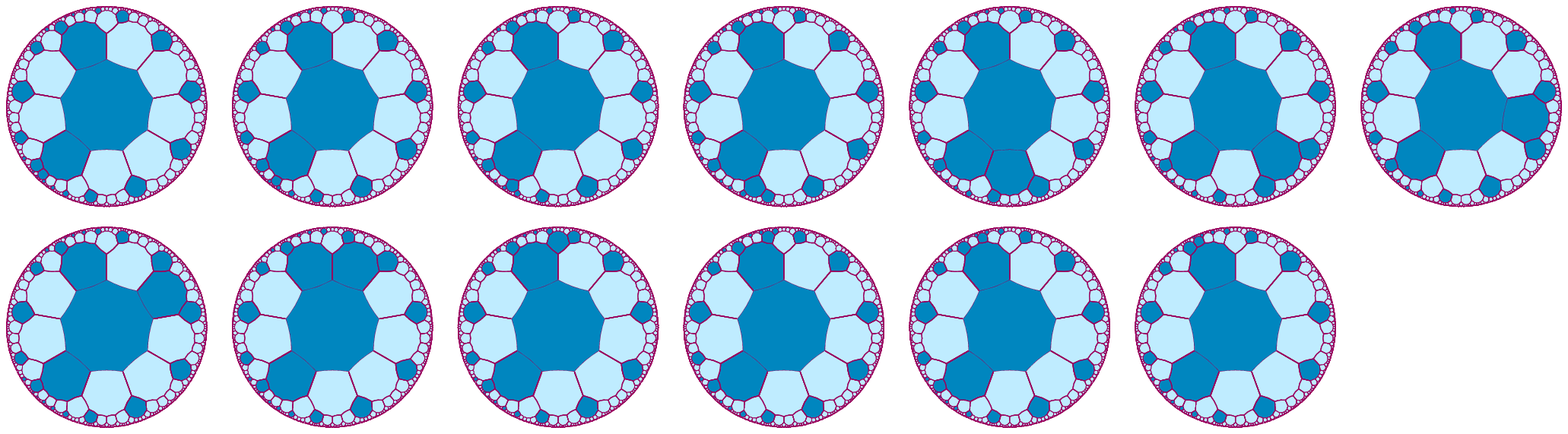} 
\hfill}
\vspace{-10pt}
\ligne{\hfill
\includegraphics[scale=0.55]{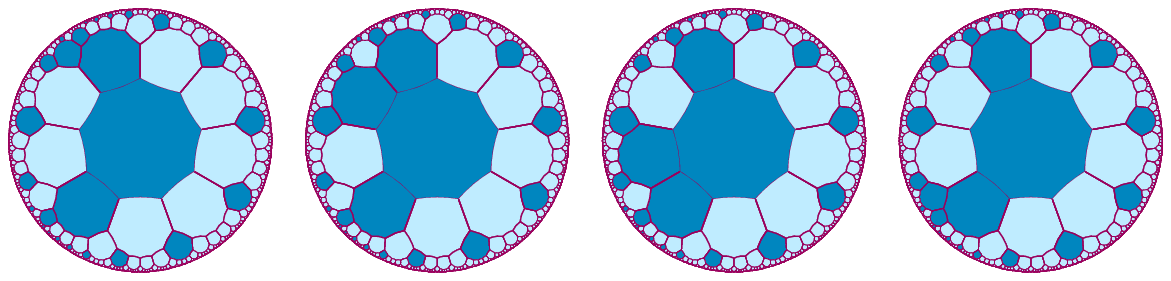} 
\hfill}
\vspace{-15pt}
\ligne{\hfill
\vtop{\leftskip 0pt\parindent 0pt\hsize=340pt
\begin{fig}\label{voieverts}
\leurre
From left to right: vertical tracks, horizontal tracks around a black node, horizontal
tracks around a white node. In each group, tracks in one direction and then in the
opposite direction.
\end{fig}
}
\hfill}
\vskip 10pt
}

The 
arriving locomotive is detected by rule~17 as it appears at the place of neighbour~8
so that the rule makes the locomotive enter the cell. Rule~23 make the cell return
to white and rule~26 witnesses that the locomotive is in neighbour~1 as required by
the numbering of an ordinary cell of the tracks. We can see this sequence of rules
for the cells 2(1), 1(8), 1(7), 1(6), 2(5) and 6(4). The cells which require other rules
are cells which connect two rounding motions, where a rounding motion is a motion
around a cell, not necessarily completely turning around the cell. As an example, we can 
see the cell 1(9) which involves the sequence rule~6, \laff {W} {WBWWWBWBB} {W} {,}
rule~29, \laff {W} {WBWWWBBBB} {B} {,} rule~18, \laff {B} {WBWWWBWBB} {W} {} and
rule~24, \laff {W} {BBWWWBWBB} {W} {.} The cell has four milestones: neighbours~2 and~9
as previously and two others: neighbours~6 and~8 as shown by rule~6. The locomotive 
enters via neighbour~7, see rule~29. It is again leaving through neighbour~1, see
rule~24. The same sequence appears for the cell~1(5). In both cases, the change in the
neighbourhood is caused by a change of direction in the motion: a kind of 
\textit{obstacle} occurs which makes the locomotive to turn around a new cell.

The fourth and fifth rows of Figure~\ref{voieverts} illustrate the motion on the same
part of the tracks but when it is performed in the reverse direction. New rules are
needed together with several rules of the previous execution, namely the rules~43 up
to~58 from Table~\ref{rulesvs}. Table~\ref{execvmd} illustrates the execution of the
rules in this case. We can see that this time, the rules which correspond to the
previous sequence 5, 17, 23, 26 are now the rules~44, \laff {W} {WBWBWWWWB} {W} {,}
47, \laff {W} {WBBBWWWWB} {B} {,} 51, \laff {B} {WBWBWWWWB} {W} {} 
and 53, \laff {W} {BBWBWWWWB} {W} {.} The milestones are now neighbours~2, 4 and~9,
see rules~44 and~51. The locomotive enters the cell through its neighbour~3, see
rule~47 and rule~53 witnesses that the locomotive leaves the cell through neighbour~1
as required by the definition of the numbering. Again, for cells with four milestones,
we have different rules: rule~45, \laff {W} {WBBWBWWWB} {W} {,}
rule~54, \laff {W} {WBBBBWWWB} {B} {,} rule~46, \laff {B} {WBBWBWWWB} {W} {}
and rule~50, \laff {W} {BBBWBWWWB} {W} {.} The milestones are now neighbours~2,3, 5
and~9, see rules~45 and~46, the locomotive enters through neighbour~4, see rule~54
and it leaves through neighbour~1, see rule~50.

\vtop{
\begin{tab}\label{execvmd}
\leurre
Execution of the rules $1$ up to~$58$ for the motion in the reverse direction
with respect to Table~{\rm\ref{execvdm}}.
\end{tab}
\vskip 2pt
\trep
\vskip 8pt
\ligne{\hfill\HH{}
\HH{{4$_4$} }\HH{{5$_4$} }\HH{{6$_4$} }\HH{{2$_5$} }\HH{{1$_5$} }\HH{{1$_6$} }
\HH{{1$_7$} }\HH{{1$_8$} }\HH{{1$_9$} }\HH{{2$_1$} }\HH{{3$_1$} }\HH{{4$_1$} }\HH{{5$_1$} }\hfill}
\ligne{\hfill\HH{1}
\HH{50}\HH{\Rr{51}}\HH{\Rr{47}}\HH{44}\HH{45}\HH{44}
\HH{44}\HH{44}\HH{45}\HH{44}\HH{44}\HH{44}\HH{45}\hfill}

\ligne{\hfill\HH{2}
\HH{45}\HH{53}\HH{\Rr{51}}\HH{\Rr{47}}\HH{45}\HH{44}
\HH{44}\HH{44}\HH{45}\HH{44}\HH{44}\HH{44}\HH{45}\hfill}

\ligne{\hfill\HH{3}
\HH{45}\HH{44}\HH{53}\HH{\Rr{51}}\HH{\Rr{54}}\HH{44}
\HH{44}\HH{44}\HH{45}\HH{44}\HH{44}\HH{44}\HH{45}\hfill}

\ligne{\hfill\HH{4}
\HH{45}\HH{44}\HH{44}\HH{53}\HH{\Rr{46}}\HH{\Rr{47}}
\HH{44}\HH{44}\HH{45}\HH{44}\HH{44}\HH{44}\HH{45}\hfill}

\ligne{\hfill\HH{5}
\HH{45}\HH{44}\HH{44}\HH{44}\HH{50}\HH{\Rr{51}}
\HH{\Rr{47}}\HH{44}\HH{45}\HH{44}\HH{44}\HH{44}\HH{45}\hfill}

\ligne{\hfill\HH{6}
\HH{45}\HH{44}\HH{44}\HH{44}\HH{45}\HH{53}
\HH{\Rr{51}}\HH{\Rr{47}}\HH{45}\HH{44}\HH{44}\HH{44}\HH{45}\hfill}

\ligne{\hfill\HH{7}
\HH{45}\HH{44}\HH{44}\HH{44}\HH{45}\HH{44}
\HH{53}\HH{\Rr{51}}\HH{\Rr{54}}\HH{44}\HH{44}\HH{44}\HH{45}\hfill}

\ligne{\hfill\HH{8}
\HH{45}\HH{44}\HH{44}\HH{44}\HH{45}\HH{44}
\HH{44}\HH{53}\HH{\Rr{46}}\HH{\Rr{47}}\HH{44}\HH{44}\HH{45}\hfill}

\ligne{\hfill\HH{9}
\HH{45}\HH{44}\HH{44}\HH{44}\HH{45}\HH{44}
\HH{44}\HH{44}\HH{50}\HH{\Rr{51}}\HH{\Rr{47}}\HH{44}\HH{45}\hfill}

\ligne{\hfill\HH{10}
\HH{45}\HH{44}\HH{44}\HH{44}\HH{45}\HH{44}
\HH{44}\HH{44}\HH{45}\HH{53}\HH{\Rr{51}}\HH{\Rr{47}}\HH{45}\hfill}

\ligne{\hfill\HH{11}
\HH{45}\HH{44}\HH{44}\HH{44}\HH{45}\HH{44}
\HH{44}\HH{44}\HH{45}\HH{44}\HH{53}\HH{\Rr{51}}\HH{\Rr{54}}\hfill}
\vskip 9pt
\trfn
\vskip 15pt
}

\vtop{
\begin{tab}\label{execvb}
\leurre
Execution on the cells~$1(2)$ and $1(3)$ of Figure~{\rm\ref{voieverts}}. To left,
clockwise, to right, counter-clockwise.
\end{tab}
\vskip 2pt
\trep
\vskip 8pt
\ligne{\hfill
\vtop{\hsize=150pt
\ligne{\hfill\HH{}
\HH{{11$_4$}}\HH{{2$_4$} }\HH{{1$_3$} }\HH{{1$_2$} }\HH{{2$_2$} }\HH{{7$_2$} }
\hfill}
\vskip 0pt
\ligne{\hfill\HH{1}
\HH{21}\HH{\Rr{18}}\HH{\Rr{29}}\HH{6}\HH{6}\HH{10}\hfill}
\vskip 0pt
\ligne{\hfill\HH{2}
\HH{10}\HH{24}\HH{\Rr{18}}\HH{\Rr{29}}\HH{6}\HH{10}\hfill}
\vskip 0pt
\ligne{\hfill\HH{3}
\HH{10}\HH{6}\HH{24}\HH{\Rr{18}}\HH{\Rr{29}}\HH{10}\hfill}
}
\hskip 20pt
\vtop{\hsize=150pt
\ligne{\hfill\HH{}
\HH{{7$_2$} }\HH{{2$_2$} }\HH{{1$_2$} }\HH{{1$_3$} }\HH{{2$_4$} }\HH{{11$_4$}}
\hfill}
\ligne{\hfill\HH{1}
\HH{21}\HH{\Rr{46}}\HH{\Rr{54}}\HH{45}\HH{45}\HH{10}\hfill}
\vskip 0pt
\ligne{\hfill\HH{2}
\HH{10}\HH{50}\HH{\Rr{46}}\HH{\Rr{54}}\HH{45}\HH{10}\hfill}
\vskip 0pt
\ligne{\hfill\HH{3}
\HH{10}\HH{45}\HH{50}\HH{\Rr{46}}\HH{\Rr{54}}\HH{10}\hfill}
}
\hfill}
\vskip 9pt
\trfn
\vskip 10pt
}

Now, consider the third and the last row of Figure~\ref{voieverts}. In one direction, for
the cells 1(3) and 1(2) in this order, we find the sequence of rules 6, 29, 18, 24 we have
already seen for cells of the track with four milestones. In the opposite direction,
we also find the sequence 45, 54, 46, 50 we observed in Table~\ref{execvmd}.

\ligne{\hfill   
\vtop{\leftskip 0pt\parindent 0pt\hsize=\tabruleli  
\ligne{
\vtop{\leftskip 0pt\parindent 0pt\hsize=300pt  
\begin{tab}\label{rulesvd}
\leurre
Rules for a double locomotive on the tracks:
\end{tab}
}  
\hfill}  
\vskip 2pt
\trep
\vskip 8pt
\ligne{\hfill  
\vtop{\leftskip 0pt\parindent 0pt\hsize=\tabrulecol
\aff { 59} {B} {WWBBBWWBW} {B}
\aff { 60} {B} {WBWWWWBBB} {B}
\aff { 61} {B} {BBWWWBWBB} {W}
\aff { 62} {B} {WBBWBWWBW} {B}
\aff { 63} {B} {BBWWWWBWB} {W}
\aff { 64} {B} {WWWWWWBBW} {B}
}\hskip 10pt
\vtop{\leftskip 0pt\parindent 0pt\hsize=\tabrulecol
\aff { 65} {B} {BBWWBWWBW} {B}
\aff { 66} {B} {WWWWWWWBW} {B}
\aff { 67} {B} {BWWWBWWBB} {B}
\aff { 68} {B} {WBWWWBBBB} {B}
\aff { 69} {W} {WWWWWWWBB} {W}
\aff { 70} {B} {WWWWWWWBB} {B}
}\hskip 10pt
\vtop{\leftskip 0pt\parindent 0pt\hsize=\tabrulecol
\aff { 71} {B} {WBBBWWBWW} {B}
\aff { 72} {B} {BBWBWWBWW} {B}
\aff { 73} {B} {BWWBWWBWB} {B}
\aff { 74} {B} {WWWBWWBBB} {B}
\aff { 75} {B} {BWWWWWWWB} {B}
\aff { 76} {B} {WWWBBBBWW} {B}
}\hskip 10pt
\vtop{\leftskip 0pt\parindent 0pt\hsize=\tabrulecol
\aff { 77} {W} {WWBBWWWWW} {W}
\aff { 78} {B} {WWWWBBBBW} {B}
\aff { 79} {B} {BBBWBWWWB} {W}
\aff { 80} {B} {WBBBWWWWB} {B}
\aff { 81} {B} {BBWBWWWWB} {W}
\aff { 82} {B} {WBBBBWWWB} {B}
}
\hfill}  
\vskip 9pt
\trfn
}  
\hfill} 

\vtop{
\begin{tab}\label{execvdmd}
\leurre
Execution for a double locomotive in the path illustrated by 
Figure~{\rm\ref{stab_tracks}} in the clockwise direction.
\end{tab}
\vskip 2pt
\trep
\vskip 8pt
\ligne{\hfill\HH{}
\HH{{5$_1$} }\HH{{4$_1$} }\HH{{3$_1$} }\HH{{2$_1$} }\HH{{1$_9$} }\HH{{1$_8$} }
\HH{{1$_7$} }\HH{{1$_6$} }\HH{{1$_5$} }\HH{{2$_5$} }\HH{{6$_4$} }\HH{{5$_4$} }\HH{{4$_4$} }\hfill}
\ligne{\hfill\HH{1}
\HH{24}\HH{\Rr{63}}\HH{60}\HH{\Rr{17}}\HH{6}\HH{5}
\HH{5}\HH{5}\HH{6}\HH{5}\HH{5}\HH{5}\HH{6}\hfill}

\ligne{\hfill\HH{2}
\HH{6}\HH{26}\HH{\Rr{63}}\HH{60}\HH{\Rr{29}}\HH{5}
\HH{5}\HH{5}\HH{6}\HH{5}\HH{5}\HH{5}\HH{6}\hfill}

\ligne{\hfill\HH{3}
\HH{6}\HH{5}\HH{26}\HH{\Rr{63}}\HH{68}\HH{\Rr{17}}
\HH{5}\HH{5}\HH{6}\HH{5}\HH{5}\HH{5}\HH{6}\hfill}

\ligne{\hfill\HH{4}
\HH{6}\HH{5}\HH{5}\HH{26}\HH{\Rr{61}}\HH{60}
\HH{\Rr{17}}\HH{5}\HH{6}\HH{5}\HH{5}\HH{5}\HH{6}\hfill}

\ligne{\hfill\HH{5}
\HH{6}\HH{5}\HH{5}\HH{5}\HH{24}\HH{\Rr{63}}
\HH{60}\HH{\Rr{17}}\HH{6}\HH{5}\HH{5}\HH{5}\HH{6}\hfill}

\ligne{\hfill\HH{6}
\HH{6}\HH{5}\HH{5}\HH{5}\HH{6}\HH{26}
\HH{\Rr{63}}\HH{60}\HH{\Rr{29}}\HH{5}\HH{5}\HH{5}\HH{6}\hfill}

\ligne{\hfill\HH{7}
\HH{6}\HH{5}\HH{5}\HH{5}\HH{6}\HH{5}
\HH{26}\HH{\Rr{63}}\HH{68}\HH{\Rr{17}}\HH{5}\HH{5}\HH{6}\hfill}

\ligne{\hfill\HH{8}
\HH{6}\HH{5}\HH{5}\HH{5}\HH{6}\HH{5}
\HH{5}\HH{26}\HH{\Rr{61}}\HH{60}\HH{\Rr{17}}\HH{5}\HH{6}\hfill}

\ligne{\hfill\HH{9}
\HH{6}\HH{5}\HH{5}\HH{5}\HH{6}\HH{5}
\HH{5}\HH{5}\HH{24}\HH{\Rr{63}}\HH{60}\HH{\Rr{17}}\HH{6}\hfill}

\ligne{\hfill\HH{10}
\HH{6}\HH{5}\HH{5}\HH{5}\HH{6}\HH{5}
\HH{5}\HH{5}\HH{6}\HH{26}\HH{\Rr{63}}\HH{60}\HH{\Rr{29}}\hfill}

\ligne{\hfill\HH{11}
\HH{6}\HH{5}\HH{5}\HH{5}\HH{6}\HH{5}
\HH{5}\HH{5}\HH{6}\HH{5}\HH{26}\HH{\Rr{63}}\HH{68}\hfill}
\vskip 9pt
\trfn
\vskip 10pt
}

Let us look at the motion of a double locomotive on the tracks defined by 
Figure~\ref{stab_tracks}. The sequence starts and ends with the same rules as with a simple
locomotive but in between three rules operate instead of two ones and they belong to
Table~\ref{rulesvd}. Accordingly, for a cell with three milestones, the sequence
consists of five rules: rule~5, \laff {W} {WBWWWWBWB} {W} {,}
rule~17, \laff {W} {WBWWWWBBB} {B} {,} rule~60, \laff {B} {WBWWWWBBB} {B} {,}
rule~63, \laff {B} {BBWWWWBWB} {W} {} and rule~26, \laff {W} {BBWWWWBWB} {W} {.}

\vtop{
\vskip 0pt
\ligne{\hfill
\includegraphics[scale=0.55]{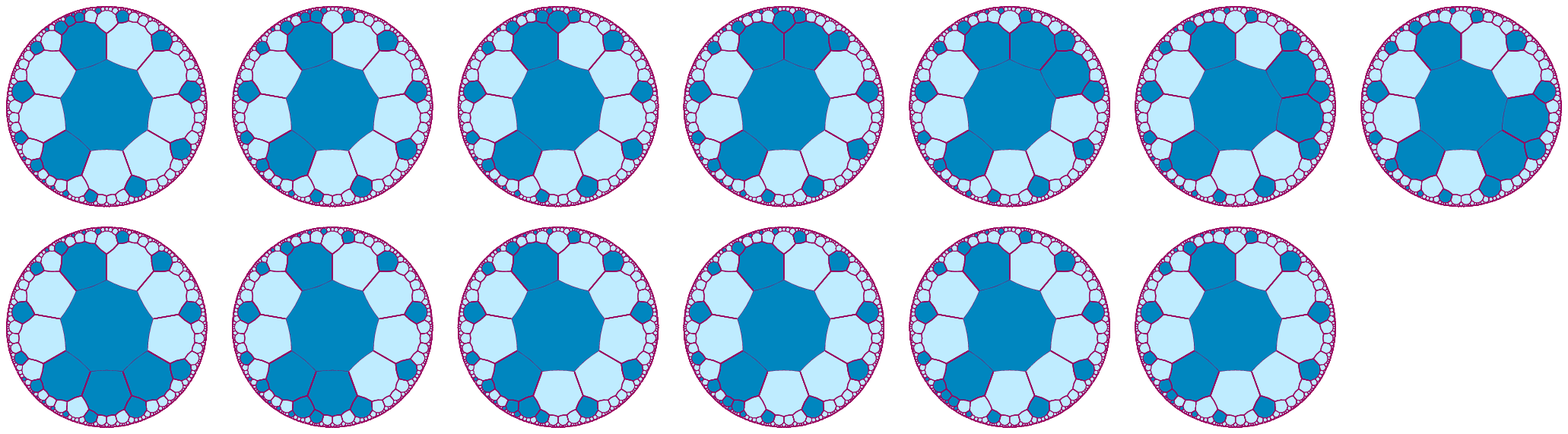} 
\hfill}
\vspace{-10pt}
\ligne{\hfill
\includegraphics[scale=0.55]{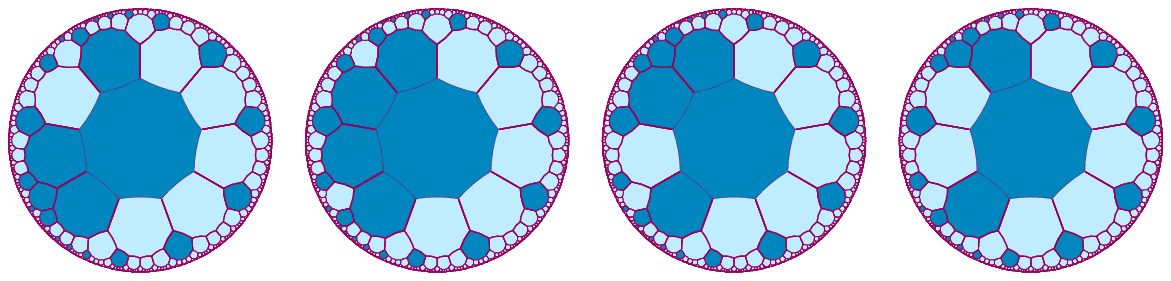} 
\hfill}
\vspace{-10pt}
\ligne{\hfill
\includegraphics[scale=0.55]{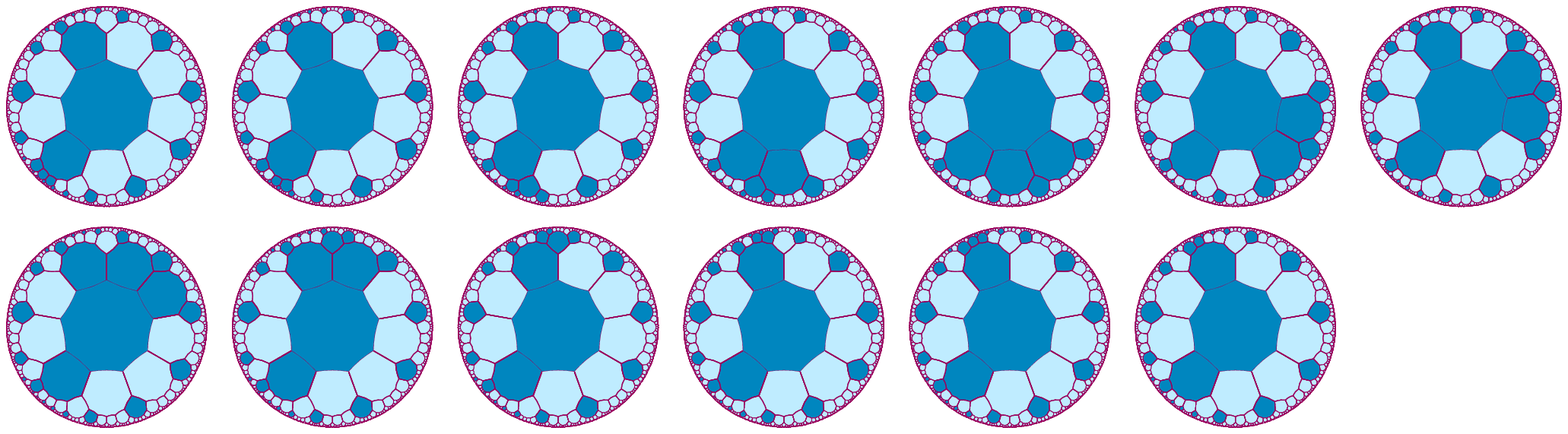} 
\hfill}
\vspace{-10pt}
\ligne{\hfill
\includegraphics[scale=0.55]{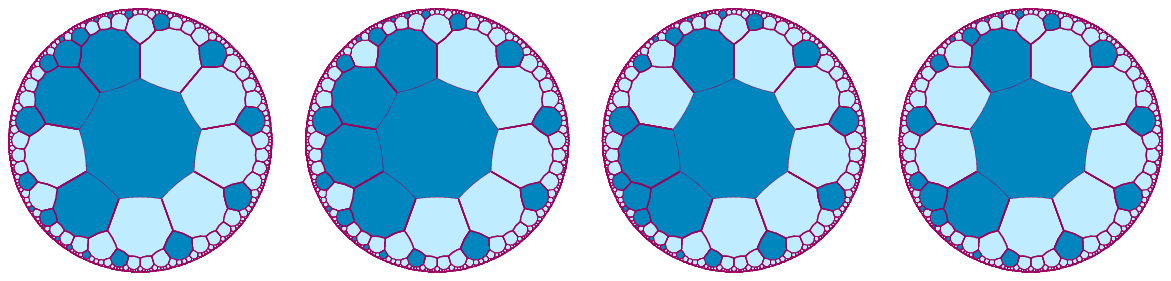} 
\hfill}
\vspace{-15pt}
\ligne{\hfill
\vtop{\leftskip 0pt\parindent 0pt\hsize=340pt
\begin{fig}\label{voievertd}
\leurre
From left to right: vertical tracks, horizontal tracks around a black node, horizontal
tracks around a white node. In each group, tracks in one direction and then in the
opposite direction.
\end{fig}
}
\hfill}
\vskip 10pt
}

\vtop{
\begin{tab}\label{execvbd}
\leurre
Execution on the cells~$1(3)$ and~$1(2)$ with a double locomotive in both directions.
\end{tab}
\vskip 2pt
\trep
\vskip 8pt
\ligne{\hfill
\vtop{\hsize=150pt
\ligne{\hfill\HH{}
\HH{{11$_4$}}\HH{{2$_4$} }\HH{{1$_3$} }\HH{{1$_2$} }\HH{{2$_2$} }\HH{{7$_2$} }
\hfill}
\ligne{\hfill\HH{1}
\HH{21}\HH{\Rr{61}}\HH{68}\HH{\Rr{29}}\HH{6}\HH{10}\hfill}
\vskip 0pt
\ligne{\hfill\HH{2}
\HH{10}\HH{24}\HH{\Rr{61}}\HH{68}\HH{\Rr{29}}\HH{10}\hfill}
\vskip 0pt
\ligne{\hfill\HH{2}
\HH{10}\HH{6}\HH{24}\HH{\Rr{61}}\HH{68}\HH{21}\hfill}
}
\hskip 20pt
\vtop{\hsize=150pt
\ligne{\hfill\HH{}
\HH{{7$_2$} }\HH{{2$_2$} }\HH{{1$_2$} }\HH{{1$_3$} }\HH{{2$_4$} }\HH{{11$_4$}}
\hfill}
\vskip 0pt
\ligne{\hfill\HH{1}
\HH{21}\HH{\Rr{79}}\HH{82}\HH{\Rr{54}}\HH{45}\HH{10}\hfill}
\vskip 0pt
\ligne{\hfill\HH{1}
\HH{10}\HH{50}\HH{\Rr{79}}\HH{82}\HH{\Rr{54}}\HH{10}\hfill}
\vskip 0pt
\ligne{\hfill\HH{2}
\HH{10}\HH{45}\HH{50}\HH{\Rr{79}}\HH{82}\HH{21}\hfill}
}
\hfill}
\vskip 9pt
\trfn
\vskip 10pt
}

\vtop{
\begin{tab}\label{execvmdd}
\leurre
Execution for a double locomotive in the path illustrated by 
Figure~{\rm\ref{stab_tracks}} in the counter-clockwise direction.
\end{tab}
\vskip 2pt
\trep
\vskip 8pt
\ligne{\hfill\HH{}
\HH{{4$_4$} }\HH{{5$_4$} }\HH{{6$_4$} }\HH{{2$_5$} }\HH{{1$_5$} }\HH{{1$_6$} }
\HH{{1$_7$} }\HH{{1$_8$} }\HH{{1$_9$} }\HH{{2$_1$} }\HH{{3$_1$} }\HH{{4$_1$} }\HH{{5$_1$} }\hfill}
\ligne{\hfill\HH{1}
\HH{50}\HH{\Rr{81}}\HH{80}\HH{\Rr{47}}\HH{45}\HH{44}
\HH{44}\HH{44}\HH{45}\HH{44}\HH{44}\HH{44}\HH{45}\hfill}

\ligne{\hfill\HH{2}
\HH{45}\HH{53}\HH{\Rr{81}}\HH{80}\HH{\Rr{54}}\HH{44}
\HH{44}\HH{44}\HH{45}\HH{44}\HH{44}\HH{44}\HH{45}\hfill}

\ligne{\hfill\HH{3}
\HH{45}\HH{44}\HH{53}\HH{\Rr{81}}\HH{82}\HH{\Rr{47}}
\HH{44}\HH{44}\HH{45}\HH{44}\HH{44}\HH{44}\HH{45}\hfill}

\ligne{\hfill\HH{4}
\HH{45}\HH{44}\HH{44}\HH{53}\HH{\Rr{79}}\HH{80}
\HH{\Rr{47}}\HH{44}\HH{45}\HH{44}\HH{44}\HH{44}\HH{45}\hfill}

\ligne{\hfill\HH{5}
\HH{45}\HH{44}\HH{44}\HH{44}\HH{50}\HH{\Rr{81}}
\HH{80}\HH{\Rr{47}}\HH{45}\HH{44}\HH{44}\HH{44}\HH{45}\hfill}

\ligne{\hfill\HH{6}
\HH{45}\HH{44}\HH{44}\HH{44}\HH{45}\HH{53}
\HH{\Rr{81}}\HH{80}\HH{\Rr{54}}\HH{44}\HH{44}\HH{44}\HH{45}\hfill}

\ligne{\hfill\HH{7}
\HH{45}\HH{44}\HH{44}\HH{44}\HH{45}\HH{44}
\HH{53}\HH{\Rr{81}}\HH{82}\HH{\Rr{47}}\HH{44}\HH{44}\HH{45}\hfill}

\ligne{\hfill\HH{8}
\HH{45}\HH{44}\HH{44}\HH{44}\HH{45}\HH{44}
\HH{44}\HH{53}\HH{\Rr{79}}\HH{80}\HH{\Rr{47}}\HH{44}\HH{45}\hfill}

\ligne{\hfill\HH{9}
\HH{45}\HH{44}\HH{44}\HH{44}\HH{45}\HH{44}
\HH{44}\HH{44}\HH{50}\HH{\Rr{81}}\HH{80}\HH{\Rr{47}}\HH{45}\hfill}

\ligne{\hfill\HH{10}
\HH{45}\HH{44}\HH{44}\HH{44}\HH{45}\HH{44}
\HH{44}\HH{44}\HH{45}\HH{53}\HH{\Rr{81}}\HH{80}\HH{\Rr{54}}\hfill}

\ligne{\hfill\HH{11}
\HH{45}\HH{44}\HH{44}\HH{44}\HH{45}\HH{44}
\HH{44}\HH{44}\HH{45}\HH{44}\HH{53}\HH{\Rr{81}}\HH{82}\hfill}
\vskip 9pt
\trfn
\vskip 10pt
}

The three milestones are in neighbours~2, 7 and~9 as witnessed by rule~5. The first
cell of the locomotive enters the cell thanks to rule~17 as in the case of a simple
locomotive: with a neighbourhood of radius~1, the cell cannot see whether the
first cell of the locomotive is followed or not by a second one. Now, when the first cell
of the locomotive is in the cell, it can see the second cell of the locomotive in
neighbour~8 which is still black. Next, rule~63 makes the cell return to white
as its neighbourhood now can see the first cell of the locomotive in its neighbour~1.
Rule~26 witnesses that the second cell of the locomotive is in neighbour~1: as the 
first cell of the locomotive is now no more visible from the cell, this why we can 
here too apply rule~26 which was also applied when a single locomotive leaves the cell.

For a cell of the track with four milestones, the rules are now
rule~29, \laff {W} {WBWWWBBBB} {B} {,} rule~68, \laff {B} {WBWWWBBBB} {B} {,}
rule~61, \laff {B} {BBWWWBWBB} {W} {} and rule~24, \laff {W} {BBWWWBWBB} {W} {.}
The milestones are in neighbours~2, 6, 8 and~9. The first cell of the locomotive 
is seen in neighbour~7 by the cell, see rule~29 which makes the first cell of the
locomotive enter the cell. Rule~68 can see the first cell of the locomotive in the
cell itself and the second cell in neighbour~7 again, so that the cell remains black:
it is the second cell of the locomotive; Rule~61 witnesses that the first cell of the
locomotive is now in neighbour~1, so that the rule makes the cell turn to white.
Rule~24 witnesses that the second cell of the locomotive is in neighbour~1.
\vskip 15pt
   We remind the reader that we can define a track leading from one cell of the
tessellation to another one.  We can choose the path leading to these cells
in order to implement the patterns defined in Section~\ref{scenar}. We can
define analogs for the Euclidean horizontal and verticals: we follow a level of the
tree for the former, a branch of the tree for the latter. We still adopt the terms
horizontal and vertical in the context of the tessellation~$\{9,3\}$.

\subsection{The fixed switch}
\label{fixedswitch}

   Figure~\ref{stabfix} illustrates the configuration of a passive fixed switch.
The same configuration is used whether the locomotive comes from the left-hand side 
or it comes from the right-hand one. Note that the locomotive may be simple as well 
as double: indeed, the fixed switch plays a role in the round-about, this is
why it may be crossed by a double locomotive.

   Table~\ref{rulesfx} displays the rules used for the passive fixed switch. More 
precisely, the table gives the rules which were not yet used. In fact several rules 
from the vertical motions are also used in this case.

\vtop{  
\begin{tab}\label{rulesfx}
\leurre
Rules for the locomotives crossing a fixed switch. 
\end{tab}
\vskip 2pt
\trep
\vskip 8pt
\ligne{\hfill Simple locomotive\hfill}
\vskip 8pt
\ligne{\hfill  
\vtop{\leftskip 0pt\parindent 0pt\hsize=\tabrulecol
\ligne{from the left:\hfill}
\vskip 4pt
\aff { 83} {W} {WBWBWBBWB} {W}
\aff { 84} {B} {WWWWWBWWB} {B}
\aff { 85} {B} {WBWWBBWWW} {B}
\aff { 86} {W} {BBBWWWWWW} {W}
\aff { 87} {B} {WWBWWWWWW} {B}
\aff { 88} {B} {WWWWWBWWW} {B}
\aff { 89} {W} {WBWBWWWWW} {W}
}
\hskip 10pt
\vtop{\leftskip 0pt\parindent 0pt\hsize=\tabrulecol
\aff { 90} {W} {WBBWWWWWW} {W}
\aff { 91} {W} {BWWWWBWBW} {W}
\aff { 92} {B} {WBWWBWBWW} {B}
\aff { 93} {W} {BWBWWWWWW} {W}
\aff { 94} {W} {WBWBWBBBB} {B}
\aff { 95} {B} {WBWWBWWWB} {B}
\aff { 96} {B} {WBWBWBBWB} {W}
\aff { 97} {B} {BWWWWBWWB} {B}
\aff { 98} {B} {BBWWBWWWW} {B}
}
\hskip 10pt
\vtop{\leftskip 0pt\parindent 0pt\hsize=\tabrulecol
\aff { 99} {W} {BBWBWBBWB} {W}
\aff {100} {B} {WBWWBWWWW} {B}
\aff {101} {B} {BWWWWBWWW} {B}
\aff {102} {B} {WBWWWBWWW} {B}
\aff {103} {W} {WBBBWWWWW} {B}
\aff {104} {B} {WWBWWBWWW} {B}
\aff {105} {B} {WBWBWWWWW} {W}
}
\hskip 10pt
\vtop{\leftskip 0pt\parindent 0pt\hsize=\tabrulecol
\ligne{from the right:\hfill}
\vskip 4pt
\aff {106} {B} {WWWWBBWWB} {B}
\aff {107} {B} {WWWBWBWWB} {B}
\aff {108} {B} {WWBWWBWWB} {B}
\aff {109} {W} {WBWBBBBWB} {B}
\aff {110} {B} {WBWWWBWWB} {B}
}
\hfill}
\vskip 8pt
\ligne{\hfill Double locomotive\hfill}
\vskip 8pt
\ligne{\hfill  
\vtop{\leftskip 0pt\parindent 0pt\hsize=\tabrulecol
\ligne{from the left:\hfill}
\vskip 4pt
\aff {111} {B} {WBWWBBBWW} {B}
\aff {112} {B} {WBWWBWBBW} {B}
\aff {113} {B} {WBWWBWWBB} {B}
\aff {114} {B} {WBWBWBBBB} {B}
}\hskip 10pt
\vtop{\leftskip 0pt\parindent 0pt\hsize=\tabrulecol
\aff {115} {B} {BBWWBWWWB} {B}
\aff {116} {B} {BBWBWBBWB} {W}
\aff {117} {B} {BBWWWBWWW} {B}
\aff {118} {B} {WBBWWBWWW} {B}
\aff {119} {B} {WBBBWWWWW} {B}
}\hskip 10pt
\vtop{\leftskip 0pt\parindent 0pt\hsize=\tabrulecol
\ligne{from the right:\hfill}
\vskip 4pt
\aff {120} {B} {WWWBBBWWB} {B}
\aff {121} {B} {WWBBWBWWB} {B}
\aff {122} {B} {WBBWWBWWB} {B}
}\hskip 10pt
\vtop{\leftskip 0pt\parindent 0pt\hsize=\tabrulecol
\aff {123} {B} {WBWBBBBWB} {B}
\aff {124} {B} {BBWWWBWWB} {B}
}
\hfill} 
\vskip 9pt
\trfn
\vskip 10pt
} 

Note that the rules~89, 103 and~105 are altered forms of the rules~44, 47 and~51
which are rules for standard cells of the track. The alteration is induced by the 
limitation of the space used for the simulation of the computation of the automaton
by the computer programm.

The switch is passive and it works both for a single and a double locomotive whatever 
the side from which they arrive to the switch. Note that the exceptional cell is the 
centre of the switch, the cell~0(0) in Figures~\ref{stabfix}. 
Rule~83, \laff {W} {WBWBWBBWB} {W} {,} is the conservative rule which is applied when 
no locomotive crosses the switch. Its milestones are neighbours~2, 4, 6, 7 and~9.
Neighbours~2 and~9 are milestones for the cell~1(5) which is the first cell of the 
track which leaves the switch. Neighbours~4 and~6 are milestones for the cell~1(9)
which belongs to the track which arrives to the cell~0(0) from the right. Neighbours~7 
and~9 are milestones for the cell~1(3) which belongs to the track which arrives to the
cell~0(0) from the left.

\vtop{
\ligne{\hfill
\includegraphics[scale=0.55]{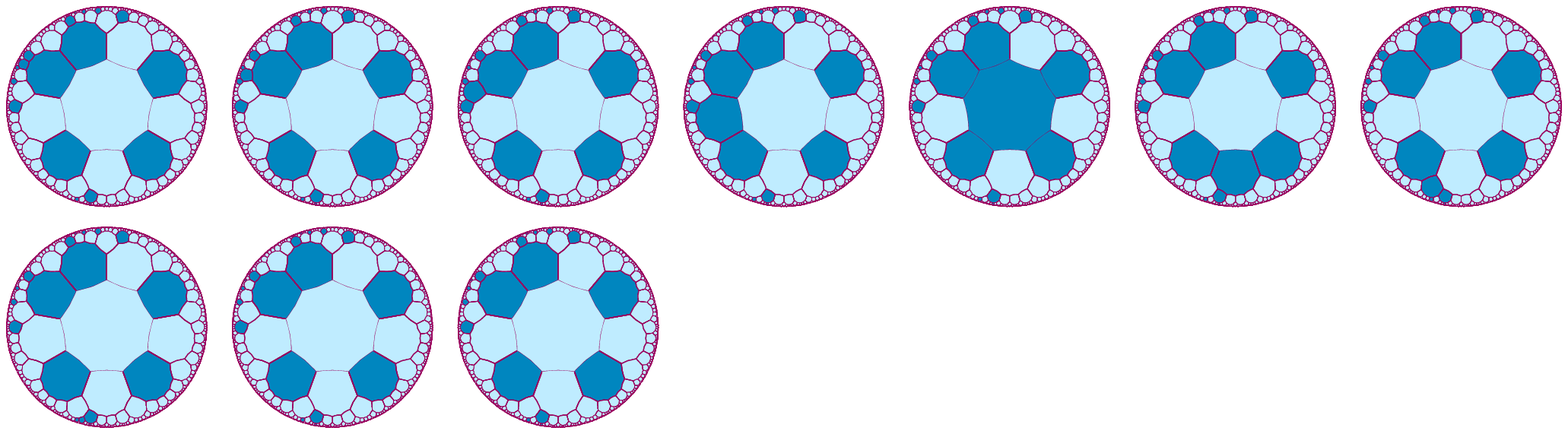} 
\hfill}
\ligne{\hfill
\includegraphics[scale=0.55]{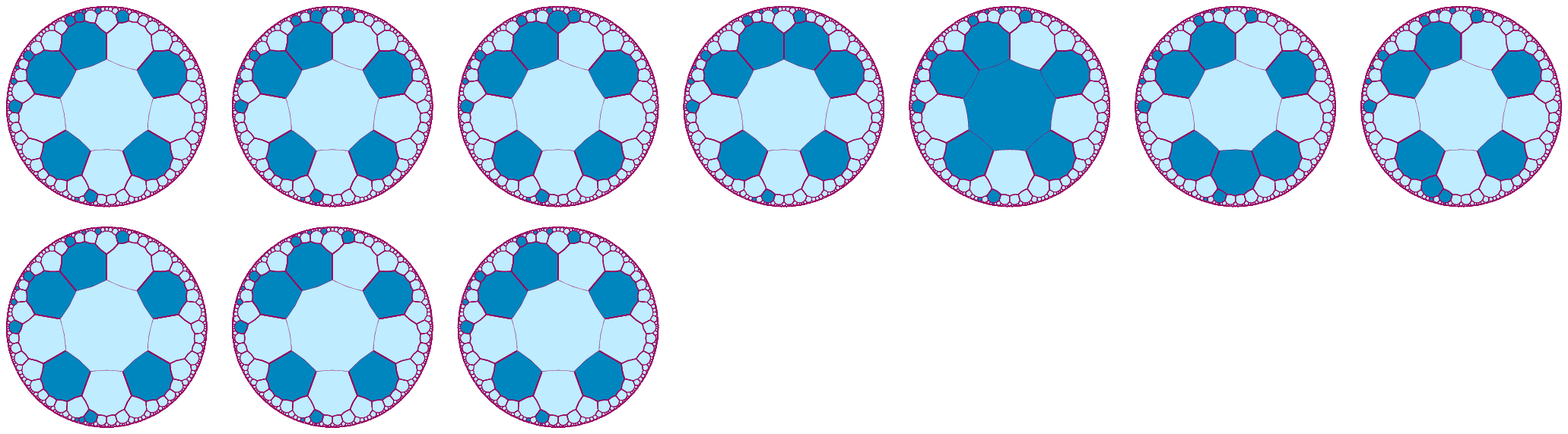} 
\hfill}
\vskip -5pt
\begin{fig}\label{fixes}
\leurre
Locomotives passively crossing a fixed switch from the left-hand side. Above : a single
locomotive; below: a double one.
\end{fig}
}

\vtop{
\ligne{\hskip 5pt
\includegraphics[scale=0.55]{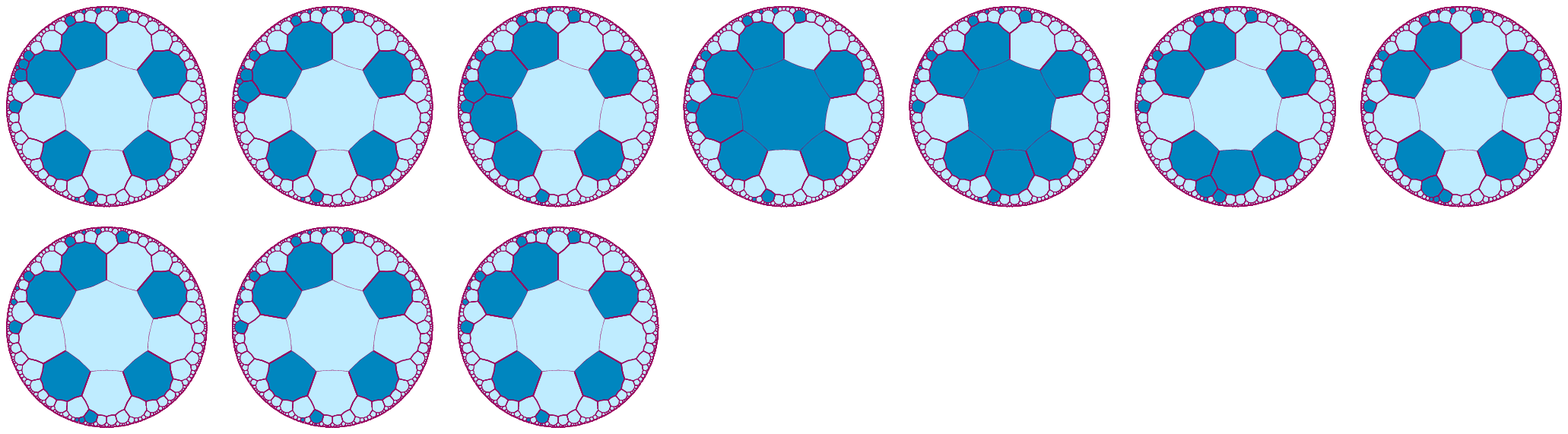} 
\hfill}
\ligne{\hskip 5pt
\includegraphics[scale=0.55]{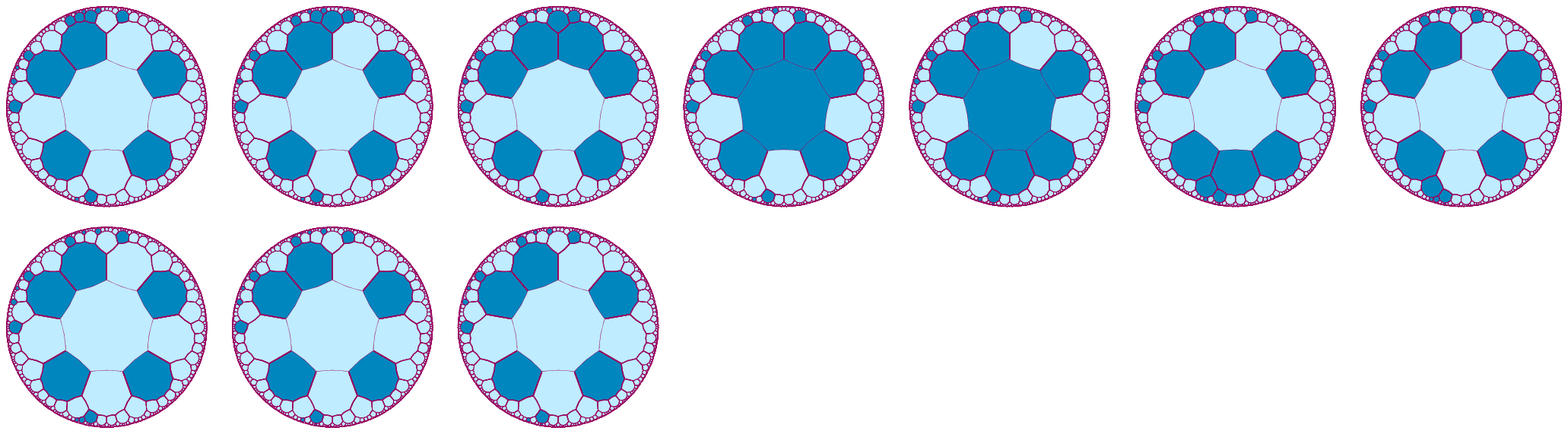} 
\hfill}
\vskip -5pt
\begin{fig}\label{fixed}
\leurre
Locomotives passively crossing a fixed switch from the right-hand side. Above : a single
locomotive; below: a double one.
\end{fig}
}
\vskip 10pt

On the path leading to the left-hand side entrance of the fixed switch,
the rules are those we have already seen: rules~44, 47, 51 and~53. In the central
cell, the rules are rule~83, we already know,
rule~94, \laff {W} {WBWBWBBBB} {B} {,} rule~96, \laff {B} {WBWBWBBWB} {W} {} and
rule~99, \laff {W} {BBWBWBBWB} {W} {.} 
The entry from the left-hand side is neighbour~8, from the right-hand side it is 
neighbour~5.
The exit from the central cell is performed through neighbour~1. Rule~94 can see the
arriving locomotive in neighbour~8, rule~96 makes it leave the cell, while rule~99
witnesses that the locomotive left the cell, being now in neighbour~1. Note the numbering
which is conformal to the way adopted for the tracks.

\ligne{\hfill
\vtop{\hsize=220pt
\begin{tab}\label{execfxgs}
\leurre
Execution of the rules for a simple locomotive crossing the passive fixed switch 
from the left.
\end{tab}
\vskip 2pt
\trep
\vskip 8pt
\ligne{\hfill\HH{}
\HH{{5$_2$} }\HH{{6$_2$} }\HH{{2$_3$} }\HH{{1$_3$} }\HH{{0$_0$} }\HH{{1$_5$} }\HH{{2$_5$} }\HH{{11$_5$}}
\hfill}
\ligne{\hfill\HH{1}
\HH{50}\HH{\Rr{51}}\HH{\Rr{47}}\HH{44}\HH{83}\HH{5}\HH{44}\HH{89}\hfill}
\vskip 0pt
\ligne{\hfill\HH{2}
\HH{45}\HH{53}\HH{\Rr{51}}\HH{\Rr{47}}\HH{83}\HH{5}\HH{44}\HH{89}\hfill}
\vskip 0pt
\ligne{\hfill\HH{3}
\HH{45}\HH{44}\HH{53}\HH{\Rr{51}}\HH{\Rr{94}}\HH{5}\HH{44}\HH{89}\hfill}
\vskip 0pt
\ligne{\hfill\HH{4}
\HH{45}\HH{44}\HH{44}\HH{53}\HH{\Rr{96}}\HH{\Rr{17}}\HH{44}\HH{89}\hfill}
\vskip 0pt
\ligne{\hfill\HH{5}
\HH{45}\HH{44}\HH{44}\HH{44}\HH{99}\HH{\Rr{23}}\HH{\Rr{47}}\HH{89}\hfill}
\vskip 0pt
\ligne{\hfill\HH{6}
\HH{45}\HH{44}\HH{44}\HH{44}\HH{83}\HH{26}\HH{\Rr{51}}\HH{\Rr{103}}\hfill}
\vskip 0pt
\ligne{\hfill\HH{7}
\HH{45}\HH{44}\HH{44}\HH{44}\HH{83}\HH{5}\HH{53}\HH{\Rr{105}}\hfill}
\ligne{\hfill\HH{8}
\HH{45}\HH{44}\HH{44}\HH{44}\HH{83}\HH{5}\HH{44}\HH{89}\hfill}
\vskip 9pt
\trfn
\vskip 10pt
}
\hfill}

   Note that, for leaving the switch, the path makes use of the rules~5, 17, 23 and~26
we have already seen, it also makes use of the rules 44, 47, 51 and~53 for the next cell.
Also note that the rules~89, 103 and~105, in which neighbour~9 is white, are altered 
forms of the rules~44, 77 and~51 as already mentioned. In the abstract model, we can
see that either rules~5, 17, 23 and~26 are used or rules 44, 47, 51 and~53. Both of those
sequences of rules have been used for the tracks.

\ligne{\hfill
\vtop{\hsize=220pt
\begin{tab}\label{execfxds}
\leurre
Execution of the rules for a simple locomotive crossing the passive fixed switch 
from the right.
\end{tab}
\vskip 2pt
\trep
\vskip 8pt
\ligne{\hfill\HH{}
\HH{{4$_1$} }\HH{{3$_1$} }\HH{{2$_1$} }\HH{{1$_9$} }\HH{{0$_0$} }\HH{{1$_5$} }\HH{{2$_5$} }\HH{{11$_5$}}
\hfill}
\vskip 0pt
\ligne{\hfill\HH{1}
\HH{24}\HH{\Rr{23}}\HH{\Rr{17}}\HH{5}\HH{83}\HH{5}\HH{44}\HH{89}\hfill}
\vskip 0pt
\ligne{\hfill\HH{2}
\HH{6}\HH{26}\HH{\Rr{23}}\HH{\Rr{17}}\HH{83}\HH{5}\HH{44}\HH{89}\hfill}
\vskip 0pt
\ligne{\hfill\HH{3}
\HH{6}\HH{5}\HH{26}\HH{\Rr{23}}\HH{\Rr{109}}\HH{5}\HH{44}\HH{89}\hfill}
\vskip 0pt
\ligne{\hfill\HH{4}
\HH{6}\HH{5}\HH{5}\HH{26}\HH{\Rr{96}}\HH{\Rr{17}}\HH{44}\HH{89}\hfill}
\vskip 0pt
\ligne{\hfill\HH{5}
\HH{6}\HH{5}\HH{5}\HH{5}\HH{99}\HH{\Rr{23}}\HH{\Rr{47}}\HH{89}\hfill}
\vskip 0pt
\ligne{\hfill\HH{6}
\HH{6}\HH{5}\HH{5}\HH{5}\HH{83}\HH{26}\HH{\Rr{51}}\HH{\Rr{103}}\hfill}
\vskip 0pt
\ligne{\hfill\HH{7}
\HH{6}\HH{5}\HH{5}\HH{5}\HH{83}\HH{5}\HH{53}\HH{\Rr{105}}\hfill}
\vskip 0pt
\ligne{\hfill\HH{8}
\HH{6}\HH{5}\HH{5}\HH{5}\HH{83}\HH{5}\HH{44}\HH{89}\hfill}
\vskip 9pt
\trfn
\vskip 10pt
}
\hfill}

Table~\ref{execfxds} shows the rules for a simple locomotive when it arrives to
the fixed switch from the right-hand side. Again, we can see the rules 5, 17, 23 and~26
for the path arriving to the switch. Of course, for the cells~1(5), 2(5) and 11(5), 
we have the same instructions as in Table~\ref{execfxgs}. For the cell~0(0), the
sequence of rules is 83, 109, 96 and~99 instead of 83, 94, 96 and~99. The single 
difference is the rule which makes the cell~0(0) turn to black as the locomotive
arrives from different neighbours. After that moment, the rules are the same.
Note that in rule~109, \laff {W} {WBWBBBBWB} {B} {,} we can see that the locomotive
arrives through neighbour~5. Remember rule~94, \laff {W} {WBWBWBBBB} {B} {,} which
shows the locomotive in neighbour~8.

\ligne{\hfill
\vtop{\hsize=220pt
\begin{tab}\label{execfxgd}
\leurre
Execution of the rules for a double locomotive crossing the passive fixed switch 
from the left.
\end{tab}
\vskip 2pt
\trep
\vskip 8pt
\ligne{\hfill\HH{}
\HH{{5$_2$} }\HH{{6$_2$} }\HH{{2$_3$} }\HH{{1$_3$} }\HH{{0$_0$} }\HH{{1$_5$} }\HH{{2$_5$} }\HH{{11$_5$}}
\hfill}
\vskip 0pt
\ligne{\hfill\HH{1}
\HH{50}\HH{\Rr{81}}\HH{80}\HH{\Rr{47}}\HH{83}\HH{5}\HH{44}\HH{89}\hfill}
\vskip 0pt
\ligne{\hfill\HH{2}
\HH{45}\HH{53}\HH{\Rr{81}}\HH{80}\HH{\Rr{94}}\HH{5}\HH{44}\HH{89}\hfill}
\vskip 0pt
\ligne{\hfill\HH{3}
\HH{45}\HH{44}\HH{53}\HH{\Rr{81}}\HH{114}\HH{\Rr{17}}\HH{44}\HH{89}\hfill}
\vskip 0pt
\ligne{\hfill\HH{4}
\HH{45}\HH{44}\HH{44}\HH{53}\HH{\Rr{116}}\HH{60}\HH{\Rr{47}}\HH{89}\hfill}
\vskip 0pt
\ligne{\hfill\HH{5}
\HH{45}\HH{44}\HH{44}\HH{44}\HH{99}\HH{\Rr{63}}\HH{80}\HH{\Rr{103}}\hfill}
\vskip 0pt
\ligne{\hfill\HH{6}
\HH{45}\HH{44}\HH{44}\HH{44}\HH{83}\HH{26}\HH{\Rr{81}}\HH{119}\hfill}
\vskip 0pt
\ligne{\hfill\HH{7}
\HH{45}\HH{44}\HH{44}\HH{44}\HH{83}\HH{5}\HH{53}\HH{\Rr{105}}\hfill}
\vskip 0pt
\ligne{\hfill\HH{8}
\HH{45}\HH{44}\HH{44}\HH{44}\HH{83}\HH{5}\HH{44}\HH{89}\hfill}
\vskip 9pt
\trfn
\vskip 10pt
}
\hfill}

\ligne{\hfill
\vtop{\hsize=220pt
\begin{tab}\label{execfxdd}
\leurre
Execution of the rules for a double locomotive crossing the passive fixed switch 
from the right.
\end{tab}
\vskip 2pt
\trep
\vskip 8pt
\ligne{\hfill\HH{}
\HH{{4$_1$} }\HH{{3$_1$} }\HH{{2$_1$} }\HH{{1$_9$} }\HH{{0$_0$} }\HH{{1$_5$} }\HH{{2$_5$} }\HH{{11$_5$}}
\hfill}
\vskip 0pt
\ligne{\hfill\HH{1}
\HH{24}\HH{\Rr{63}}\HH{60}\HH{\Rr{17}}\HH{83}\HH{5}\HH{44}\HH{89}\hfill}
\vskip 0pt
\ligne{\hfill\HH{2}
\HH{6}\HH{26}\HH{\Rr{63}}\HH{60}\HH{\Rr{109}}\HH{5}\HH{44}\HH{89}\hfill}
\vskip 0pt
\ligne{\hfill\HH{3}
\HH{6}\HH{5}\HH{26}\HH{\Rr{63}}\HH{123}\HH{\Rr{17}}\HH{44}\HH{89}\hfill}
\vskip 0pt
\ligne{\hfill\HH{4}
\HH{6}\HH{5}\HH{5}\HH{26}\HH{\Rr{116}}\HH{60}\HH{\Rr{47}}\HH{89}\hfill}
\vskip 0pt
\ligne{\hfill\HH{5}
\HH{6}\HH{5}\HH{5}\HH{5}\HH{99}\HH{\Rr{63}}\HH{80}\HH{\Rr{103}}\hfill}
\vskip 0pt
\ligne{\hfill\HH{6}
\HH{6}\HH{5}\HH{5}\HH{5}\HH{83}\HH{26}\HH{\Rr{81}}\HH{119}\hfill}
\vskip 0pt
\ligne{\hfill\HH{7}
\HH{6}\HH{5}\HH{5}\HH{5}\HH{83}\HH{5}\HH{53}\HH{\Rr{105}}\hfill}
\vskip 0pt
\ligne{\hfill\HH{8}
\HH{6}\HH{5}\HH{5}\HH{5}\HH{83}\HH{5}\HH{44}\HH{89}\hfill}
\vskip 9pt
\trfn
\vskip 10pt
}
\hfill}

Tables~\ref{execfxgd} and~\ref{execfxdd} show the rules applied when a double locomotive
passively crosses the fixed switch, the former when the locomotive comes from the
left, the latter, when it comes from the right. We find the rules for the tracks
from Table~\ref{rulesvd}. The rules for the cell~0(0) are a bit different from
those of Tables~\ref{execfxgs} and~\ref{execfxds}. They have a common 'end'
with rules~116,
and then rule~99 as in Tables~\ref{execfxgs} and~\ref{execfxds}. Of course, the 
conservative rule~83 is also present. Now, for a double locomotive coming from the left,
rule~96 is replaced by two rules: rule~114, \laff {B} {WBWBWBBBB} {B} {}
and rule~116, \laff {B} {BBWBWBBWB} {W} {.} The former can see the first cell
of the locomotive in~0(0) and the second one in neighbour~8 while the latter
can see the second cell of the locomotive in~0(0) and the first cell in neighbour~1.
Rule~114 does not change the state of~0(0) while rule~116 does it. We have a similar 
situation in Table~\ref{execfxdd}. After rule~109 which was present in 
Table~\ref{execfxds} for a simple locomotive, rule~123, \laff {B} {WBWBBBBWB} {B} {}
is used and then rule~116 is applied as, starting from that time, there is no difference
with a double locomotive coming from the left. Note that rule~123 can see the first
cell of the locomotive in~0(0) and the second cell of the locomotive in its
neighbour~5.

\subsection{The round-about}

   In Section~\ref{scenar}, we have seen the idle configurations of the other pieces of the
round-about: the doubler, see Sub-subsection~\ref{subsdoubler} and the selector,
see Sub-subsection~\ref{subssel}.

\subsubsection{The doubler}
\label{subsdoubler}

   The left-hand side picture of Figure~\ref{stab_doublsel} illustrates the configuration
of the doubler. The locomotive arrives as a simple one.
The creation of the second cell of the locomotive happens in a different way
as in~\cite{mmarXiv1512}. Here, a kind of shock is organised: when the simple locomotive
arrives at 1(9). It is then replicated at the same time in~1(1) and~1(8), so that
two simple locomotive start at the same time. They meet four steps later when
one of them arrives at~1(4) and the other at~1(5). Then, the locomotive in 1(5) goes
to the way defined by the cells 2(5), 11(5), 12(5) and~13(5). As the locomotive in~1(4)
is still there, the bloc constituted by the cells in 1(4) and~2(5) behaves like a double
locomotive which goes along the above indicated track. Now, the locomotive in 1(4)
remains there one more step. It vanishes when the second cell of the double locomotive
has left the cell 2(5). See Table~\ref{execdbl} which gives traces of execution
of the rules already given together with the new ones dispatched by Table~\ref{rulesdbl}.

\vtop{  
\begin{tab}\label{rulesdbl}
\leurre
Rules for the locomotive traversing the doubler.
\end{tab}
\vskip 2pt
\trep
\vskip 8pt
\ligne{\hfill  
\vtop{\leftskip 0pt\parindent 0pt\hsize=\tabrulecol
\aff {125} {B} {WWWWBBWWW} {B}
\aff {126} {B} {BBWBWWWWW} {B}
\aff {127} {W} {BBWWWWWBB} {W}
\aff {128} {W} {WBWBWWWBW} {W}
\aff {129} {W} {WBWWWBWBW} {W}
\aff {130} {W} {WBWWBWWWB} {W}
\aff {131} {W} {WBWWBWBWB} {W}
\aff {132} {B} {WWWBWBWWW} {B}
}\hskip 10pt
\vtop{\leftskip 0pt\parindent 0pt\hsize=\tabrulecol
\aff {133} {B} {BWWBWWWWW} {B}
\aff {134} {W} {WBWWWWWBB} {W}
\aff {135} {W} {BWWWWBBBW} {B}
\aff {136} {W} {WBWWBBBWB} {B}
\aff {137} {B} {BWWWWBWBW} {W}
\aff {138} {W} {WWWWWWBBW} {W}
\aff {139} {B} {WBWWBWBWB} {W}
\aff {140} {W} {BWWWWBWBB} {W}
}\hskip 10pt
\vtop{\leftskip 0pt\parindent 0pt\hsize=\tabrulecol
\aff {141} {W} {BBWWBWBBB} {W}
\aff {142} {B} {WWBWWWBWW} {B}
\aff {143} {B} {WBWWWWWBW} {B}
\aff {144} {W} {WBBBWWWBW} {B}
\aff {145} {W} {WBWWWBBBW} {B}
\aff {146} {B} {BBWBWWWBW} {B}
\aff {147} {B} {WBWBWWWBB} {B}
\aff {148} {B} {WBWBBWWWB} {B}
}\hskip 10pt
\vtop{\leftskip 0pt\parindent 0pt\hsize=\tabrulecol
\aff {149} {B} {BBWBBWWWB} {W}
\aff {150} {B} {WBWBWWWBW} {W}
\aff {151} {W} {BBWBBWWWB} {W}
\aff {152} {B} {WWBBWBWWW} {B}
\aff {153} {B} {WWWBBBWWW} {B}
}
\hfill} 
\vskip 9pt
\trfn
\vskip 10pt
} 

We can note that rules 44, 47, 51 and~53 apply to the cells of the track which
counter-clockwise turn around the central cell and the rules~5, 17, 23 and~26
for those which clockwise turn around~0(0). Note that 
rule~131, \laff {W} {WBWWBWBWB} {W} {,} is the conservative rule for~1(9) where
the locomotive is duplicated into two independent simple locomotives. Its milestones
are neighbours~2, 5, 7 and~9, see also rule~139, \laff {B} {WBWWBWBWB} {W} {.}
The locomotive arrives through neighbour~6 and the
new locomotives leave the cell through neighbours~8 and~1 as it is witnessed by 
rule~136, \laff {W} {WBWWBBBWB} {B} {,} which makes the locomotive arriving through
neighbour~6 enter the cell, and by rule~141, \laff {W} {BBWWBWBBB} {W} {,} 
where neighbours~8 and~1 are black: they are the new locomotives.

\vtop{  
\begin{tab}\label{execdbl}
\leurre
Execution of the rules for the locomotive traversing the doubler.
\end{tab}
\vskip 2pt
\trep
\vskip 8pt
\ligne{\hfill\HH{}
\HH{{8$_1$} }\HH{{7$_1$} }\HH{{1$_9$} }\HH{{1$_1$} }\HH{{1$_2$} }\HH{{1$_3$} }\HH{{1$_4$} }
\HH{{1$_8$} }\HH{{1$_7$} }\HH{{1$_6$} }\HH{{1$_5$} }\HH{{2$_5$} }\HH{{11$_5$}}\HH{{12$_5$}}\hfill}
\ligne{\hfill\HH{1}
\HH{24}\HH{\Rr{23}}\HH{131}\HH{44}\HH{44}\HH{44}\HH{128}
\HH{5}\HH{5}\HH{5}\HH{129}\HH{130}\HH{44}\HH{44}\hfill}
\ligne{\hfill\HH{2}
\HH{6}\HH{26}\HH{\Rr{136}}\HH{44}\HH{44}\HH{44}\HH{128}
\HH{5}\HH{5}\HH{5}\HH{129}\HH{130}\HH{44}\HH{44}\hfill}
\ligne{\hfill\HH{3}
\HH{6}\HH{5}\HH{\Rr{139}}\HH{\Rr{47}}\HH{44}\HH{44}\HH{128}
\HH{\Rr{17}}\HH{5}\HH{5}\HH{129}\HH{130}\HH{44}\HH{44}\hfill}
\ligne{\hfill\HH{4}
\HH{6}\HH{5}\HH{141}\HH{\Rr{51}}\HH{\Rr{47}}\HH{44}\HH{128}
\HH{\Rr{23}}\HH{\Rr{17}}\HH{5}\HH{129}\HH{130}\HH{44}\HH{44}\hfill}
\ligne{\hfill\HH{5}
\HH{6}\HH{5}\HH{131}\HH{53}\HH{\Rr{51}}\HH{\Rr{47}}\HH{128}
\HH{26}\HH{\Rr{23}}\HH{\Rr{17}}\HH{129}\HH{130}\HH{44}\HH{44}\hfill}
\ligne{\hfill\HH{6}
\HH{6}\HH{5}\HH{131}\HH{44}\HH{53}\HH{\Rr{51}}\HH{\Rr{144}}
\HH{5}\HH{26}\HH{\Rr{23}}\HH{\Rr{145}}\HH{130}\HH{44}\HH{44}\hfill}
\ligne{\hfill\HH{7}
\HH{6}\HH{5}\HH{131}\HH{44}\HH{44}\HH{53}\HH{146}
\HH{5}\HH{5}\HH{26}\HH{\Rr{18}}\HH{\Rr{54}}\HH{44}\HH{44}\hfill}
\ligne{\hfill\HH{8}
\HH{6}\HH{5}\HH{131}\HH{44}\HH{44}\HH{53}\HH{147}
\HH{5}\HH{5}\HH{5}\HH{24}\HH{148}\HH{\Rr{47}}\HH{44}\hfill}
\ligne{\hfill\HH{9}
\HH{6}\HH{5}\HH{131}\HH{44}\HH{44}\HH{53}\HH{147}
\HH{5}\HH{5}\HH{5}\HH{24}\HH{\Rr{149}}\HH{80}\HH{\Rr{47}}\hfill}
\ligne{\hfill\HH{10}
\HH{6}\HH{5}\HH{131}\HH{44}\HH{44}\HH{53}\HH{\Rr{150}}
\HH{5}\HH{5}\HH{5}\HH{6}\HH{151}\HH{\Rr{81}}\HH{80}\hfill}
\ligne{\hfill\HH{11}
\HH{6}\HH{5}\HH{131}\HH{44}\HH{44}\HH{44}\HH{128}
\HH{5}\HH{5}\HH{5}\HH{129}\HH{130}\HH{53}\HH{\Rr{81}}\hfill}
\vskip 9pt
\trfn
\vskip 10pt
}

\vtop{
\ligne{\hskip 5pt
\includegraphics[scale=0.55]{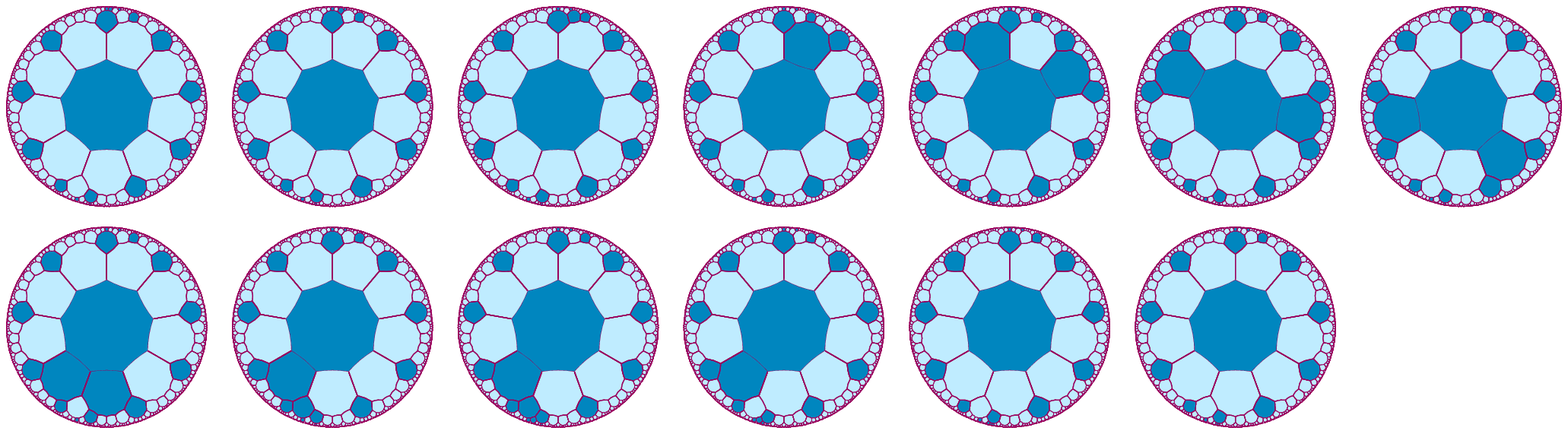} 
\hfill}
\vspace{-5pt}
\begin{fig}\label{doubleur}
\leurre
The structure which doubles a locomotive: a single locomotive enters the structure;
a double one leaves it.
\end{fig}
}

Note that the cell 1(5) is ruled by already seen rules, 18, 24 and~6 and two new ones,
rule 129, \laff {W} {WBWWWBWBW} {W} {,}
and rule~145, \laff {W} {WBWWWBBBW} {B} {.} The former one is the conservative rule
of the cell while the second one makes the cell turn to black as it can see the
locomotive which clockwise turned around 0(0) in neighbour~7. Rule~18 makes the cell
white again and it has the configuration of the cell of the track with four milestones,
see Table~\ref{execvdm}. Then, rule~24 and then rule~6 apply, as already seen in the 
case of such a cell of the track in the same table. Then the conservative rule~129
applies as the locomotives are no more visible from the cell~1(5). Note the particular
situation of the cell 2(5) which has a specific conservative rule: 
rule~130, \laff {W} {WBWWBWWWB} {W} {.}
Rule~54 is then applied, a rule we already met in Table~\ref{execvmd} which allows the
cell coming from 1(5) to enter ~2(5). Then, rule~148, \laff {B} {WBWBBWWWB} {B} {,} 
allows to copy the black cell lying in~1(4), the locomotive which came from the
other exit of~1(9), into the cell~2(5): this constitute the second cell of the new
double locomotive. Then rule~149, \laff {B} {BBWBBWWWB} {W} {,} makes the second cell
of the double locomotive leave the cell: in this configuration, the cell 2(5) has
an additional milestone in the presence of the duplicated locomotive in 1(4) which
is the neighbour~4 of~2(5). Next, rule~151, \laff {W} {BBWBBWWWB} {W} {,} witnesses
that the second cell of the double locomotive left~2(5), being now in its neighbour~1.
The rules executed in~11(5) and 12(5) show the rules we have studied in
Table~\ref{execvmdd} for a double locomotive. 

Let us look at the cell~1(4). Table~\ref{execdbl} shows us that the conservative rule
is rule~128, \laff {W} {WBWBWWWBW} {W} {,} whose orientation is conformal to
the configuration illustrated by Figures~\ref{stab_doublsel} and~\ref{doubleur}.
Then, rule~144, \laff {W} {WBBBWWWBW} {B} {,} makes the cell turn to black as the arriving
simple locomotive is in neighbour~3.  
Rules~146, \laff {B} {BBWBWWWBW} {B} {,}
and~147, \laff {B} {WBWBWWWBB} {B} {,} keep the locomotive in~1(4) as long as
there is a locomotive in its neighbours~1 or~9 corresponding to the cells~1(5)
and~2(5) respectively. When both cells are white, the corresponding locomotive
is destroyed, see rule~150, \laff {B} {WBWBWWWBW} {W} {.} The table together
with Figure~\ref{doubleur} allow
us to check that this vanishing has no influence on the neighbouring of the cell.
Table~\ref{execdbl} allows us to check that the double locomotive leaves the doubler 
which remains in the same configuration as when the simple locomotive approached it.
This can be seen in Figure~\ref{doubleur}, provided that the reader has at its disposal
a PostScript version of the paper: it will be required to magnify the last pictures
of the figure.

\subsubsection{The fork}

   Although the fork is not connected with the round-about, we place it here as its
structure, see Figure~\ref{stab_active} is a kind of simplification of
the doubler, see Figure~\ref{stab_doublsel}. We already have mentioned this property
in Subsection~\ref{forcontrol}.

\vtop{  
\begin{tab}\label{rulesfk}
\leurre
Rules for the locomotive traversing the fork. 
\end{tab}
\vskip 2pt
\trep
\vskip 8pt
\ligne{\hfill  
\vtop{\leftskip 0pt\parindent 0pt\hsize=\tabrulecol
\aff {154} {W} {WBWBWBWWB} {W}
\aff {155} {W} {BBWBWWWWW} {W}
\aff {156} {W} {BWWWWWWBW} {W}
\aff {157} {W} {BWBWWBWBW} {W}
\aff {158} {W} {BWBWWWWBW} {W}
\aff {159} {W} {BWBWWWWWB} {W}
\aff {160} {W} {BWWBWBWBW} {W}
}\hskip 10pt
\vtop{\leftskip 0pt\parindent 0pt\hsize=\tabrulecol
\aff {161} {W} {WWWWBWBWB} {W}
\aff {162} {W} {WWWWBBBWB} {B}
\aff {163} {W} {BWWBBBWBW} {B}
\aff {164} {B} {WWWWBWBWB} {W}
\aff {165} {W} {BWBWWWWBB} {B}
\aff {166} {B} {BWWBWBWBW} {W}
\aff {167} {W} {WWWWBWBBB} {W}
}\hskip 10pt
\vtop{\leftskip 0pt\parindent 0pt\hsize=\tabrulecol
\aff {168} {W} {WBWBBBWWB} {B}
\aff {169} {W} {BWBWWBWBB} {B}
\aff {170} {B} {BWBWWWWBW} {W}
\aff {171} {B} {BWWWWWBWW} {B}
\aff {172} {W} {BWWBWBBBB} {W}
\aff {173} {B} {WBWBWBWWB} {W}
\aff {174} {B} {BWBWWBWBW} {W}
}\hskip 10pt
\vtop{\leftskip 0pt\parindent 0pt\hsize=\tabrulecol
\aff {175} {B} {WBWWWWBWW} {B}
\aff {176} {W} {BWWWWWWBB} {W}
\aff {177} {W} {BBBWWWWBW} {W}
\aff {178} {W} {BBWBWBWWB} {W}
\aff {179} {W} {BWBWWBBBW} {W}
\aff {180} {B} {WWWWBWBWW} {B}
}
\hfill}
\vskip 9pt
\trfn
\vskip 10pt
} 

The locomotive arriving to the cell~1(9) follows the same path as in the case of
the doubler. Here, the side~1 of the cell~1(9) of the fork is the side shared
with~2(9) while, in the doubler, it was the side shared with 1(8). So that
rules~160, 163, 166 and~172 which control the crossing of~1(9) by the locomotive
are rotated forms of the rules~131, 136, 139 and~141 performing the same actions
in the doubler:
\vskip 10pt
\ligne{\hfill
\vtop{\leftskip 0pt\parindent 0pt\hsize=\tabrulecol
\aff {160} {W} {BWWBWBWBW} {W}
\aff {163} {W} {BWWBBBWBW} {B}
\aff {166} {B} {BWWBWBWBW} {W}
\aff {172} {W} {BWWBWBBBB} {W}
}\hskip 20pt
\vtop{\leftskip 0pt\parindent 0pt\hsize=\tabrulecol
\aff {131} {W} {WBWWBWBWB} {W}
\aff {136} {W} {WBWWBBBWB} {B}
\aff {139} {B} {WBWWBWBWB} {W}
\aff {141} {W} {BBWWBWBBB} {W}
}
\hfill}
\vskip 10pt
In particular, we can see in rule~172 that two locomotives leave the cell~1(9)
as neighbours~7 and~9 which are the cells~1(1) and~1(8) respectively. Note that 
1(1) and~1(8) have different neighbourhoods: 1(1) has a standard neighbourhood involving
the rules~44, 47, 51 and~53 already seen in Tables~\ref{rulesvs}
and~\ref{execvmd}. 
\vskip 5pt
\vtop{
\ligne{\hskip 5pt
\includegraphics[scale=0.55]{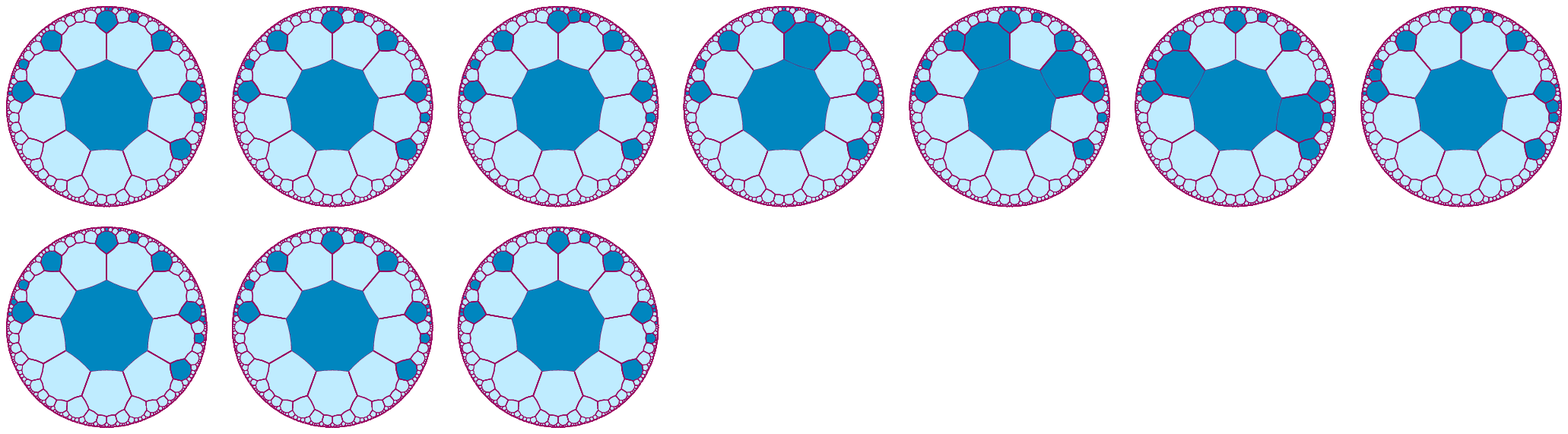} 
\hfill}
\vspace{-5pt}
\begin{fig}\label{fourche}
\leurre
The structure which doubles a locomotive: a single locomotive enters the structure;
a double one leaves it.
\end{fig}
}
\vskip 5pt

\ligne{\hfill 
\vtop{  
\hsize=280pt
\begin{tab}\label{execfk}
\leurre
Execution of the rules for the locomotive traversing the fork.
\end{tab}
\vskip 2pt
\trep
\vskip 8pt
\ligne{\hfill\HH{}
\HH{{8$_1$} }\HH{{7$_1$} }\HH{{1$_9$} }\HH{{1$_1$} }\HH{{1$_2$} }
\HH{{6$_2$} }\HH{{7$_3$} }\HH{{1$_8$} }\HH{{1$_7$} }\HH{{6$_7$} }\HH{{7$_8$} }\hfill}
\ligne{\hfill\HH{1}
\HH{24}\HH{\Rr{23}}\HH{160}\HH{44}\HH{154}
\HH{44}\HH{44}\HH{158}\HH{157}\HH{44}\HH{44}\hfill}
\vskip 0pt
\ligne{\hfill\HH{2}
\HH{6}\HH{26}\HH{\Rr{163}}\HH{44}\HH{154}
\HH{44}\HH{44}\HH{158}\HH{157}\HH{44}\HH{44}\hfill}
\ligne{\hfill\HH{3}
\HH{6}\HH{5}\HH{\Rr{166}}\HH{\Rr{47}}\HH{154}
\HH{44}\HH{44}\HH{\Rr{165}}\HH{157}\HH{44}\HH{44}\hfill}
\ligne{\hfill\HH{4}
\HH{6}\HH{5}\HH{172}\HH{\Rr{51}}\HH{\Rr{168}}
\HH{44}\HH{44}\HH{\Rr{170}}\HH{\Rr{169}}\HH{44}\HH{44}\hfill}
\ligne{\hfill\HH{5}
\HH{6}\HH{5}\HH{160}\HH{53}\HH{\Rr{173}}
\HH{\Rr{47}}\HH{44}\HH{177}\HH{\Rr{174}}\HH{\Rr{47}}\HH{44}\hfill}
\ligne{\hfill\HH{6}
\HH{6}\HH{5}\HH{160}\HH{44}\HH{178}
\HH{\Rr{51}}\HH{\Rr{47}}\HH{158}\HH{179}\HH{\Rr{51}}\HH{\Rr{47}}\hfill}
\ligne{\hfill\HH{7}
\HH{6}\HH{5}\HH{160}\HH{44}\HH{154}
\HH{53}\HH{\Rr{51}}\HH{158}\HH{157}\HH{53}\HH{\Rr{51}}\hfill}
\ligne{\hfill\HH{8}
\HH{6}\HH{5}\HH{160}\HH{44}\HH{154}
\HH{44}\HH{53}\HH{158}\HH{157}\HH{44}\HH{53}\hfill}
\vskip 9pt
\trfn
\vskip 10pt
} 
\hfill} 

The cell~1(8)
has the same neighbourhood, up to rotation, but the place of side~1 involves 
rules~158, \laff {W} {BWBWWWWBW} {W} {,} the conservative rule of the cell,
rule~165, \laff {W} {BWBWWWWBB} {B} {,} making the locomotive enter the cell,
rule~170, \laff {B} {BWBWWWWBW} {W} {,} making it leave the cell and
rule~177, \laff {W} {BBBWWWWBW} {W} {} witnessing that the locomotive is going
away through neighbour~2, see Table~\ref{execfk}. The cell 1(2), has not an 
orientation analogous to that
of the cell~1(1) because the track goes through~6(2) so that the locomotive enters
through 1(1), its neighbour~5. The rules are rule 154, \laff {W} {WBWBWBWWB} {W} {,}
rule 168, \laff {W} {WBWBBBWWB} {B} {,}
rule 173, \laff {B} {WBWBWBWWB} {W} {,}
and rule~178, \laff {W} {BBWBWBWWB} {W} {.} The cell~1(7) is a rotated image of~1(2)
around~0(0). The rules~157 and~174 are rotated form of the rules~154 and~173
respectively, but the rules~169 and~179 are not
rotated forms of the rules~168, and~178 respectively, as the locomotive 
enters a neighbour of~1(7) which is not the rotated image of the neighbour through
which the locomotive enters the cell~1(2) as can be seen on rule~169,
\laff {W} {BWBWWBWBB} {B} {,} and on rule~179, \laff {W} {BWBWWBBBW} {W} {.}

   Note that Figure~\ref{fourche} and Table~\ref{execfk} illustrate the fact
that the structure we just described with the help of the rules from Table~\ref{rulesfk}
performs the action which we expected from the fork: two simple locomotives leave the
fork following different tracks.

\subsubsection{The selector}
\label{subssel}

    The right-hand side picture of Figure~\ref{stab_doublsel} illustrates the configuration
of the selector. Tables~\ref{rulessels} and~\ref{rulesseld} give the rules for the 
locomotive and for the cells of the selector, Table~\ref{rulessels} for a simple
locomotive, Table~\ref{rulesseld} for a double one.

\vtop{  
\begin{tab}\label{rulessels}
\leurre
Rules for a simple locomotive crossing the selector.
\end{tab}
\vskip 2pt
\trep
\vskip 8pt
\ligne{\hfill  
\vtop{\leftskip 0pt\parindent 0pt\hsize=\tabrulecol
\aff {181} {W} {WBBWWWBWB} {W}
\aff {182} {W} {BWWWWWBWB} {W}
\aff {183} {B} {BWWWWBBWW} {B}
\aff {184} {W} {WBWWWBBWB} {W}
\aff {185} {W} {WBBWWBWBB} {W}
\aff {186} {B} {BWWBBWWWW} {B}
\aff {187} {W} {WBWBBWWBB} {W}
\aff {188} {B} {BWWWBWBWW} {B}
\aff {189} {W} {WBBWWBBBB} {B}
}\hskip 10pt
\vtop{\leftskip 0pt\parindent 0pt\hsize=\tabrulecol
\aff {190} {B} {BWBWWWBWW} {B}
\aff {191} {W} {WBBBWBWWB} {B}
\aff {192} {B} {BBWWWBBWW} {B}
\aff {193} {B} {WBBWWBWBB} {W}
\aff {194} {B} {BWWBBWWWB} {B}
\aff {195} {B} {BBWWWWBWW} {B}
\aff {196} {B} {BWBWWWWWW} {B}
\aff {197} {B} {BWWWWWWBW} {B}
\aff {198} {W} {BBBWWWBWB} {B}
}\hskip 10pt
\vtop{\leftskip 0pt\parindent 0pt\hsize=\tabrulecol
\aff {199} {B} {BWBWWBBWW} {W}
\aff {200} {W} {BBBWWBWBB} {W}
\aff {201} {B} {BWWBBWWBW} {B}
\aff {202} {W} {BWWBBBWWB} {W}
\aff {203} {B} {WBBWWWBWW} {W}
\aff {204} {W} {BWWBWBBWW} {B}
\aff {205} {W} {WWBWWBWBB} {W}
\aff {206} {B} {BWWBBWBWW} {B}
\aff {207} {B} {WBWWWBBWB} {W}
}\hskip 10pt
\vtop{\leftskip 0pt\parindent 0pt\hsize=\tabrulecol
\aff {208} {B} {BWWBBBWWW} {B}
\aff {209} {B} {BBWWWWWWB} {B}
\aff {210} {W} {BBWWWBBWB} {W}
\aff {211} {W} {WBBBBWWBB} {B}
\aff {212} {B} {WBWBBWWBB} {W}
\aff {213} {W} {BBWBBWWBB} {W}
\aff {214} {B} {WBWWWWWWB} {B}
}
\hfill}
\vskip 9pt
\trfn
\vskip 10pt
} 

\ligne{\hfill 
\vtop{  
\hsize=280pt
\begin{tab}\label{execsels}
\leurre
Execution of the rules for a simple locomotive crossing the selector.
\end{tab}
\vskip 2pt
\trep
\vskip 8pt
\ligne{\hfill\HH{}
\HH{{8$_8$} }\HH{{7$_8$} }\HH{{6$_7$} }\HH{{1$_7$} }\HH{{0$_0$} }
\HH{{1$_9$} }\HH{{2$_9$} }\HH{{11$_9$}}\HH{{12$_9$}}\HH{{1$_6$} }\HH{{1$_8$} }\hfill}
\ligne{\hfill\HH{2}
\HH{26}\HH{\Rr{23}}\HH{\Rr{135}}\HH{185}\HH{154}
\HH{184}\HH{187}\HH{44}\HH{44}\HH{183}\HH{186}\hfill}
\ligne{\hfill\HH{3}
\HH{5}\HH{26}\HH{\Rr{137}}\HH{\Rr{189}}\HH{154}
\HH{184}\HH{187}\HH{44}\HH{44}\HH{183}\HH{186}\hfill}
\ligne{\hfill\HH{4}
\HH{5}\HH{5}\HH{140}\HH{\Rr{193}}\HH{\Rr{191}}
\HH{184}\HH{187}\HH{44}\HH{44}\HH{192}\HH{194}\hfill}
\ligne{\hfill\HH{5}
\HH{5}\HH{5}\HH{91}\HH{200}\HH{\Rr{173}}
\HH{\Rr{29}}\HH{187}\HH{44}\HH{44}\HH{\Rr{199}}\HH{201}\hfill}
\ligne{\hfill\HH{6}
\HH{5}\HH{5}\HH{91}\HH{205}\HH{202}
\HH{\Rr{207}}\HH{\Rr{211}}\HH{44}\HH{44}\HH{\Rr{204}}\HH{206}\hfill}
\ligne{\hfill\HH{7}
\HH{5}\HH{5}\HH{91}\HH{185}\HH{154}
\HH{210}\HH{\Rr{212}}\HH{\Rr{47}}\HH{44}\HH{183}\HH{208}\hfill}
\ligne{\hfill\HH{8}
\HH{5}\HH{5}\HH{91}\HH{185}\HH{154}
\HH{184}\HH{213}\HH{\Rr{51}}\HH{\Rr{47}}\HH{183}\HH{186}\hfill}
\ligne{\hfill\HH{9}
\HH{5}\HH{5}\HH{91}\HH{185}\HH{154}
\HH{184}\HH{187}\HH{53}\HH{\Rr{51}}\HH{183}\HH{186}\hfill}
\ligne{\hfill\HH{10}
\HH{5}\HH{5}\HH{91}\HH{185}\HH{154}
\HH{184}\HH{187}\HH{44}\HH{53}\HH{183}\HH{186}\hfill}
\vskip 9pt
\trfn
\vskip 10pt
} 
\hfill}

The cells 8(8) and 7(8) on the track leading to the selector make use of rules of 
Table~\ref{rulesvs}. The cell~6(7) makes use of other rules: its conservative rule 
is rule~91, \laff {W} {BWWWWBWBW} {W} {,}, then rule~135, \laff {W} {BWWWWBBBW} {B} {,}
can see the locomotive in neighbour~7, so that it makes it enter the cell,
then rule~137, \laff {B} {BWWWWBWBW} {W} {,} makes the cell go turn back to white
and rule~140, \laff {W} {BWWWWBWBB} {W} {} witnesses that the locomotive left the
cell through neighbour~9.
 
The conservative rule for the central cell is rule~154 which we already know from the 
fork where it was used for the cell 1(2). The detection of the arriving locomotive 
through neighbour~3 is performed by rule~191, \laff {W} {WBBBWBWWB} {B} {.} The locomotive
leaves the cell thanks to rule~173 which we already noticed in Table~\ref{execfk},
were it operated on the cell 1(2). Then, we have rule~202, \laff {W} {BWWBBBWWB} {W} {,}
which shows three important features when compared with
rule 154, \laff {W} {WBWBWBWWB} {W} {.} Indeed, two locomotives appear in 
neighbours~1 and~5, which corresponds to the cells~1(5) and~1(9) and the 
milestone in 1(6), neighbour~2, vanished: also see rule~199, further. At the next time,
rule~154 again applies which  means that the milestone was restored and that the
locomotives vanished. As can be seen on Table~\ref{execsels} and in
the first two rows of Figure~\ref{select}, the locomotive
which was in the cell~1(9) goes on its way out of the selector and, also, out of the
round-about, and the locomotive which was in the cell~1(5) vanishes. These properties
can also be notice on Table~\ref{execsels}: the motion of the locomotive created
in 1(9) can be seen with the columns of execution for the cells~2(9), 11(9) and~12(9)
of the tracks. The vanishing of the locomotive in 1(5) can be seen by the
evolution of the cell 1(6). Its conservative rule is 
rule~183, \laff {B} {BWWWWBBWW} {B} {,} where its neighbours~1, 2 3, and~4 are 
the cells~2(7), 1(7), 0(0) and 1(5) respectively.
Rule~192, \laff {B} {BBWWWBBWW} {B} {,} can see that a locomotive approaches: its 
first cell is in neighbour~2, the cell 1(7), and a possible second cell if any 
is not visible from~1(5). Next, rule~199, \laff {B} {BWBWWBBWW} {W} {,} can see both 
the central cell 0(0) and the cell 1(7). As neighbour~3 only is black, the cell knows 
that the arrived locomotive is simple, triggering that the milestone at 1(6) vanishes. 
Next, the rule 204, \laff {W} {BWWBWBBWW} {B} {,} shows that there is a locomotive in 1(5)
and the return of rule~183 from the next time show that the neighbour~5 always remained
white and this neighbour is the cell~2(6), the next cell on the track after 1(5).
Note that, symmetrically, the cell 1(8) always remained black and it witnesses the
passage of the locomotive created in~1(9). The conservative rule of the cell~1(8) is
rule~186, \laff {B} {BWWBBWWWW} {B} {,} whose neighbourhood is a rotated form of the 
neighbourhood of the rule~183. Rule~194, \laff {B} {BWWBBWWWB} {B} {,} shows that the
locomotive appears in 1(7) which is neighbour~9 of~1(8). Next, 
rule~201, \laff {B} {BWWBBWWBW} {B} {,} can see that the locomotive is in its 
neighbour~8, {\it i.e.} the cell 0(0) and, as neighbour~9 is now white, the cell 1(8) 
knows that the locomotive is simple. Accordingly, it remains black. Note that
the neighbourhood of the rule~201 is a rotated form of the neighbourhood of rule~199.
The rules have the same current state, but the new state is different: rules~199 
and~201 are not rotationnally compatible. However, as we did not require our automaton
to be rotation invariant, we can keep both those rules. Next, 
rule~206, \laff {B} {BWWBBWBWW} {B} {,} and rule~208, \laff {B} {BWWBBBWWW} {B} {,}
witness the motion of the locomotive: rule~206 can see it in its neighbour~7, which is
the cell 1(9) and rule 208 can see it in its neighbour~6 which is the cell 2(9).

\vtop{  
\begin{tab}\label{rulesseld}
\leurre
Rules for a double locomotive crossing the selector.
\end{tab}
\vskip 2pt
\trep
\vskip 8pt
\ligne{\hfill  
\vtop{\leftskip 0pt\parindent 0pt\hsize=\tabrulecol
\aff {215} {B} {BWWWBBBWW} {B}
\aff {216} {B} {BWWWWBBBW} {B}
\aff {217} {B} {BWBBWWBWW} {B}
\aff {218} {B} {WBBWWBBBB} {B}
\aff {219} {B} {BWWWWBWBB} {W}
\aff {220} {B} {BBBWWWBWW} {B}
\aff {221} {B} {WBBBWBWWB} {B}
}\hskip 10pt
\vtop{\leftskip 0pt\parindent 0pt\hsize=\tabrulecol
\aff {222} {B} {BBBWWBBWW} {B}
\aff {223} {B} {BBBWWBWBB} {W}
\aff {224} {B} {BWWBBWWBB} {W}
\aff {225} {B} {BBWWBBWWB} {W}
\aff {226} {B} {BBBWWWWWW} {B}
\aff {227} {B} {BWWWWWWBB} {B}
\aff {228} {B} {BBBWWWBWB} {W}
}\hskip 10pt
\vtop{\leftskip 0pt\parindent 0pt\hsize=\tabrulecol
\aff {229} {B} {BWBBWBBWW} {B}
\aff {230} {W} {BBBWWBWBW} {W}
\aff {231} {W} {BWWBBWBBW} {B}
\aff {232} {B} {WBWWWBBBW} {W}
\aff {233} {W} {WBBWBWWBB} {W}
\aff {234} {W} {WBBWWWBBB} {W}
\aff {235} {B} {WWBWWWWWB} {B}
}\hskip 10pt
\vtop{\leftskip 0pt\parindent 0pt\hsize=\tabrulecol
\aff {236} {B} {WWBWWWWBW} {B}
\aff {237} {B} {WWBWBWWWW} {B}
\aff {238} {B} {WWBBWWWWW} {B}
\aff {239} {B} {BWWWWWBWB} {B}
\aff {240} {W} {BWWWWWBBB} {B}
}
\hfill}
\vskip 9pt
\trfn
\vskip 10pt
} 

Table~\ref{rulesseld} show us the rules for the case when a double locomotive
crosses the selector. We can see new rules for the tracks with cell~6(7) corresponding
to the presence of a double locomotive: as previously, rule~135 can see the first cell of
a locomotive, ignoring whether it is a simple or a double one. Rule 216, 
\laff {B} {BWWWWBBBW} {B} {,} can
see the first cell of the locomotive in~6(7) and the second one in its neighbour~7.
Then rule~219, \laff {B} {BWWWWBWBB} {W} {,} makes the cell return to white and 
rule~140 witnesses that the second cell of the locomotive is already on the track for
going away.

\ligne{\hfill 
\vtop{  
\hsize=280pt
\begin{tab}\label{execseld}
\leurre
Execution of the rules for a double locomotive crossing the selector.
\end{tab}
\vskip 2pt
\trep
\vskip 8pt
\ligne{\hfill\HH{}
\HH{{8$_8$} }\HH{{7$_8$} }\HH{{6$_7$} }\HH{{1$_7$} }\HH{{0$_0$} }
\HH{{1$_5$} }\HH{{2$_6$} }\HH{{7$_6$} }\HH{{1$_6$} }\HH{{1$_8$} }\hfill}
\ligne{\hfill\HH{2}
\HH{26}\HH{\Rr{63}}\HH{216}\HH{\Rr{189}}\HH{154}
\HH{181}\HH{184}\HH{5}\HH{183}\HH{186}\hfill}
\ligne{\hfill\HH{3}
\HH{5}\HH{26}\HH{\Rr{219}}\HH{218}\HH{\Rr{191}}
\HH{181}\HH{184}\HH{5}\HH{192}\HH{194}\hfill}
\ligne{\hfill\HH{4}
\HH{5}\HH{5}\HH{140}\HH{\Rr{223}}\HH{221}
\HH{\Rr{198}}\HH{184}\HH{5}\HH{222}\HH{\Rr{224}}\hfill}
\ligne{\hfill\HH{5}
\HH{5}\HH{5}\HH{91}\HH{230}\HH{\Rr{225}}
\HH{\Rr{228}}\HH{\Rr{29}}\HH{5}\HH{229}\HH{\Rr{231}}\hfill}
\ligne{\hfill\HH{6}
\HH{5}\HH{5}\HH{91}\HH{185}\HH{154}
\HH{234}\HH{\Rr{207}}\HH{\Rr{17}}\HH{215}\HH{186}\hfill}
\ligne{\hfill\HH{7}
\HH{5}\HH{5}\HH{91}\HH{185}\HH{154}
\HH{181}\HH{210}\HH{\Rr{23}}\HH{183}\HH{186}\hfill}
\ligne{\hfill\HH{8}
\HH{5}\HH{5}\HH{91}\HH{185}\HH{154}
\HH{181}\HH{184}\HH{26}\HH{183}\HH{186}\hfill}
\ligne{\hfill\HH{9}
\HH{5}\HH{5}\HH{91}\HH{185}\HH{154}
\HH{181}\HH{184}\HH{5}\HH{183}\HH{186}\hfill}
\ligne{\hfill\HH{10}
\HH{5}\HH{5}\HH{91}\HH{185}\HH{154}
\HH{181}\HH{184}\HH{5}\HH{183}\HH{186}\hfill}
\ligne{\hfill\HH{11}
\HH{5}\HH{5}\HH{91}\HH{185}\HH{154}
\HH{181}\HH{184}\HH{5}\HH{183}\HH{186}\hfill}
\vskip 9pt
\trfn
\vskip 10pt
} 
\hfill}

The rules for the central cell are the same as for a simple locomotive as long as the
first cell of the locomotive is not in the cell~0(0). When this is the case, as previously,
rule~191 detects the first cell of the locomotive, letting it enter the cell.
Next, rule~221, \laff {B} {WBBBWBWWB} {B} {,} let the second cell enter 0(0)
and then, rule~225, \laff {B} {BBWWBBWWB} {W} {,} makes the central cell go back to
white and it shows us two locomotives, one in its neighbour~1, the cell 1(5), and one
in its neighbour~5, the cell 1(9). The rule also shows us that the cell 1(8), its 
neighbour~4 turned to white. At the next time, the cell 0(0) is again under its
conservative rule, rule~154. Table~\ref{execseld} shows us that the locomotive which
appeared in 1(5) goes on its way. The cell 2(6) needs new rules, due to its numbering:
we have the sequence rule 184, \laff {W} {WBWWWBBWB} {W} {,} then rule~29, which
we have already seen in Table~\ref{execvdm} for cells of the track with four
milestones, then rule~207, \laff {B} {WBWWWBBWB} {W} {,} makes the cell
turn back to white and then rule~210, \laff {W} {BBWWWBBWB} {W} {,} witnessing
that the locomotive, now in neighbour~1, will go away.

The cells 1(6) and~1(8) behave in the reverse way, with respect to what we have
seen in Table~\ref{execsels}. The conservative rules are, of course, the same.
For the cell 1(8), the sequence is now rule~194, already seen for the simple locomotive,
rule~224, \laff {B} {BWWBBWWBB} {W} {,} which can see a double locomotive in its
neighbours~8 and~9, the cells 0(0) and 1(7) respectively, and then
rule~231, \laff {W} {BWWBBWBBW} {B} {,} which can see two locomotives: the second cell
of the double locomotive in~0(0) and the new simple locomotive in its neighbour~7,
the cell~1(9), and then, the conservative rule, rule~186. This proves that the
locomotive in 1(9) vanished and none of them is running on the track which can be seen
by the cell~1(8). 

\vtop{
\ligne{\hskip 5pt
\includegraphics[scale=0.55]{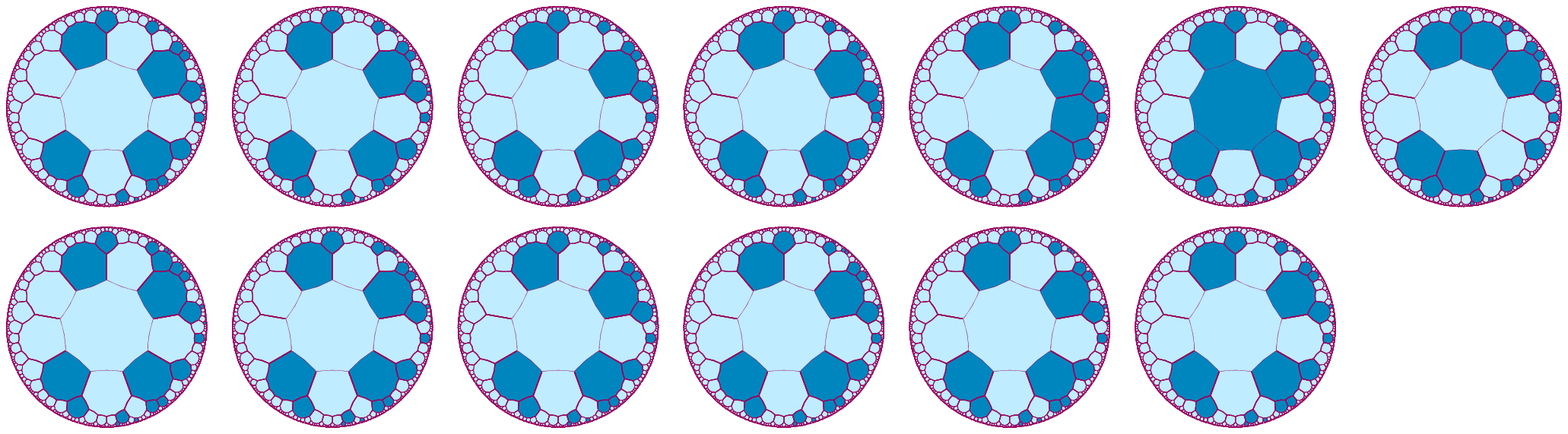} 
\hfill}
\ligne{\hfill
\includegraphics[scale=0.55]{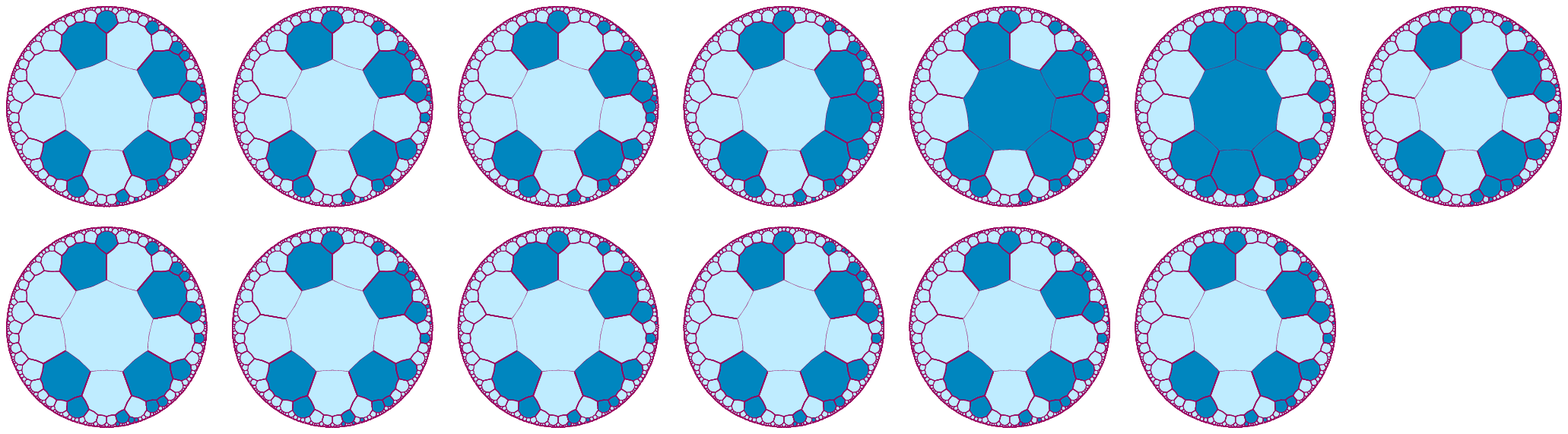} 
\hfill}
\begin{fig}\label{select}
\leurre
The selector of the round about. Above: a single locomotive enters the selector. It leaves
it to go on its way on the right track. Below: a double locomotive enters the selector. 
It leaves it on the way going to the next selector.
\end{fig}
}

For the cell 1(6), we have an opposite situation. As in the case of a single locomotive,
the arrival of the locomotive is detected by rule~192: it is in its neighbour~2, the 
cell 1(7). The cell 1(6) cannot see whether the locomotive is simple or double. 
This is why this is clear at the next step, when rule~222, \laff {B} {BBBWWBBWW} {B} {,}
is applied. The rule witnesses that the cell 1(6) can see that the locomotive is
double as its first cell is in neighbour~2, cell 1(7), and the second one is in
neighbour~3, cell 0(0). The rule makes the cell 1(6) remain black. The consequence is
that the locomotive created in~1(5), see rule~229, \laff {B} {BWBBWBBWW} {B} {,}
where this simple locomotive appears in neighbour~4, goes on its way on the track
starting from~1(5). The rule also shows that the second locomotive of that which 
arrived to the selector is now in 0(0), the neighbour~3
of~1(6). At the next time, rule~215, \laff {B} {BWWWBBBWW} {B} {,} shows that
the new locomotive, a single one is now in 2(6), the neighbour~5 of~1(6).
After that, the locomotive no more turns around the cell 1(6). As shown by 
Table~\ref{execseld}, this can be seen in both cells 2(6) and 7(6).

Let us look at what happens in the cells of the track going to the next selector,
the track starting from the cell~1(5). Its conservative rule is 
rule 181, \laff {W} {WBBWWWBWB} {W} {,} which is not a rotated form of
rule~6 we have seen for cells of the track with four milestones: the neighbourhood  
of~1(5) is a reflection of those cells, not a rotated image of them.
Next, rule~198, \laff {W} {BBBWWWBWB} {B} {,} can see the locomotive in its
neighbour~1, the cell~0(0). Then, rule~228, \laff {B} {BBBWWWBWB} {W} {,} makes it 
leave the cell and does not make the second cell of the double locomotive, which is now 
in its neighbour~1, enter 1(5). This is why the side~1 of the cell~1(5) is not
turned to the next cell of the track, in order to allow a behaviour which is different
from that of an ordinary cell of the track with respect to a double locomotive which
is the situation seen by rule~228. After that, rule~234, \laff {W} {WBBWWWBBB} {W} {,}
witnesses that the new simple locomotive goes on its way on the track as its now
in neighbour~8, the cell 2(6). It also witnesses that the second cell of the
arrived double locomotive vanished as neighbour~1 is white as well as cell~1(5)
itself. Then, the cell is controlled by rule~181 again.

Note that the cell~2(6) is applied the same rules as the cell~1(9) in the case of a 
simple locomotive crossing the selector.

\subsection{Flip-flop and active memory switch}
\label{flflp}

    In Section~\ref{scenar} we have seen that we can implement both a flip-flop and
the active memory switch by using the same basic devices provided that they are
suitably arranged. The last two pictures of Figure~\ref{stab_active} show the two possible
configurations of the controller. The difference is made by cell~1(3) whose state we call 
the \textbf{colour} of the controller. 

\vtop{  
\begin{tab}\label{rulesctrl}
\leurre
Rules for the controller of the active switches.
\end{tab}
\vskip 2pt
\trep
\vskip 8pt
\ligne{\hfill\hbox to 150pt{\hfill passage of a locomotive\hfill}
\hskip 10pt\hbox to 150pt{\hfill arrival of a signal\hfill}
\hfill}
\vskip 4pt
\ligne{\hfill  
\vtop{\leftskip 0pt\parindent 0pt\hsize=\tabrulecol
\ligne{\hfill black control\hfill}
\vskip 5pt
\aff {241} {W} {WBBWBWBWB} {W}
\aff {242} {W} {BWBBWWWWW} {W}
\aff {243} {W} {BBWBWBWWW} {W}
\aff {244} {B} {WBBBBBWWW} {B}
\aff {245} {W} {BBWWWWWBW} {W}
\aff {246} {W} {WBBBBWBWB} {B}
\aff {247} {B} {WBWWWWWBB} {B}
\aff {248} {B} {WBBWBWBWB} {W}
\aff {249} {B} {BWWWBWWWW} {B}
\aff {250} {W} {BBBBWBWWW} {W}
}\hskip 10pt
\vtop{\leftskip 0pt\parindent 0pt\hsize=\tabrulecol
\aff {251} {B} {WBBBBBWWB} {B}
\aff {252} {B} {BBWWWWWBW} {B}
\aff {253} {W} {BBBWBWBWB} {W}
\aff {254} {B} {WBBBBBWBW} {B}
\aff {255} {B} {WWWWBWWWB} {B}
\vskip 5pt
\ligne{\hfill white control\hfill}
\vskip 5pt
\aff {256} {W} {WBBWBWBWW} {W}
\aff {257} {W} {BWWBWBWWW} {W}
\aff {258} {W} {WBBBBBWWW} {W}
\aff {259} {W} {WBBBBWBWW} {W}
}\hskip 10pt
\vtop{\leftskip 0pt\parindent 0pt\hsize=\tabrulecol
\ligne{\hfill white control\hfill}
\vskip 5pt
\ligne{\hfill white control\hfill}
\aff {260} {W} {BBBBWWWWW} {W}
\aff {261} {W} {BWWBBBWWW} {B}
\aff {262} {W} {WBBWBWBBW} {W}
\aff {263} {B} {BWWBWBWWW} {W}
\aff {264} {W} {BBBBBBWWW} {B}
\aff {265} {W} {WBBWBWBBB} {W}
}\hskip 10pt
\vtop{\leftskip 0pt\parindent 0pt\hsize=\tabrulecol
\ligne{\hfill black control\hfill}
\vskip 5pt
\ligne{\hfill black control\hfill}
\aff {260} {W} {BBBBWWWWW} {W}
\aff {266} {W} {BBWBBBWWW} {B}
\aff {267} {B} {BBWBWBWWW} {W}
\aff {268} {B} {BBBBBBWWW} {W}
}
\hfill}
\vskip 9pt
\trfn
\vskip 10pt
} 

The colour of the controller is changed by the 
arrival of the appropriate signal: we remind the reader that the signal has the form of 
a simple locomotive. Table~\ref{rulesctrl} gives the rules corresponding to the crossing 
of the controller by the locomotive and to the arrival of a signal for changing the 
colour of the controller. As we shall see, the same signal changes the black colour 
of~1(3) to white and conversely.

\vtop{
\begin{tab}\label{execctrl}
\leurre
Execution of the rules
for the motion of the locomotive through the controller.
\end{tab}
\vskip 2pt
\trep
\vskip 8pt
\ligne{\hfill locomotive crossing the black controller: \hfill
\hbox to 120pt{\hfill the white one:\hfill}\hfill}
\vskip 5pt
\ligne{\hfill
\vtop{\hsize=200pt
\ligne{\hfill\HH{}
\HH{{7$_8$} }\HH{{6$_7$} }\HH{{1$_7$} }\HH{{0$_0$} }\HH{{1$_4$} }\HH{{2$_5$} }
\HH{{7$_5$} }\HH{{1$_3$} }
\hfill}
\ligne{\hfill\HH{2}
\HH{\Rr{23}}\HH{\Rr{17}}\HH{185}\HH{241}\HH{44}\HH{5}\HH{158}\HH{244}
\hfill}
\ligne{\hfill\HH{3}
\HH{26}\HH{\Rr{23}}\HH{\Rr{189}}\HH{241}\HH{44}\HH{5}\HH{158}\HH{244}
\hfill}
\ligne{\hfill\HH{4}
\HH{5}\HH{26}\HH{\Rr{193}}\HH{\Rr{246}}\HH{44}\HH{5}\HH{158}\HH{244}
\hfill}
\ligne{\hfill\HH{5}
\HH{5}\HH{5}\HH{200}\HH{\Rr{248}}\HH{\Rr{47}}\HH{5}\HH{158}\HH{251}
\hfill}
\ligne{\hfill\HH{6}
\HH{5}\HH{5}\HH{185}\HH{253}\HH{\Rr{51}}\HH{\Rr{17}}\HH{158}\HH{254}
\hfill}
\ligne{\hfill\HH{7}
\HH{5}\HH{5}\HH{185}\HH{241}\HH{53}\HH{\Rr{23}}\HH{\Rr{165}}\HH{244}
\hfill}
\ligne{\hfill\HH{8}
\HH{5}\HH{5}\HH{185}\HH{241}\HH{44}\HH{26}\HH{\Rr{170}}\HH{244}
\hfill}
\ligne{\hfill\HH{9}
\HH{5}\HH{5}\HH{185}\HH{241}\HH{44}\HH{5}\HH{177}\HH{244}
\hfill}
}\hskip 10pt
\vtop{\hsize=120pt
\ligne{\hfill\HH{}
\HH{{7$_8$} }\HH{{6$_7$} }\HH{{1$_7$} }\HH{{0$_0$} }
\hfill}
\ligne{\hfill\HH{2}
\HH{\Rr{23}}\HH{\Rr{17}}\HH{185}\HH{256}
\hfill}
\ligne{\hfill\HH{3}
\HH{26}\HH{\Rr{23}}\HH{\Rr{189}}\HH{256}
\hfill}
\ligne{\hfill\HH{4}
\HH{5}\HH{26}\HH{\Rr{193}}\HH{259}
\hfill}
\ligne{\hfill\HH{5}
\HH{5}\HH{5}\HH{185}\HH{256}
\hfill}
}
\hfill}
\vskip 9pt
\trfn
\vskip 10pt
}

Table~\ref{rulesctrl} gives the rules which control the motion of the locomotive
whatever the colour of the controller. It also contains the rules which
control the change of colour of the controller. Table~\ref{execctrl} indicates for each
cell of the track crossing the controller which rule is applied at each time. The table
is split into two sub-tables: one for the case when the controller is black, the
other for the case when it is white.

Consider the case of a passage of the locomotive when the locomotive passes through a
black controller. The passage through the cells~7(8) and 6(7) makes use of ordinary
rules of the tracks, namely the sequence of rules~5, 17, 23 and~26. The cell~1(7)
makes use of the rules~185, 189, 193 and 200 as the same cell in the selector. 
Here, as the milestones of the cell~1(7) do not change, rule~205 is not used
here.

The conservative rule for the central cell 0(0) is rule~241, 
\laff {W} {WBBWBWBWB} {W} {,}. The side~1 is here the next cell on the track, {\it i.e.}
the cell~1(4). Next, rule~246, \laff {W} {WBBBBWBWB} {B} {,} 
can see the locomotive in neighbour~4, {\it i.e.} the cell 1(7), letting it enter 
the cell. Then, rule~248, \laff {B} {WBBWBWBWB} {W} {,} changes the cell~0(0) to
white and rule~253, \laff {W} {BBBWBWBWB} {W} {,} witnesses that the locomotive,
now in neighbour~1, {\it i.e.} the cell~1(4), is leaving the central cell, so that
it goes on on the track. We can see that the cell~0(0) behaved like a cell of the
tracks: the locomotive crossed it and its milestones were unchanged. After this time,
rule~241 again manages the controller. The cell 2(7) is again applied the rules of the
tracks, and the cell 1(7) is applied the rules which were applied to the cell~1(8)
in the fork. This is illustrated by the first two rows of Figure~\ref{control}.

When the locomotive passes through a white controller, the arrival is identical
for the cells~7(8), 6(7) and 1(7). As the controller is white, the conservative
rule for the cell~0(0) is now rule~256, \laff {W} {WBBWBWBWW} {W} {,} where
we can see that its neighbour~9 is white while it is black in rule~241. When the
locomotive is in the cell~1(7), rule~259, \laff {W} {WBBBBWBWW} {W} {,}
applies to~0(0). The locomotive is in neighbour~4 as expected and we can see that
the cell remains white. This cancels the locomotive, which is conformal to the
expected behaviour of the controller. This can be checked in the third row of
Figure~\ref{control}.

\vtop{
\ligne{\hfill
\includegraphics[scale=0.55]{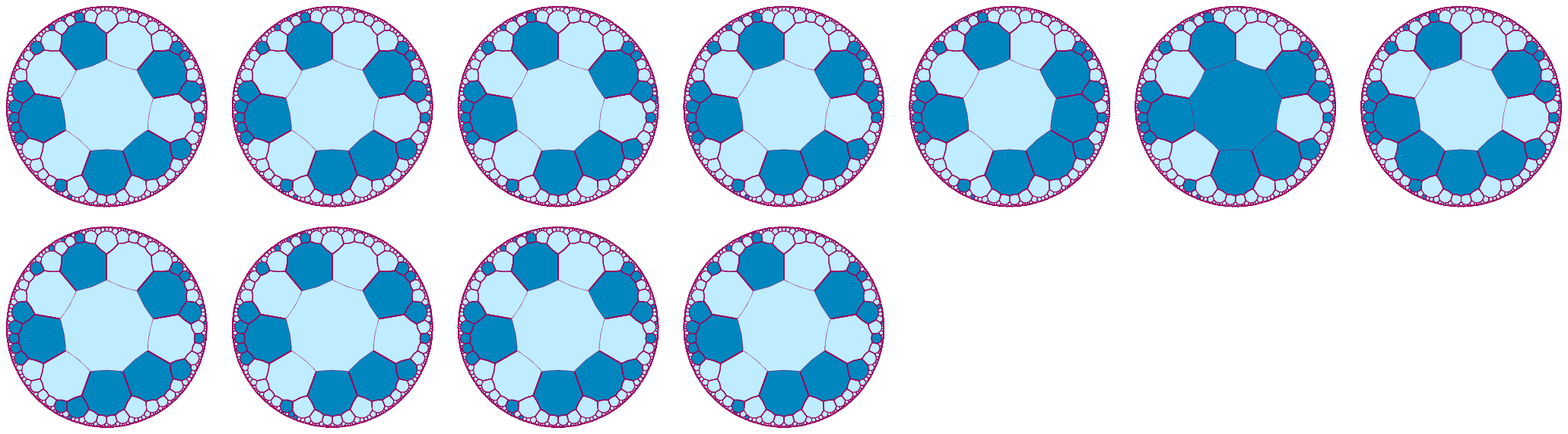} 
\hfill}
\ligne{\hfill
\includegraphics[scale=0.55]{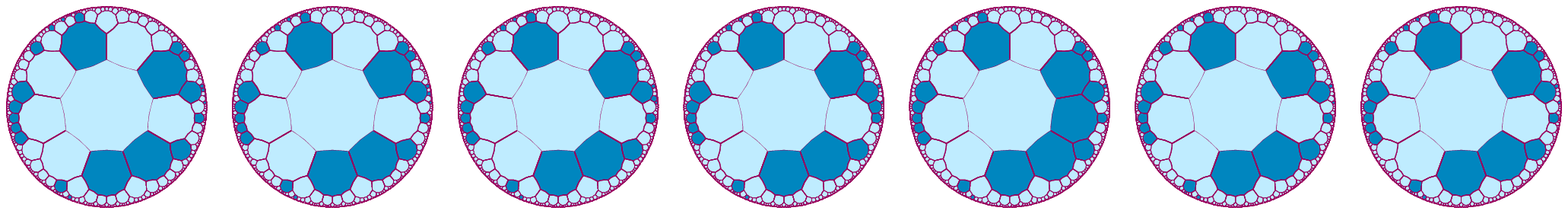} 
\hfill}
\vspace{-5pt}
\begin{fig}\label{control}
\leurre
The controller of the flip-flop and of the active memory switch. Above, the
controller is black: the locomotive passes without problem. Below, the controller is
white: it stops the locomotive which vanishes.
\end{fig}
}

\ligne{\hfill 
\vtop{  
\hsize=250pt
\begin{tab}\label{execctrls}
\leurre
Execution of the
rules for the signal changing the colour in the controller of the active switches.
\end{tab}
\vskip 2pt
\trep
\vskip 8pt
\ligne{\hfill 
\vtop{\hsize=120pt 
\ligne{\hfill from white to black\hfill}
\vskip 4pt
\ligne{\hfill\HH{}
\HH{{6$_1$} }\HH{{2$_2$} }\HH{{1$_2$} }\HH{{1$_3$} }
\hfill}
\ligne{\hfill\HH{2}
\HH{53}\HH{\Rr{51}}\HH{\Rr{261}}\HH{258}
\hfill}
\ligne{\hfill\HH{3}
\HH{44}\HH{53}\HH{\Rr{263}}\HH{\Rr{264}}
\hfill}
\ligne{\hfill\HH{4}
\HH{44}\HH{44}\HH{243}\HH{244}
\hfill}
\ligne{\hfill\HH{5}
\HH{44}\HH{44}\HH{243}\HH{244}
\hfill}
}\hskip 10pt
\vtop{\hsize=120pt
\ligne{\hfill from black to white\hfill}
\vskip 4pt
\ligne{\hfill\HH{}
\HH{{6$_1$} }\HH{{2$_2$} }\HH{{1$_2$} }\HH{{1$_3$} }
\hfill}
\ligne{\hfill\HH{2}
\HH{53}\HH{\Rr{51}}\HH{\Rr{266}}\HH{244}
\hfill}
\ligne{\hfill\HH{3}
\HH{44}\HH{53}\HH{\Rr{267}}\HH{\Rr{268}}
\hfill}
\ligne{\hfill\HH{4}
\HH{44}\HH{44}\HH{257}\HH{258}
\hfill}
\ligne{\hfill\HH{5}
\HH{44}\HH{44}\HH{257}\HH{258}
\hfill}
}
\hfill} 
\vskip 9pt
\trfn
\vskip 10pt
} 
\hfill}

Now, consider the case when a signal arrives to the controller. In both cases,
this arrival means that the colour of the controller has to be changed to the opposite
one. Table~\ref{rulesctrl} contains the rules which manage this working of the 
controller. Table~\ref{execctrls} shows where the rules are applied and at which
time. Figure~\ref{controls} illustrates the arrival of the signal in both cases:
when the controller is white and when it is black. In both cases, the figure shows
that the signal makes the controller change its colour.

The track through which the signal arrives in the form of a simple locomotive
passes through the cells 21(1), 5(1), 6(1), 2(2), arriving at 1(2). Table~\ref{execctrls}
displays the rules for the cells~6(1), 2(2) and 1(2) only. On the cells~6(1) and~2(2),
we have the standard rules for the tracks 44, 47, 51 and 53. The rules for
the cell~1(2) are 243, \laff {W} {BBWBWBWWW} {W} {,} the conservative rule when the 
controller is black, and then rule~266, \laff {W} {BBWBBBWWW} {B} {,} can see the
locomotive in its neighbour~5, {\it i.e.} the cell 2(2), making it enter the cell,
and then rule~267, \laff {B} {BBWBWBWWW} {W} {,} which makes the cell turn back to white.
At last, rule~257, \laff {W} {BWWBWBWWW} {W} {,} witnesses that its neighbour~2,
the cell~1(3), turned to white, so that rule~257 is the conservative rule of the
cell~1(2) when the controller is white. Note that the conservative rule for the 
cell~1(3) is rule~244, \laff {B} {WBBBBBWWW} {B} {,} when the controller is black. 
When the signal is in 1(2), rule~268, \laff {B} {BBBBBBWWW} {W} {,} makes the
cell turn to white. Then, rule~258, \laff {W} {WBBBBBWWW} {W} {,} applies, which
is the conservative rule of 1(3) when the controller is white.

When the controller is white, the same rules apply to the cells 6(1) and 2(2) when they
are crossed by the signal. As the signal is white, the conservative rule for 1(2)
is rule~257. Next, rule~261, \laff {W} {BWWBBBWWW} {B} {,} is the rule which allows 
the locomotive to enter the cell~1(2) as its neighbour~2, the cell 1(3), 
is now white. Then, when the signal reached the cell 1(2), rule~263, 
\laff {B} {BWWBWBWWW} {W} {,} applies, restoring the white colour in that cell.
Next, the rule~243 witnesses that its neighbour~2, {\it i.e.} the cell 1(3), is
now black, and so that rule applies, as it is the conservative rule for 1(2)
when the controller is black, as already noted. As seen previously, the conservative 
rule for~1(3) when the controller is white is rule~258. Table~\ref{execctrls} show us 
that when the signal is
in the cell 1(2), rule~264, \laff {W} {BBBBBBWWW} {B} {,} applies, which makes the
cell~1(3) turn to black. After that time, rule~244 applies as it is the conservative
rule for 1(3) when the controller is black.

\vtop{
\ligne{\hfill
\includegraphics[scale=0.55]{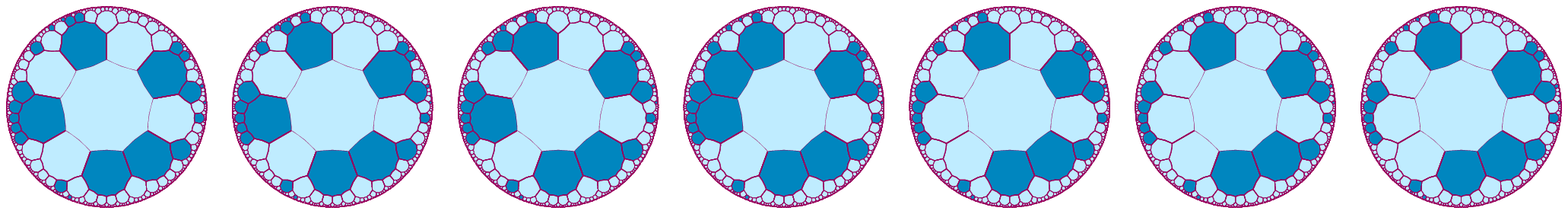} 
\hfill}
\ligne{\hfill
\includegraphics[scale=0.55]{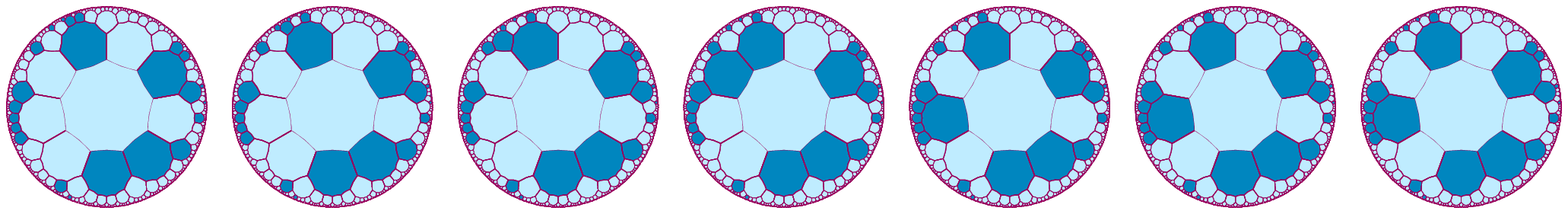} 
\hfill}
\begin{fig}\label{controls}
\leurre
The controller of the flip-flop and of the active memory switch. The signal 
arrives to change the selection. Above, the controller is changed to white. Below, 
it is changed to black.
\end{fig}
}

\subsection{The passive memory switch}

We arrive to the last structure of the circuit. As seen in Section~\ref{scenar},
it is enough to implement the specific controlling device which we called there
\textbf{sensor}. The structure is illustrated by Figure~\ref{stab_ctrlsgn}.
Below, Table~\ref{rulescaptcr} gives the rules for managing both the passage of 
a locomotive and the change of the colour of the sensor. Remember that this colour 
is given by the cell~1(1) in Figure~\ref{stab_ctrlsgn}.

The structure of the sensor tries to re-use as most as possible of the rules applied
to the controller of the active switches. We try to do that although the 
working of the sensor is opposite to that of the controller. Indeed, when the controller
is black, it let the locomotive go while this is not the case when the sensor is black
as can be seen in Figure~\ref{captcr} and in Table~\ref{execcaptcrb}.

\ligne{\hfill
\vtop{  
\hsize=250pt
\begin{tab}\label{rulescaptcr}
\leurre
Rules for the sensor of the passive memory switch.
\end{tab}
\vskip 2pt
\trep
\vskip 8pt
\ligne{\hfill\hbox to 150pt{\hfill passage of a locomotive\hfill}
\hskip 10pt\hbox to \tabrulecol{\hfill signal\hfill}
\hfill}
\vskip 4pt
\ligne{\hfill  
\vtop{\leftskip 0pt\parindent 0pt\hsize=\tabrulecol
\ligne{\hfill black sensor\hfill}
\aff {269} {W} {BWBBWBWBW} {W}
\aff {270} {W} {WBBWBWWWW} {W}
\aff {271} {W} {BBBBWBWBW} {W}
\aff {272} {B} {WBBWWWWWB} {B}
\aff {273} {B} {BWBWWWWWB} {B}
}\hskip 10pt
\vtop{\leftskip 0pt\parindent 0pt\hsize=\tabrulecol
\ligne{\hfill white sensor\hfill}
\aff {274} {W} {BWBBWWWBW} {W}
\aff {275} {W} {BBBBWWWBW} {B}
\aff {276} {B} {BWBBWWWBW} {W}
\aff {277} {W} {WBBBBBWWB} {B}
\aff {278} {W} {BWBBWBWWW} {W}
\aff {279} {W} {BWBBWBWBB} {W}
}\hskip 10pt
\vtop{\leftskip 0pt\parindent 0pt\hsize=\tabrulecol
\aff {280} {B} {WBBWBWWWW} {W}
\aff {281} {W} {BWBBBBWBW} {W}
}
\hfill} 
\vskip 9pt
\trfn
\vskip 10pt
} 
\hfill}

\ligne{\hfill 
\vtop{
\hsize=250pt
\begin{tab}\label{execcaptcrb}
\leurre
Execution of the rules managing the black sensor.
\end{tab}
\vskip 2pt
\trep
\vskip 8pt
\ligne{\hfill 
\vtop{\hsize=105pt
\ligne{\hfill locomotive\hfill}
\vskip 4pt
\ligne{\hfill\HH{}
\HH{{7$_7$} }\HH{{6$_6$} }\HH{{1$_6$} }\HH{{0$_0$} }\HH{{1$_1$} }\hfill}
\ligne{\hfill\HH{2}
\HH{\Rr{23}}\HH{\Rr{17}}\HH{6}\HH{269}\HH{244}
\hfill}
\ligne{\hfill\HH{3}
\HH{26}\HH{\Rr{23}}\HH{\Rr{29}}\HH{269}\HH{244}
\hfill}
\ligne{\hfill\HH{4}
\HH{5}\HH{26}\HH{\Rr{18}}\HH{271}\HH{244}
\hfill}
\ligne{\hfill\HH{5}
\HH{5}\HH{5}\HH{6}\HH{269}\HH{244}
\hfill}
}
\hskip 20pt
\vtop{\hsize=105pt
\ligne{\hfill signal\hfill}
\vskip 4pt
\ligne{\hfill\HH{}
\HH{{6$_8$} }\HH{{2$_9$} }\HH{{1$_9$} }\HH{{1$_1$} }\hfill}
\ligne{\hfill\HH{2}
\HH{53}\HH{\Rr{51}}\HH{\Rr{266}}\HH{244}
\hfill}
\ligne{\hfill\HH{3}
\HH{44}\HH{53}\HH{\Rr{267}}\HH{\Rr{268}}
\hfill}
\ligne{\hfill\HH{4}
\HH{44}\HH{44}\HH{257}\HH{258}
\hfill}
\ligne{\hfill\HH{5}
\HH{44}\HH{44}\HH{257}\HH{258}
\hfill}
}
\hfill} 
\vskip 9pt
\trfn
\vskip 10pt
} 
\hfill}

\ligne{\hfill
\vtop{\hsize=200pt
\begin{tab}\label{execcaptcrw}
\leurre
Execution of the rules controlling the
passage of the locomotive through a white sensor.
\end{tab}
\vskip 2pt
\trep
\vskip 8pt
\ligne{\hfill\HH{}
\HH{{7$_7$} }\HH{{6$_6$} }\HH{{1$_6$} }\HH{{0$_0$} }\HH{{1$_4$} }\HH{{2$_5$} }\HH{{7$_5$} }\HH{{1$_1$} }\hfill}
\ligne{\hfill\HH{2}
\HH{\Rr{23}}\HH{\Rr{17}}\HH{6}\HH{274}\HH{44}\HH{5}\HH{158}\HH{258}
\hfill}
\ligne{\hfill\HH{3}
\HH{26}\HH{\Rr{23}}\HH{\Rr{29}}\HH{274}\HH{44}\HH{5}\HH{158}\HH{258}
\hfill}
\ligne{\hfill\HH{4}
\HH{5}\HH{26}\HH{\Rr{18}}\HH{\Rr{275}}\HH{44}\HH{5}\HH{158}\HH{258}
\hfill}
\ligne{\hfill\HH{5}
\HH{5}\HH{5}\HH{24}\HH{\Rr{276}}\HH{\Rr{47}}\HH{5}\HH{158}\HH{\Rr{277}}
\hfill}
\ligne{\hfill\HH{6}
\HH{5}\HH{5}\HH{6}\HH{279}\HH{\Rr{51}}\HH{\Rr{17}}\HH{158}\HH{244}
\hfill}
\ligne{\hfill\HH{7}
\HH{5}\HH{5}\HH{6}\HH{269}\HH{53}\HH{\Rr{23}}\HH{\Rr{165}}\HH{244}
\hfill}
\ligne{\hfill\HH{8}
\HH{5}\HH{5}\HH{6}\HH{269}\HH{44}\HH{26}\HH{\Rr{170}}\HH{244}
\hfill}
\ligne{\hfill\HH{9}
\HH{5}\HH{5}\HH{6}\HH{269}\HH{44}\HH{5}\HH{177}\HH{244}
\hfill}
\vskip 9pt
\trfn
\vskip 10pt
}
\hfill}

However, 
the rules controlling the signal in the controller can be formally used in the
same way with another interpretation. The point is that the cell~1(1) must behave
in an opposite way with respect to the cell 1(3) in the controller when the
locomotive is present. Note that the conservative rules for 1(1) are the same
as those for 1(3) in the controller, namely rule~244 for a black sensor and rule~258
for a white one. From Table~\ref{execcaptcrb} we can see that the black sensor
is turned to a white one by rule~268 which detect the presence of the signal
in neighbour~1, exactly as the action of the same rule on 1(3) in the controller.
From Table~\ref{execcaptcrw}, we can see that the white sensor turns to black under
the action of rule~277, \laff {W} {WBBBBBWWB} {B} {,} which is triggered by the
occurrence of the locomotive in neighbour~9, {\it i.e.} the cell 0(0) by which 
the locomotive actually passes. To conclude with the passage of the signal, only
in the case when the sensor is black, we note from Tables~\ref{execcaptcrb}
and~\ref{execctrls} that the rules for the cells 6(8), 2(9) and~1(9) in the sensor
and those used for the cells 6(1), 2(2) and~1(2) respectively in the controller,
see page~\pageref{execctrls},
are exactly the same. Also see the last row of Figure~\ref{captcr}.

\vtop{
\ligne{\hfill
\includegraphics[scale=0.55]{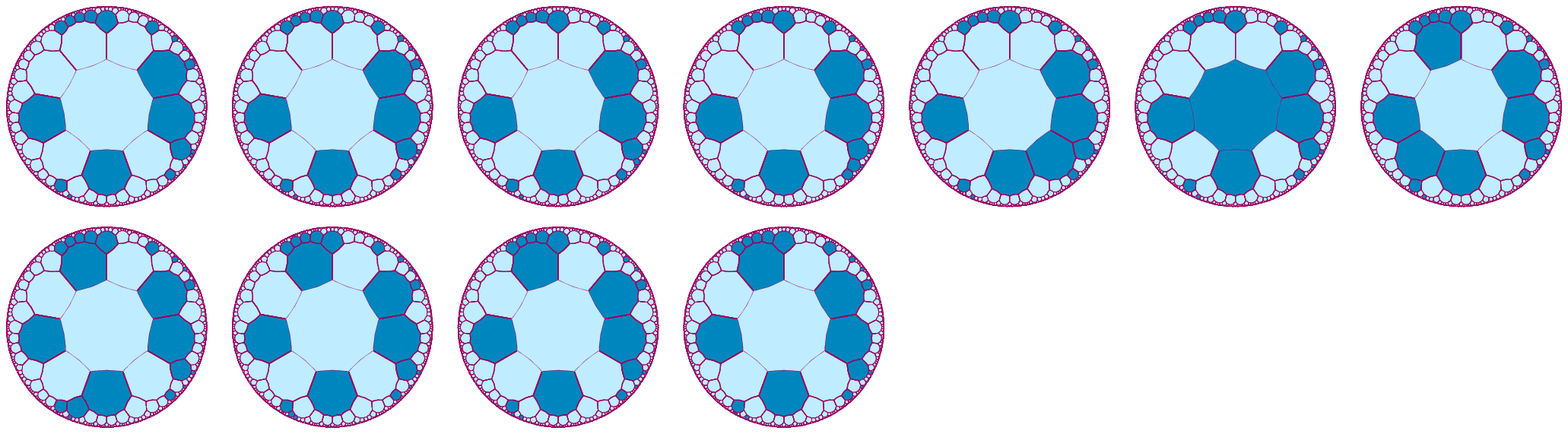}
\hfill}
\vspace{-10pt}
\ligne{\hfill
\includegraphics[scale=0.55]{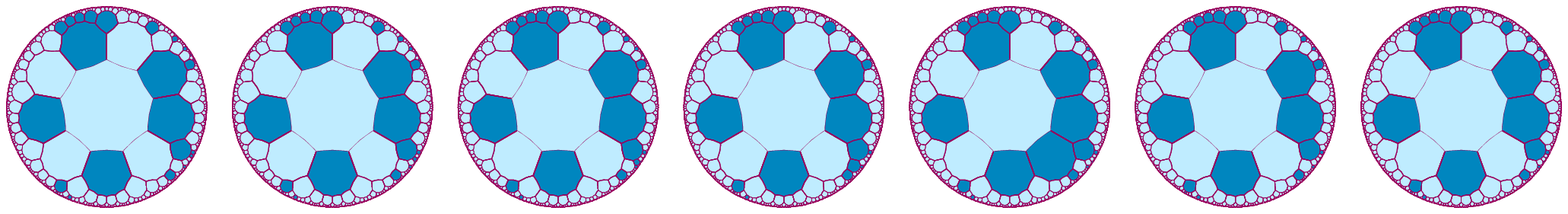}
\hfill}
\vspace{-10pt}
\ligne{\hfill
\includegraphics[scale=0.55]{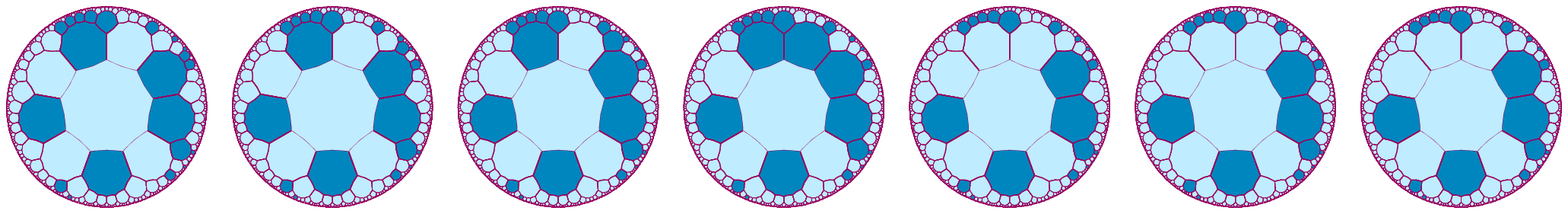}
\hfill}
\begin{fig}\label{captcr}
\leurre
The controller-sensor of the passive memory switch. Above: a single locomotive passively
crossed the switch through its selected track. Middle row: the locomotive crossed the passive
switch through the non-selected track. Below: the change of signal induced by a passive crossing
through the non-selected track.
\end{fig}
}
\vskip 10pt

Let us now have a look at the passage of the locomotive, which is illustrated by the 
first two rows of Figure~\ref{captcr} when the sensor is black and by its third row
when the sensor is white.
The arrival to the cell~0(0) is controlled by the rules for the tracks on the
cells 7(7), 6(6) and 1(6), see both Tables~\ref{execcaptcrb} and~\ref{execcaptcrw}.
For the cell~0(0) there are two conservative rules~: one for the black sensor,
rule~269, \laff {W} {BWBBWBWBW} {W} {,} the other for the white sensor, 
rule~274, \laff {W} {BWBBWWWBW} {W} {.} When the sensor is black, after rule~269,
rule~271, \laff {W} {BBBBWBWBW} {W} {,} is triggered by the occurrence of the
locomotive in the cell~1(6) which is neighbour~2 of the cell 0(0). That rule keeps
the cell~0(0) white. Accordingly, the locomotive which arrived to the sensor
is stopped, which corresponds to the expected working of the
passive switch. When the sensor is white, after rule~274, when the locomotive is
in the cell 1(6), the neighbour~2 of~0(0), rule~275, \laff {W} {BBBBWWWBW} {B},
is applied which makes the locomotive enter the cell. Next, rule~276,
\laff {B} {BWBBWWWBW} {W} {,} makes the cell return to white, then rule~279,
\laff {W} {BWBBWBWBB} {W} {,} witnesses that the locomotive is now in neighbour~9,
the cell 1(4), which is the next cell of the track. Rule~279 also shows us that 
the neighbour~6 of 0(0), {\it i.e.} the cell 1(1), is now black, which proves that
after the passage of the locomotive, the sensor is now black. Consequently, 
rule~269 is from now applied as it is the conservative rule for~0(0) when the sensor
is black. This is the case until a signal again changes the colour of~1(1). Notice 
that the locomotive which goes on is now a signal which goes to the active memory switch
in order to trigger the change of colour in the controllers of the active switch.
The rest of the track involves rules which we have already studied. 
As we already know the rules involved by the cell 1(1) whatever its colour,
our study is now complete. Accordingly, Theorem~\ref{letheo} is proved.
\hfill\boxempty

\section*{Conclusion}

Several questions are raised by this result. The number of rules is not very high. It 
could probably be reduced by eliminating the cases when rotated forms of the same rule
appear among the rules. However, the number of such cases is not very high. Most probably,
even by eliminating such cases we remain with a number of rules which should
be bigger than 256. This completely rules out the possibility of implementing the
same construction in the tessellation $\{7,3\}$. 
Another question is how small can be the number of neighbours for a simulation which 
would keep rotation invariance? Still hard work ahead.


\begin{thebibliography}{5}
\newcount\bibi\bibi=1

\bibitem{mmbook2}
M. Margenstern,
{\it Cellular Automata in Hyperbolic Spaces, vol. $2$, Implementation and computations},
Collection: {\it Advances in Unconventional Computing and Cellular Automata},
Editor: Andrew Adamatzky,
Old City Publishing, Philadelphia, (2008), 360p.

\bibitem{mmbook3}
M. Margenstern,
{\it Small Universal Cellular Automata in Hyperbolic Spaces: A Collection of Jewels},
Springer Verlag, (2013), 331p.

\bibitem{mmarXiv1510}
M. Margenstern,
A weakly universal cellular automaton in the pentagrid with three states,
{\it arXiv}:1510.09129, (2015), 40p.

\bibitem{mmarXiv1512}
M. Margenstern,
A weakly universal cellular automaton on the pentagrid with two states,
{\it arXiv}:1512.07988v1, (2015), 38p.

\bibitem{mmJCA2016}
M. Margenstern,
A weakly universal cellular automaton with 2 states in the tiling $\{11,3\}$,
{\it Journal of Cellular Automata}, {\bf 11}(2-3), (2016), 113-144. 

\bibitem{mmchapter}
M. Margenstern,
Cellular Automata in Hyperbolic Spaces,
chapter in {\it Advances in Unconventional Computing}, Springer, to appear.

\bibitem{mmarXiv1605}
M. Margenstern,
A new system of coordinates for the tilings $\{p,3\}$ and $\{p$$-$$2,4\}$,
{\it arXiv}:1605.03753, (2016), 33p.

\bibitem{mmysENTCS}
M. Margenstern, Y. Song,
A universal cellular automaton on the ternary heptagrid,
{\it Electronic Notes in Theoretical Computer Science},
{\bf 223}, (2008), 167-185.

\bibitem{mmysPPL}
M. Margenstern, Y. Song,
A new universal cellular automaton on the pentagrid,
{\it Parallel Processing Letters},
{\bf 19}(2), (2009), 227-246.

\bibitem{minsky}
M.L. Minsky,
{\it Computation: Finite and Infinite Machines}, Prentice-Hall,
Englewood Cliffs, NJ, 1967.


\bibitem{stewart}
I.~Stewart, A Subway Named Turing, Mathematical Recreations in {\it Scientific
American}, (1994), 90-92.


\end{thebibliography}
\end{document}